\documentclass[a4paper,fleqn,usenatbib,useAMS]{mnras}
\pdfoutput=1

\usepackage[T1]{fontenc}
\usepackage{ae,aecompl}
\usepackage{multirow}
\usepackage[pdftex]{graphicx}
\usepackage{ctable}
\usepackage{bm}

\newcommand{\acknowledgments}{\begin{small}\section*{Acknowledgments}\end{small}}

\title[Rotation curve fitting and its attraction to cores]{Rotation curve fitting and its fatal attraction to cores in realistically simulated galaxy observations}

\author[Juan C. B. Pineda et al.]{%
\parbox[t]{\textwidth}{Juan C.B. Pineda$^{1}$\thanks{Contact e-mail: basto-pineda@usp.br\newline juan.basto.pineda@gmail.com}, Christopher C. Hayward$^{2,3,4}$, Volker Springel$^{5,6}$ and Claudia Mendes de Oliveira$^1$}
\vspace*{6pt} \\
$^{1}$Instituto de Astronomia, Geof\'isica e Ci\^encias Atmosf\'ericas, Universidade de S\~ao Paulo, R. do Mat\~ao 1226, 05508-090 S\~ao Paulo, Brazil \\
$^2$Center for Computational Astrophysics, Flatiron Institute, 162 Fifth Avenue, New York, NY 10010, USA\\
$^{3}$TAPIR 350-17, California Institute of Technology, 1200 E. California Boulevard, Pasadena, CA 91125, USA \\
$^4$Harvard--Smithsonian Center for Astrophysics, 60 Garden Street, Cambridge, MA 02138, USA \\
$^{5}$Heidelberger Institut f\"ur Theoretische Studien, Schloss--Wolfsbrunnenweg 35, 69118 Heidelberg, Germany \\
$^6$Zentrum f\"ur Astronomie der Universit\"at Heidelberg,
  Astronomisches Recheninstitut, M\"{o}nchhofstr. 12-14, 69120
  Heidelberg, Germany
}

\date{Submitted to MNRAS} 
\pubyear{2016}

\begin{document}
      
\label{firstpage}
\pagerange{\pageref{firstpage}--\pageref{lastpage}}
\maketitle

\begin{abstract}
We study the role of systematic effects in observational studies of
the cusp-core problem under the minimum disc approximation using a
suite of high-resolution (25-pc softening length) hydrodynamical
simulations of dwarf galaxies. We mimic realistic kinematic
observations and fit the mock rotation curves with two analytic models
commonly used to differentiate cores from cusps in the dark matter
distribution.  We find that the cored pseudo-isothermal sphere
(ISO) model is strongly favoured by the reduced $\chi^2_\nu$
of the fits in spite of the fact that our simulations contain cuspy
Navarro-Frenk-White profiles (NFW).  We show that even idealized
measurements of the gas circular motions can lead to the incorrect
answer if velocity underestimates induced by pressure support,
with a typical size of order $\sim$5 km s$^{-1}$ in the central
kiloparsec, are neglected. Increasing the spatial resolution of
the mock observations leads to more misleading results because the
inner region, where the effect of pressure support is most
significant, is better sampled.  Fits to observations with a
spatial resolution of 100 pc (2 arcsec at 10 Mpc) favour the ISO
model in 78-90 per cent of the cases, while at
800-pc resolution, 40-78 per cent of the galaxies indicate the
fictitious presence of a dark matter core.  The coefficients of our
best-fit models agree well with those reported in observational
studies; therefore, we conclude that NFW haloes can not be ruled out
reliably from this type of analysis.

\end{abstract}


\section{Introduction}
\label{sec:Introduction}

In the last decades, cosmological numerical simulations based on the
$\Lambda$CDM concordance model have substantially improved our
understanding of the dynamical evolution of the Universe on large
scales \citep{Davis1985, MillenniumI, MillenniumII, MillenniumXXL,
  Multidark}.  More recently, it has also become possible to include
baryons in large cosmological volumes and directly follow galaxy
formation, with very promising results \citep{IllustrisII,Schaye2015}.
However, a number of well documented small-scale discrepancies between
$\Lambda$CDM and observations still remain to be understood,
including the so-called \emph{cusp-core} problem concerning the inner
structure of galactic dark matter haloes \citep{Flores1994,Moore1994}.
On one hand, dark matter (DM) haloes assembled in cosmological simulations
exhibit \emph{cuspy} radial density profiles which steeply increase
towards the center \citep{Navarro1996, Navarro1997, Moore1999b,
  Klypin2001, Navarro2004, Diemand2005, Stadel2009}.  They are fairly
well represented by the Navarro-Frenk-White (NFW) fitting formula \citep{Navarro1996},
with an asymptotic behaviour of
$\rho_{\rm inner}\propto r^{-1}$.  On the other hand, kinematic
observations of disc galaxies (via rotation curves) and dwarf
spheroidals (via stellar velocity dispersions) often seem to be
more compatible with core-like DM haloes 
ranging from $\rho_{\rm inner}\propto r^0$ to $\rho_{\rm inner}\propto r^{-0.3}$
\citep{Flores1994,deBlok2001,Salucci2001,deBlok2002,Kuzio2006,Spano2008,
Oh2011,Walker2011,Oh2015}. A number of studies have also inferred
intermediate slopes, which do not evidence constant-density cores but
still are substantially shallower than canonical cusps from simulations 
\citep{Simon2005,Adams2014}.

Several physical mechanisms have been proposed to reconcile these
findings and explain the origin of DM cores.  The current leading
picture invokes repetitive starburst episodes and the associated supernovae
feedback to blow out central baryons and induce rapid changes in the
gravitational potential \citep{Navarro1996b, Gelato1999, Read2005,
Mashchenko2008, Governato2010, Governato2012, Pontzen2014, Madau2014,
Chan2015,Onorbe2015}.  However, many feedback implementations do not form
cores \citep{Lia2000,Gnedin2002,Ceverino2009}, which has not precluded
the formation of realistic galaxies \citep{Sawala2010,Marinacci2014}
as well as realistic populations of galaxies at $z=0$ in fully cosmological
runs \citep{IllustrisI,Schaller2015}. Also, there are lingering doubts
\citep{Garrison-Kimmel2013} regarding whether the available supernova
energy is actually sufficient to create cores of the alleged size.

From the observational point of view, a problematic issue of using
rotation curves to probe the DM density profiles is to accurately
account for the stellar gravitational potential, mainly because of the
uncertain mass-to-light conversion factor
\citep{vanAlbada1985,Bell2001,Bershady2010}.  One way around this
problem lies in studying dark matter-dominated systems, such as
late-type dwarf irregulars and low surface brightness (LSB) galaxies. For
these systems, it is acceptable to use the minimum disc approximation,
i.e. to ignore the existence of baryons and use the observed rotation curves at
face value to derive an indicative upper limit on the amount of DM
\citep{deBlok1997,deBlok2001,Spekkens2005,Kuzio2006}.  Furthermore,
different observations suggest that LSBs are characterized by a
comparatively quiescent evolution, which likely implies relatively
unperturbed DM haloes, making them very interesting for cosmology
\citep{deBlok1995,Impey1997,Du2015}.  Indeed, these kinds of galaxies
have been a main target of observational studies and represent some of
the most acute challenges for the $\Lambda$CDM cosmogony, as their
rotation curves are interpreted by several authors as strong evidence
for DM cores \citep{deBlok2001,deBlok2002,Simon2003,Oh2011,Oh2015}.
However, some studies
find some dwarfs and some LSB galaxies to be compatible with CDM cusps or
claim that available data simply does not allow to differentiate cusps
from cores \citep[e.g.][]{VandenBosch2000, VandenBosch2001,
  Swaters2003, Spekkens2005,Simon2005,Valenzuela2007}.
The systematic uncertainties involved in studying this problem are a
matter of active debate in the literature.

A number of effects that may lower the inner rotation curves and
potentially mask cusps and make them appear as cores have been
investigated over the years. 
For instance, early H$\,${\sc i} rotation curves
had poor spatial resolution and were considerably affected by beam
smearing. This motivated the gathering of high-resolution optical
data for which beam smearing is expected to no longer be a problem
\citep{B-Ouellette1999,Swaters2000}, though it may still play a role
in the very inner measurements, specially for distant galaxies \citep{Swaters2003}.
More recently, some surveys of H$\,${\sc i} in very nearby
galaxies, like THINGS \citep{Walter2008} and LITTLE THINGS
\citep{Hunter2012} \citep[amongst others, e.g.,][]{Begum2008, Ott2012},
have also reached the necessary resolution to alleviate
beam smearing concerns.
The first H~$\alpha$
rotation curves were obtained from long-slit spectroscopy, with the
associated risk of missing the kinematic center of the galaxy or
its major axis; fortunately, this is no longer an issue since
high-resolution velocity fields have become available
\citep{B-Ouellette2001,Garrido2002}.  Other problems that can be
assessed by means of 2D velocity maps are possible offsets between
the photometric and kinematic centers and the presence of non-circular
motions \citep{Simon2005,Oh2008}.  A detailed analysis of these
effects for a sample of 19 galaxies from THINGS was
presented by \citet{Trachternach2008}, who concluded that these effects
are rather small and the sample is hence suitable for the mass modeling
studies presented in \citet{deBlok2008}.

Projection effects related to the thickness of gaseous discs
are also potentially problematic because mixing of material along the
line-of-sight tends to lower the inferred circular velocity
\citep{Rhee2004}.  Additionally, pressure exerted by the gas can
effectively lower the gravitational radial acceleration, thus lowering
the rotational velocity needed for support.  With very few exceptions
\citep[e.g.][]{Oh2011}, pressure support corrections are usually
neglected because they are expected to be small
\citep{deBlok2002}.  Halo triaxiality has also been considered as a
possible explanation for the cusp-core discrepancy \citep{Hayashi2007},
but using both observations and numerical modelling, other authors
have argued that this is not likely to be the case
\citep{Simon2005, Kuzio2009, Kazantzidis2010, Kuzio2011}. Besides all
the potential complications already mentioned, we note that galaxies
are often irregular and present substructures such as bars, bulges,
and spiral arms. In addition, rotation curves are often wiggled,
warped, or lopsided. The interpretation of these features and their
impact on the cusp-core problem is not clear and cannot be modelled
from first principles.

A powerful approach to study these systematic effects is by means of
controlled experiments with simulated data. The first attempts in this
direction mimicked long-slit observations of analytic velocity
fields, including some uncertainties.  In this way, \citet{deBlok2003}
concluded that it should be possible to recognize the real steepness
of a DM halo from its measured rotation curve and that no single
systematic effect can account for the cusp-core difference.  However,
following the same approach, \citet{Swaters2003} concluded that systematic
effects tend to erase the signature of cusp-like haloes and that
rotation curve analyses cannot compellingly rule out the presence of cusps.
Similarly, \citet{Spekkens2005} found cuspy DM haloes to be consistent
with the observed distribution of slopes once systematic effects are
taken into account.  \citet{Dutton2005} pointed out that uncertainties
inherent in mass modelling studies prevent them from setting firm
constraints on the shape of DM haloes.  \citet{Kuzio2009} did not use
analytic velocity fields but instead integrated orbits numerically.
They mimicked 2D integral field unit (IFU) data using a reference sample of
observations and concluded that if present at all, NFW haloes should be still
recognizable.  However, although these kinds of models provided useful
insights into the problem, they were clearly oversimplifications.  For
instance, with the exception of \citet{Dutton2005}, these studies
assumed infinitely thin massless discs, and none included
hydrodynamics.

\citet{Rhee2004} brought analyses of observations
and simulations closer together. They performed N-body simulations of
stellar discs inside cuspy haloes and \emph{observed} them in a
realistic manner, concluding that projection effects, small bulges, and
bars can often lead to false detections of DM cores.
\citet{Valenzuela2007} confirmed these results using simulated analogs
of the dwarf galaxies NGC 3109 and NGC 6822.  They also compared a pure
N-body simulation with hydrodynamical runs and suggested that pressure
support related to stellar and supernova feedback can also produce the
illusion of cores.  A different result from similar simulations was
presented by \citet{Kuzio2011}. They concluded that the signatures of
cores, cusps, and triaxiality in DM haloes should be clearly
detectable in observed velocity fields.  In recent work,
\citet{Oh2011b} analysed mock observations of two dwarf galaxies formed
in zoom-in cosmological simulations that undergo the supernovae-driven
cusp-to-core transformation \citep{Governato2010} and compared them with a
sample of dwarfs from the THINGS survey. \citet{Oh2011b} found that
their mock observations trace the true rotation curves and true surface
density profiles of the simulated galaxies fairly well. They also state
that their simulations are a good match to real galaxies regarding these
quantities, but \citet{Oman2015} showed that the alleged agreement is only
apparent in some cases. Moreover, \citet{Oman2015} demonstrated that
the diversity of dwarf galaxy rotation curves is much greater than that
of galaxies formed in cosmological simulations, posing a new challenge
to any model trying to solve the cusp-core problem. Recently, \citet{Read2016}
addressed this question using mock observations from a suit of very high-resolution
(4 pc) simulations and indicate that at least part of the observed diversity
can be explained from the starburst cycles of galaxies and their influence
on the dynamical state of the galaxy.

Given the body of in part contradictory conclusions in the literature,
it is apparent that further investigations of potential systematic
effects in observational inferences about the cusp-core problem are
worthwhile. In this work, we carry out a comprehensive theoretical
study of the kinematics of a set of simulated dwarf galaxies by
carefully creating synthetic observations that are then analysed
in exactly the same manner as real data. In this way, we can
determine the effects of different sources of error in the context of
the cusp-core problem and assess to what extent these errors can lead to
misleading inferences about the structure of the analysed galaxies.
Here we focus on the minimum disc approximation, exploting the 
fact that our models are dark matter dominated at all radii,
and we only use rotation curve fitting to classify cusps and cores;
a complementary analysis using rotation curve inversion methods
will be presented in a separate paper, as they demand a different approach.

The outline of the paper is as follows.  In Sec.~\ref{sec:Simulations}
we present our simulated galaxy sample and summarize the
simulation methodology.  In Sec.~\ref{sec:Analysis} we introduce the
methods we use to analyse the information from the snapshots.  In
Sec.~\ref{sec:Results} we present the dynamical evolution of the different
components, the mock observed rotation curves, and the results from the
rotation curve fitting methods that aim to distinguish cusps from
cores.  In Sec.~\ref{sec:Discussion} we discuss the systematic
effects that influence these results, aiming to disentangle the
effects of spatial resolution, inclination, pressure support, etc.
Finally, in Sec.~\ref{sec:Conclusions} we present our conclusions
and a summary of our results.

\section{Simulations}							%
\label{sec:Simulations}							%

\subsection{Physical characteristics of the simulated galaxies}		%
\label{subsec:Galaxies}							%

We simulate six dwarf galaxies in isolation at high resolution using
the {\small N}-body+smoothed-particle hydrodynamics (SPH)
code {\sc gadget-2} \citep{Springel2005}.  Each
simulation is composed of a dark matter halo plus an exponential disc
of stars and gas, as summarised in Table~\ref{tab:simulations}.  The
methods for creating the numerical realizations of the initial
conditions are essentially the same as described in
\citet{Springel2005} and \citet{Cox2004}. Here we hence focus on the
motivation for choosing the physical parameters of the galaxies and
only briefly mention the most relevant technical details.

Galaxy models `Dwarf1' to `Dwarf4' (or simply D1,..., D4) are
representative of the bulk of the properties of observed late-type dwarfs
and LSBs
\citep[e.g.][]{deBlok2002,Spekkens2005,deBlok2008,Kuzio2008,Kuzio2009,Stark2009,Oh2011}.
These four galaxies are constructed to lie on the stellar and baryonic
Tully-Fisher relations (TF) of \citet{Bell2001}.
Notice that if both TF relations are required to be satisfied
simultaneously, this puts a constraint on the gas fraction as a
function of the stellar mass.  Using representative values from
\citet{Bell2001}, we define the following target relations:
\begin{eqnarray}
  \log{(M_{\star})} & = & 0.83 + 4.34\log{(V_{\rm flat})}  \label{eq:TF}, \\
  f_g & = & \frac{M_{\rm gas}}{M_{\rm gas}+M_\star} = 1-\frac{M_\star^{0.21}}{170} \label{eq:gasfrac} ,
\end{eqnarray}
where $M_\star$ and $M_{\rm gas}$ are the total masses of the stellar
and the gaseous discs, $f_g$ is the gas fraction, and $V_{\rm flat}$ is
the maximum circular velocity.  The radial stellar and gaseous
distributions drop exponentially, with surface density profiles given
by
\begin{eqnarray}
  \Sigma_\star(r) & = & \frac{M_\star}{2\pi h_0^2}e^{-r/h_0}  ,\\
  \Sigma_{\rm gas}(r) & = & \frac{M_{\rm gas}}{2\pi h_{\rm gas}^2}e^{-r/h_{\rm gas}} .
\end{eqnarray}
The stellar scale lengths are chosen to be relatively large (but still
realistic) in order to give our galaxies a low surface brightness.  We
make the gaseous discs three times more extended than the stellar ones
as suggested by observations \citep{Broeils1994}.

The vertical structure of the stellar discs follows a typical
${\rm sech}^2(z/z_0)$ model with the scale height $z_0$ equal to one fifth
of the radial scale length $h_0$, so their 3D density fields read
\begin{equation}
  \rho_\star(r,z) = \frac{M_\star}{4\pi h_0^2 z_0}e^{-r/h_0}{\rm sech}^2(z/z_0) .
\end{equation}
For gaseous discs the vertical structure is self-consistently
calculated considering the full gravitational potential and assuming
hydrodynamic equilibrium under a given equation of state that we
choose to be the multiphase model of \citet{Springel2003}.

Dark matter haloes are modelled with the cuspy NFW
density profile \citep{Navarro1996}.  We did not simulate cored haloes
because several mechanisms that might cause cusps
to appear as cores have been proposed, but no mechanisms that can
cause cores to appear as cusps are known. Moreover, to provide a
potential solution to the cusp-core problem, it is only necessary to
demonstrate that cusps can be mistaken for cores. 
The original NFW
formulation is
\begin{equation}\label{eq:nfw_dens}
  \rho_{\raisebox{-2pt}{\textrm{{\tiny NFW}}}}(r) = \frac{\rho_0}{(r/R_s)[1+(r/R_s)]^2}.
\end{equation}
In equation~(\ref{eq:nfw_dens}), $\rho_0$ is a characteristic density
and $R_s$ represents a transition radius between an inner and an outer
exponential law ($\rho_{\rm inner}\sim r^{-1}$;
$\rho_{\rm outer}\sim r^{-3}$).  An alternative parametrization
that provides easier comprehension of the halo structure is obtained by
casting the enclosed mass into a circular velocity profile,
\begin{equation}\label{eq:vnfw_v200}
  v_{\raisebox{-2pt}{\textrm{{\tiny NFW}}}}(r) = v_{\raisebox{-2pt}{\textrm{{\scriptsize 200}}}}\sqrt{\frac{\log{(1+cx)}-cx/(1+cx)}{x[\log{(1+c)}-c/(1+c)]}} , 
\end{equation}
with $v_{200}$ representing the circular velocity at $r_{200}$, the
radius at which the halo mean density is 200 times the critical density
for closure\footnote{Notice that $v_{200}$ encodes
  the halo mass through $M_{200}={v_{200}^3}/{(10GH_0)}$; we use
  $H_0=70$ km s$^{-1}$ Mpc$^{-1}$.}.  The parameter $c \equiv r_{200}/R_s$ measures the
central concentration of the mass distribution, and
$x \equiv r/r_{200}$.

For each galaxy, the halo mass is chosen as a function of the stellar
mass following the abundance matching relation of \citet{Guo2010}.
The concentrations are determined using the halo mass-concentration relation
at redshift zero \citep{Ludlow2014}, which is a fundamental outcome of
large cosmological simulations.  \citet{Guo2010}
used the \citet{Chabrier2003} initial mass function (IMF), whereas \citet{Bell2001}
used a scaled-down
\citet{Salpeter1955} IMF, giving rise to a systematic difference of 0.15 dex in
stellar mass.  For this reason, we add 0.15 dex to the actual
stellar mass when we use equations (\ref{eq:TF}) and
(\ref{eq:gasfrac}) to define the target $V_{\rm flat}$ and $f_g$,
respectively.  Compared to the stellar masses at face value, this
increment raises the target circular velocities by 8 per cent and the target
gaseous masses by 10 to 25 per cent.

Once we have defined an initial configuration, we create the corresponding
initial conditions file and check if $V_{\rm flat}$ actually
satisfies the TF relations inside the expected scatter, slightly
adjusting the halo parameters otherwise.  As a side effect of this
tuning, the less massive galaxies, D1 and D2, end up with DM haloes
that are less massive than predictions from the abundance matching
target relation.  Nevertheless, through a comparison with different
samples from the literature, \citet{Oh2011} has noted that this is not
at odds with observations of low-mass galaxies, for which the
abundance matching relation of \citet{Guo2010} is not directly
constrained but rather represents an extrapolation to smaller masses.

\begin{figure}
\begin{minipage}{\linewidth}
\centering
\includegraphics[width=4cm,height=4cm]{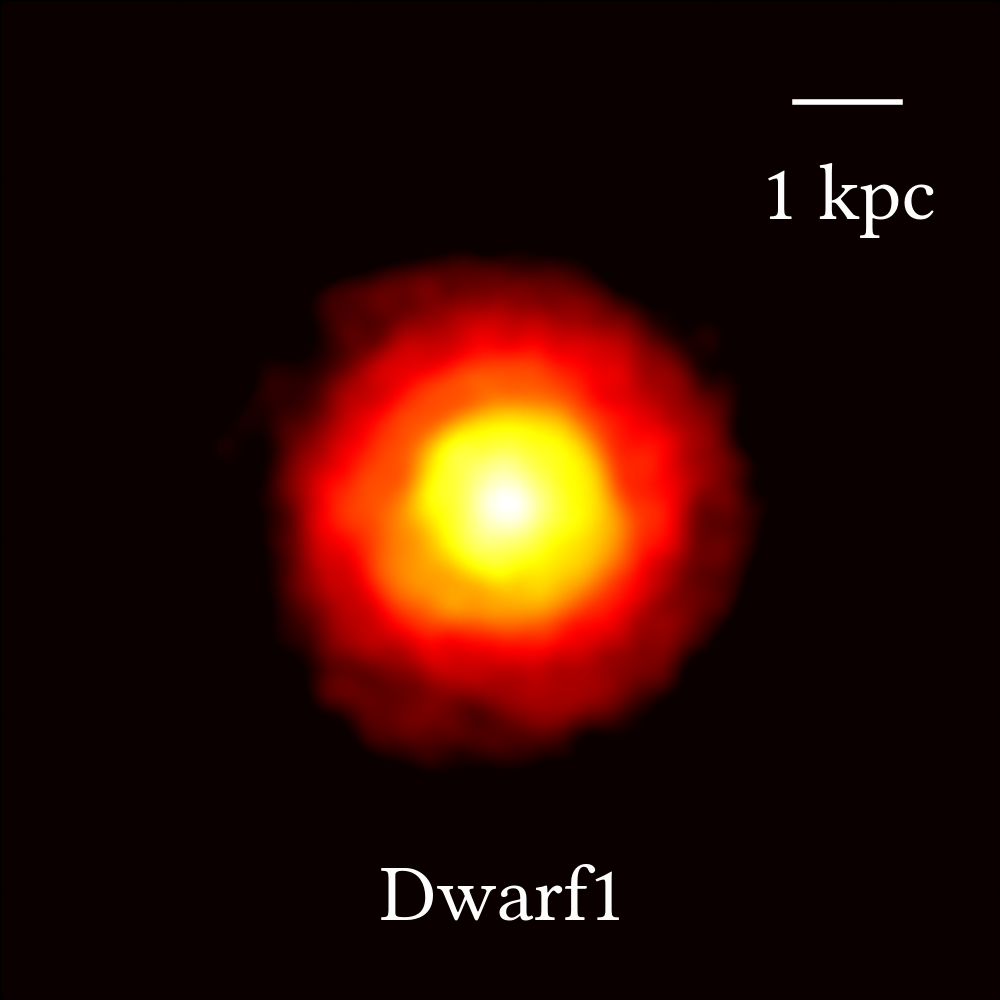}
\includegraphics[width=4cm,height=4cm]{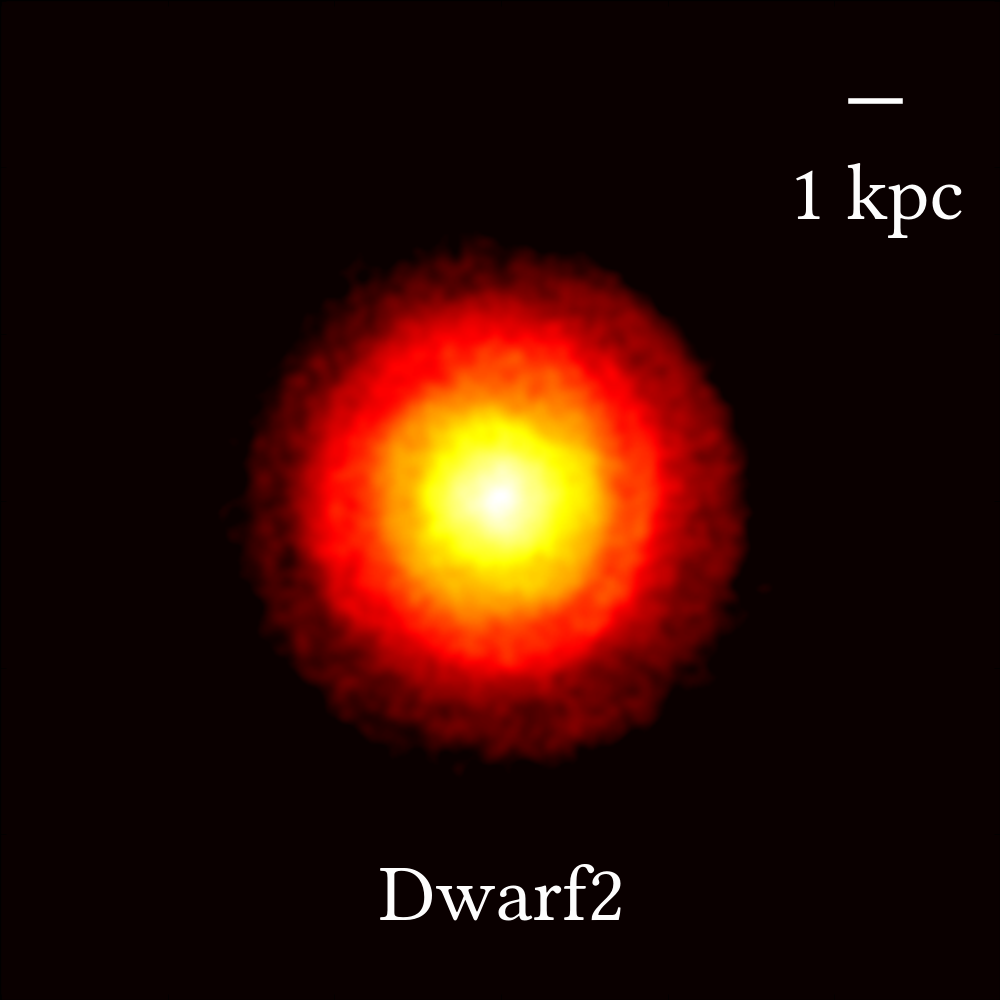}
\includegraphics[width=4cm,height=4cm]{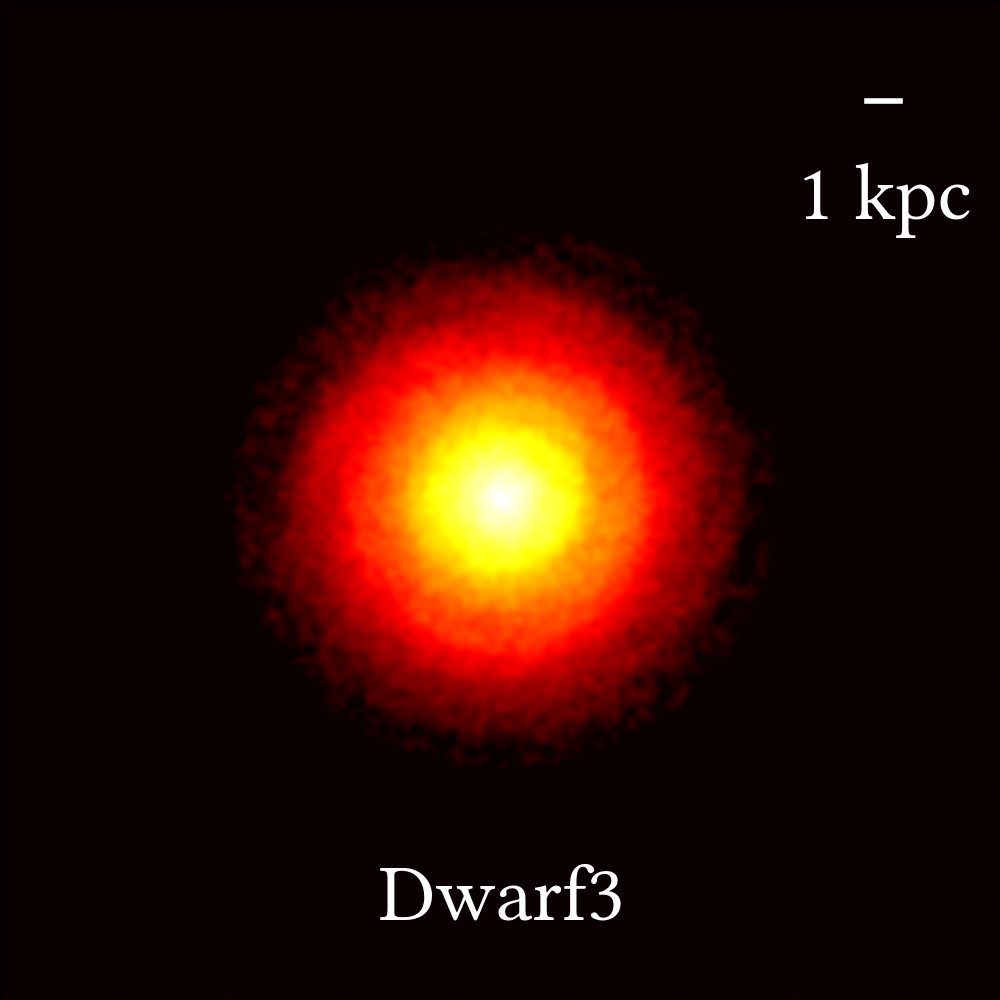}
\includegraphics[width=4cm,height=4cm]{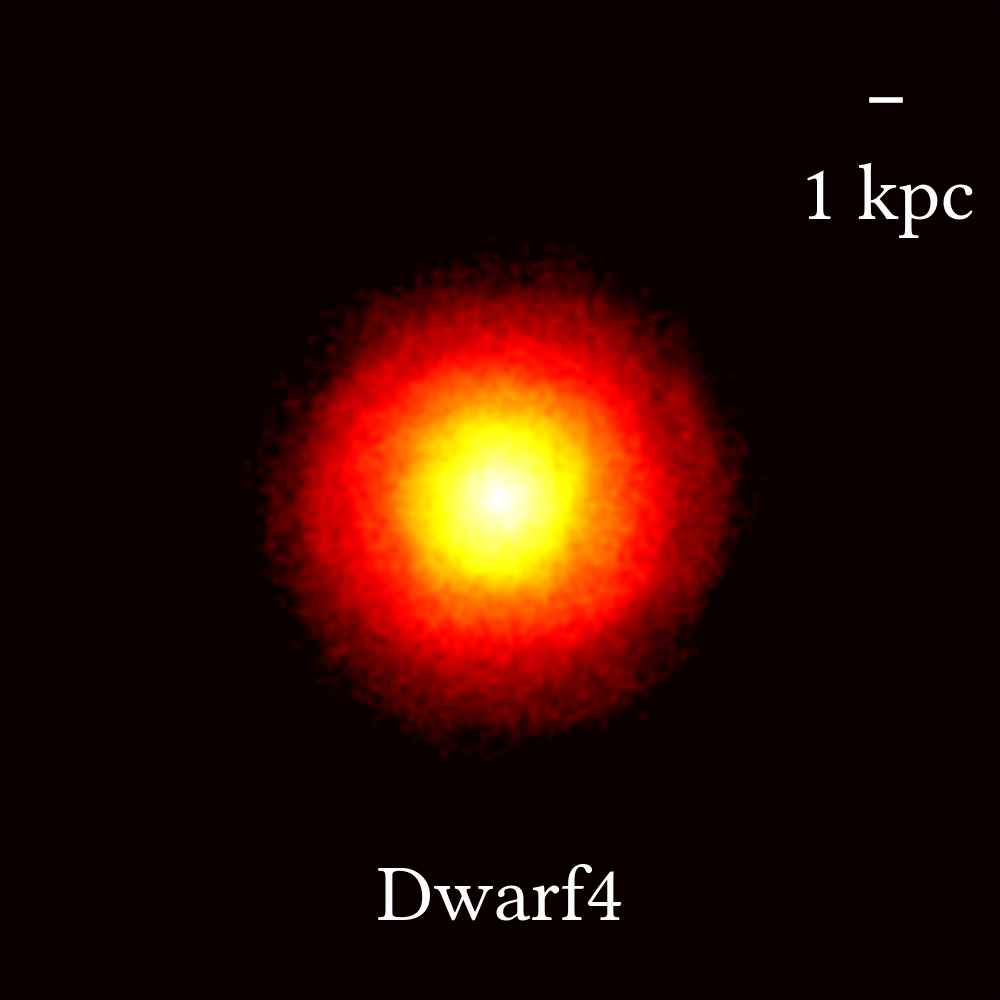}
\includegraphics[width=4cm,height=4cm]{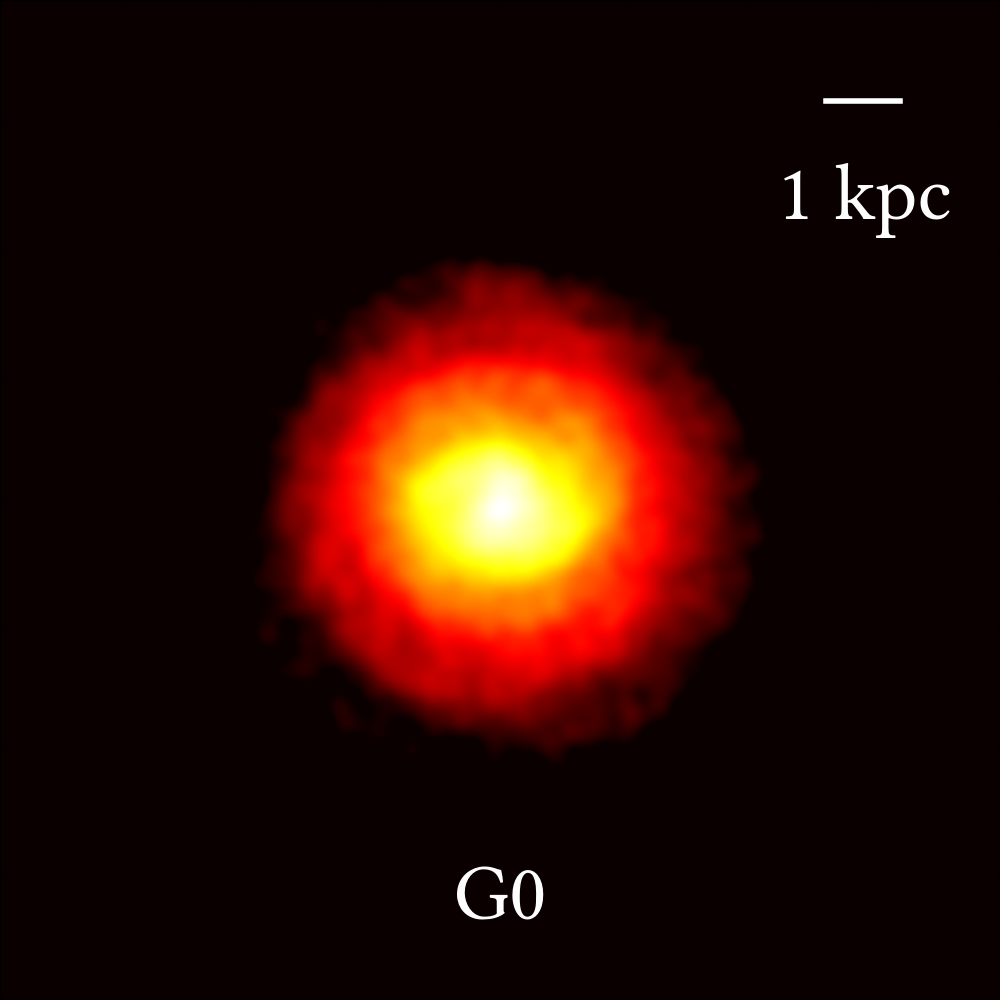}
\includegraphics[width=4cm,height=4cm]{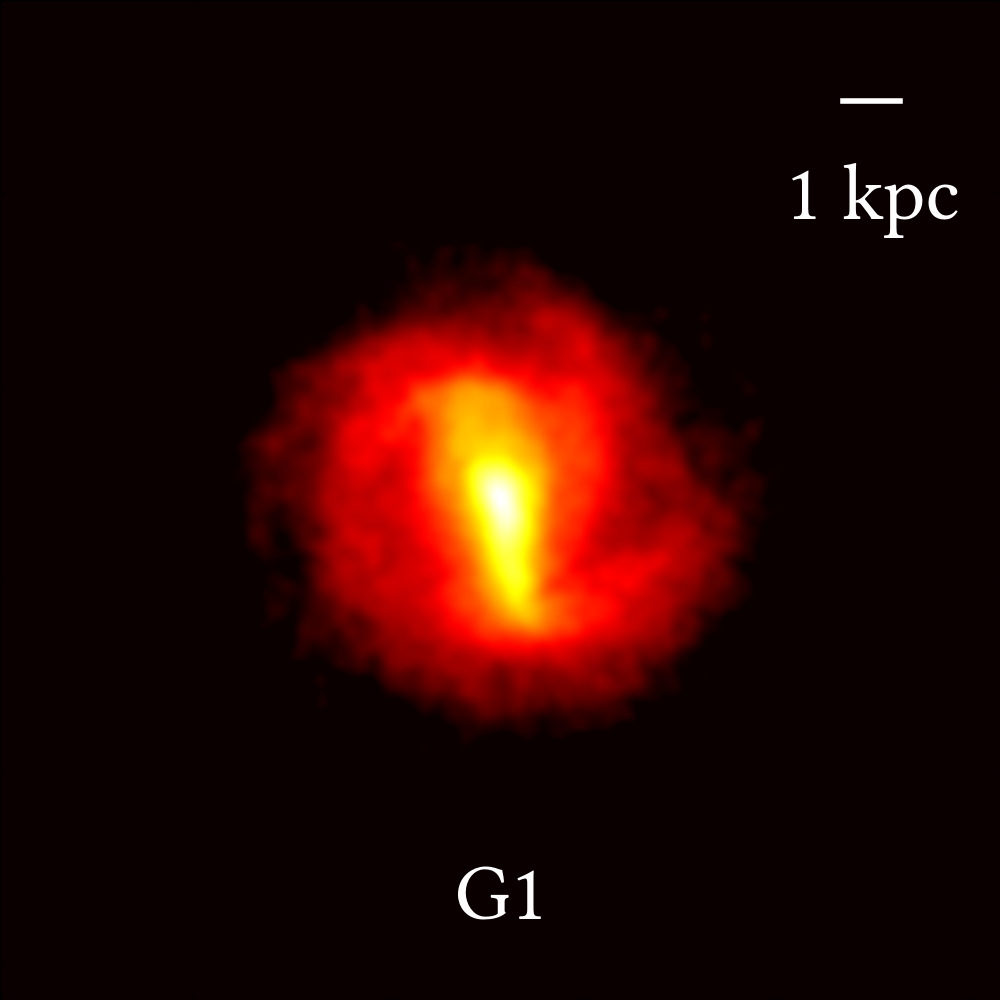}
\end{minipage}
\caption{Face-on stellar density maps of our simulated galaxies at
  half the simulation time (3 Gyr). Note that with the exception of G1,
  the simulated galaxies do not exhibit non-axisymmetric structures such
  as spiral arms or bars.}
\label{galaxies}
\end{figure}

\begin{figure}
\begin{minipage}{\linewidth}
\centering
\includegraphics[width=4cm,height=4cm]{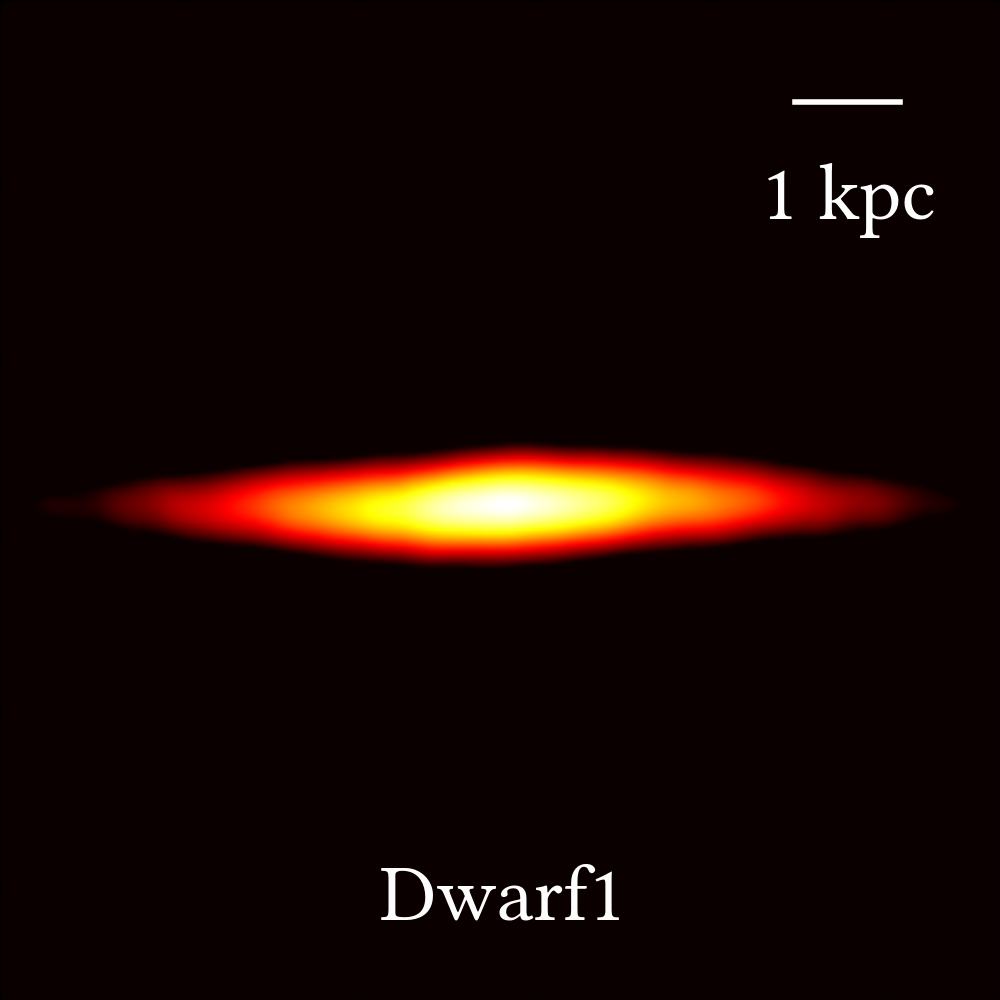}
\includegraphics[width=4cm,height=4cm]{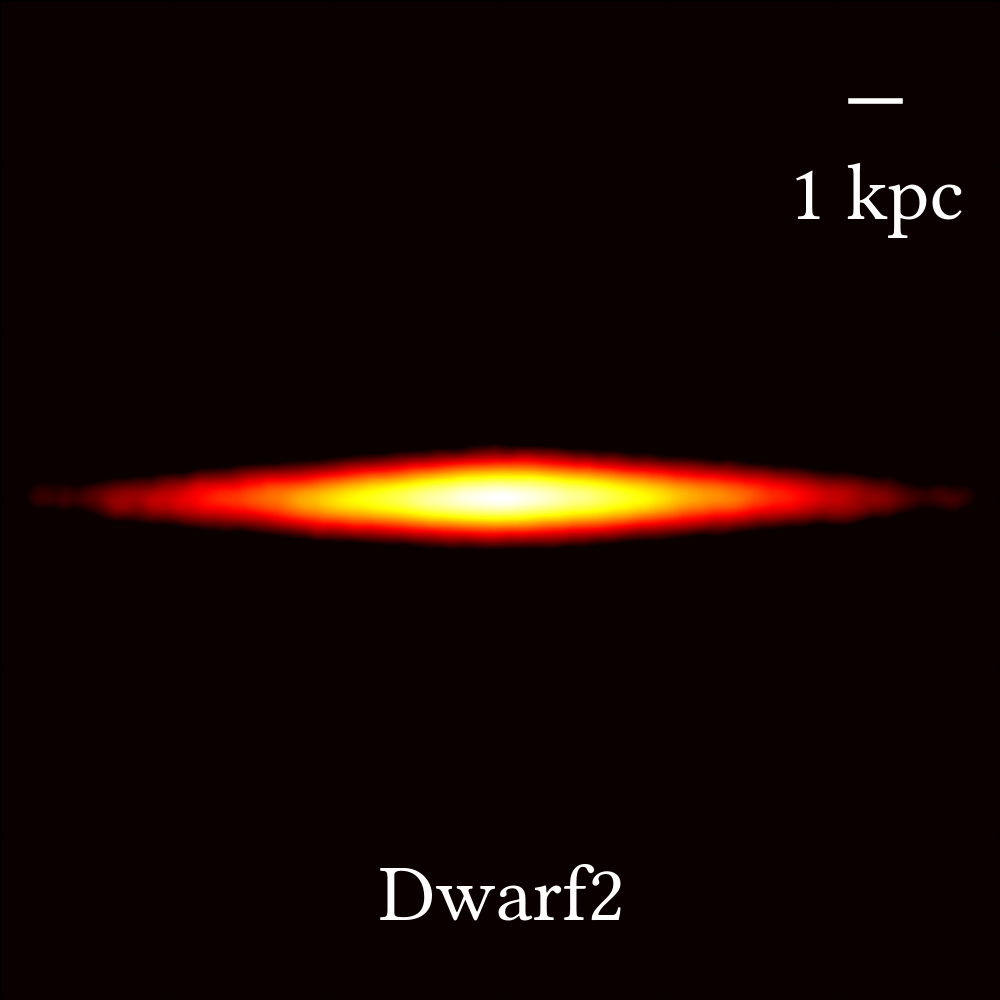}
\includegraphics[width=4cm,height=4cm]{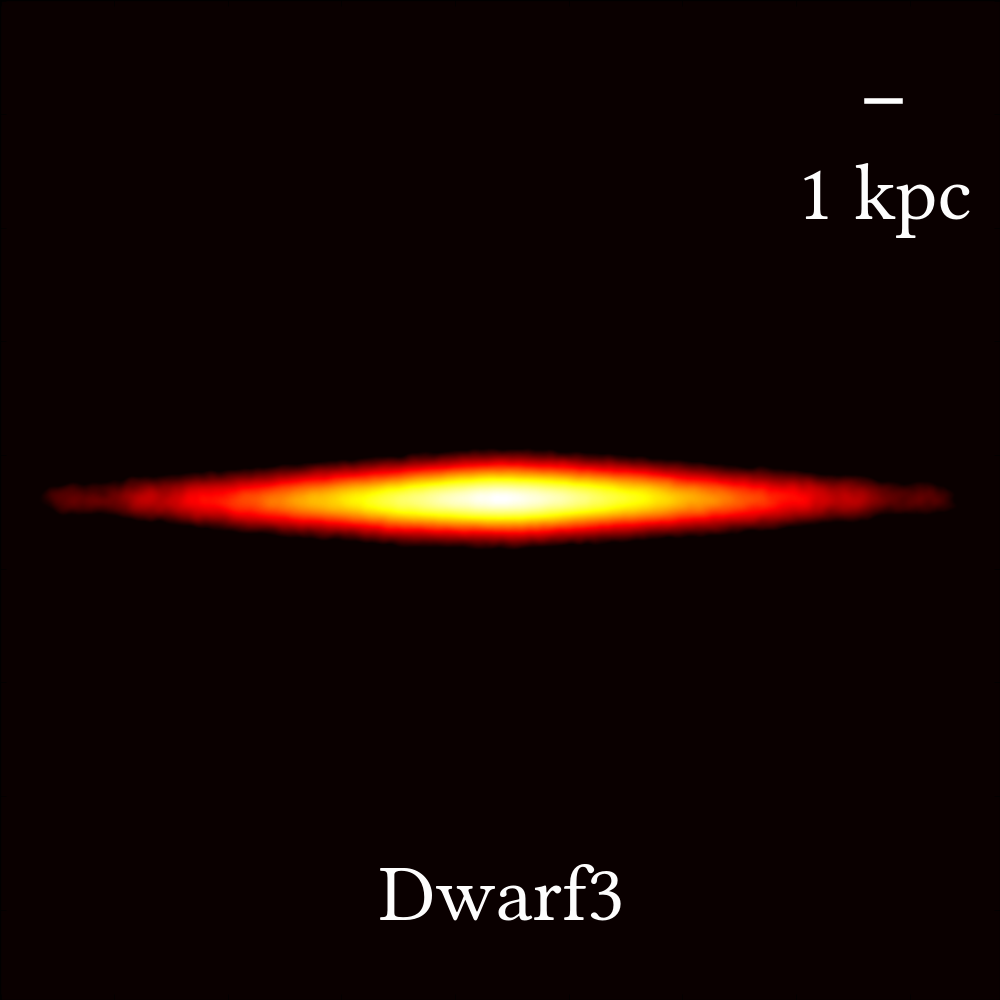}
\includegraphics[width=4cm,height=4cm]{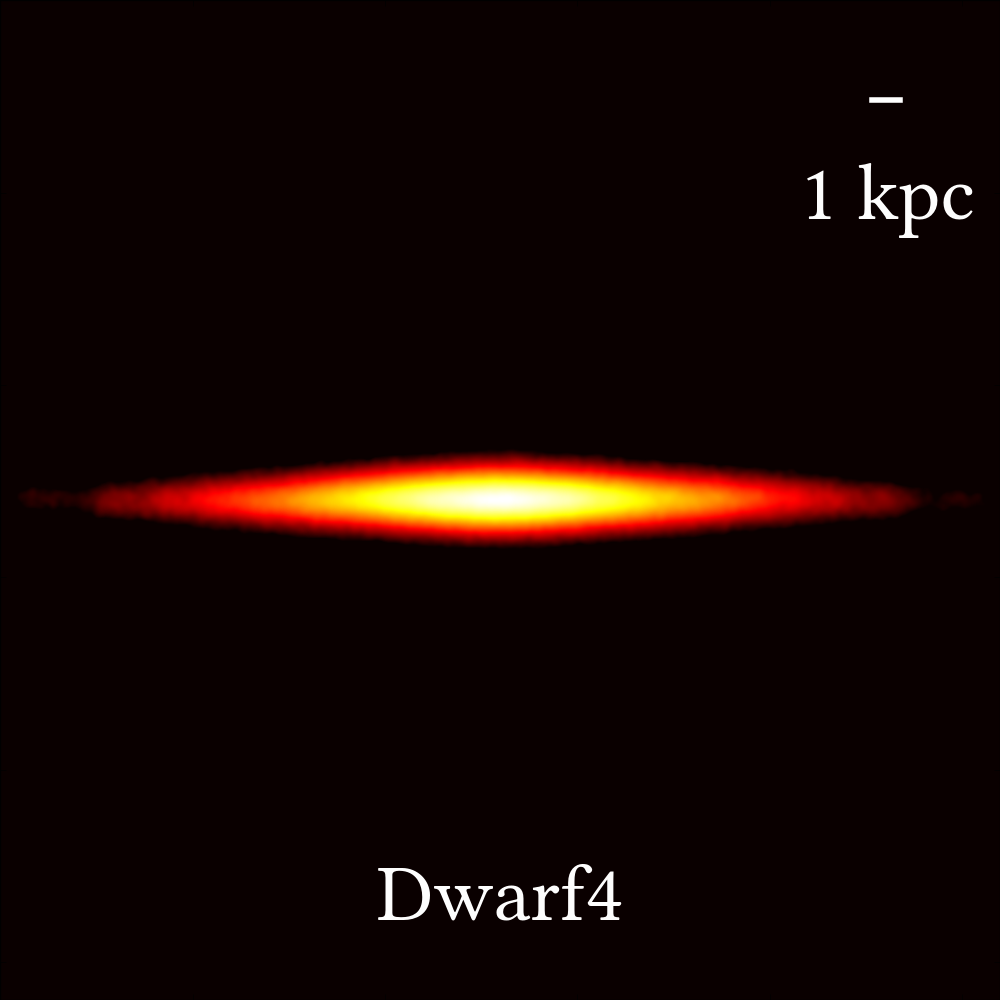}
\includegraphics[width=4cm,height=4cm]{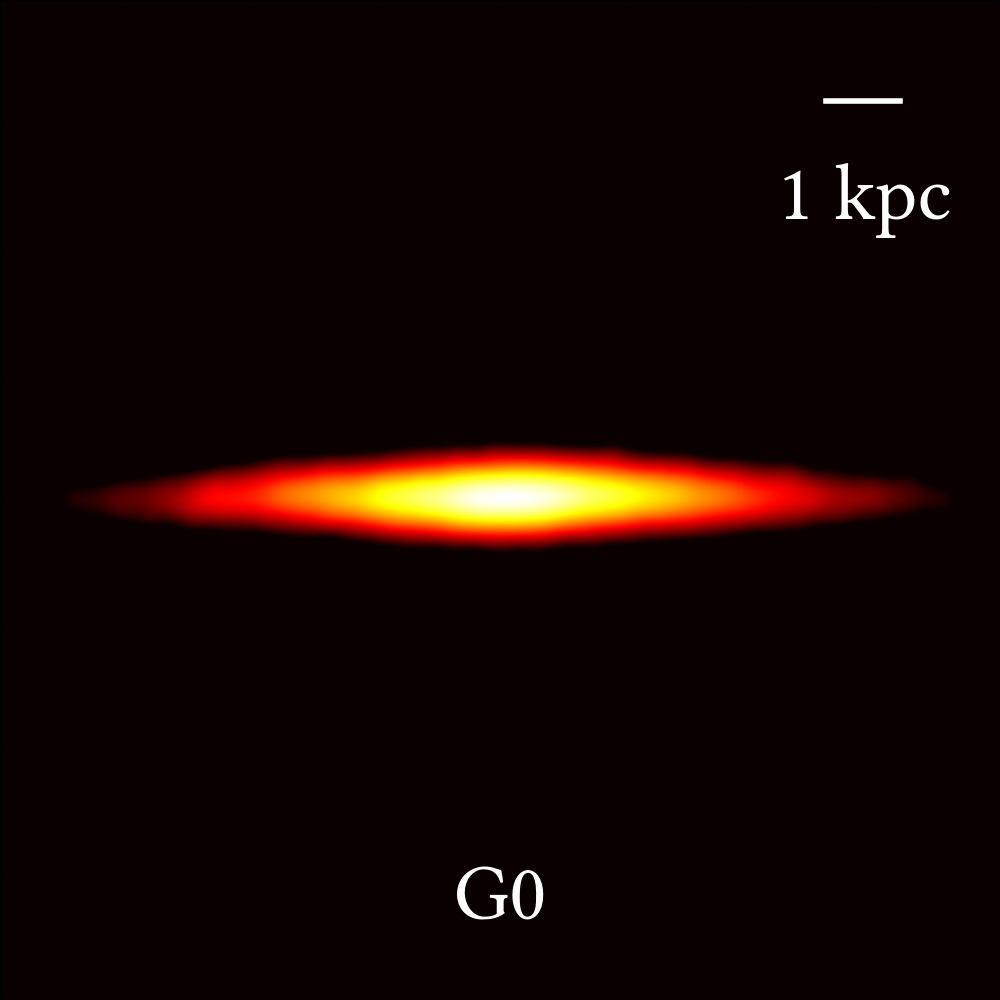}
\includegraphics[width=4cm,height=4cm]{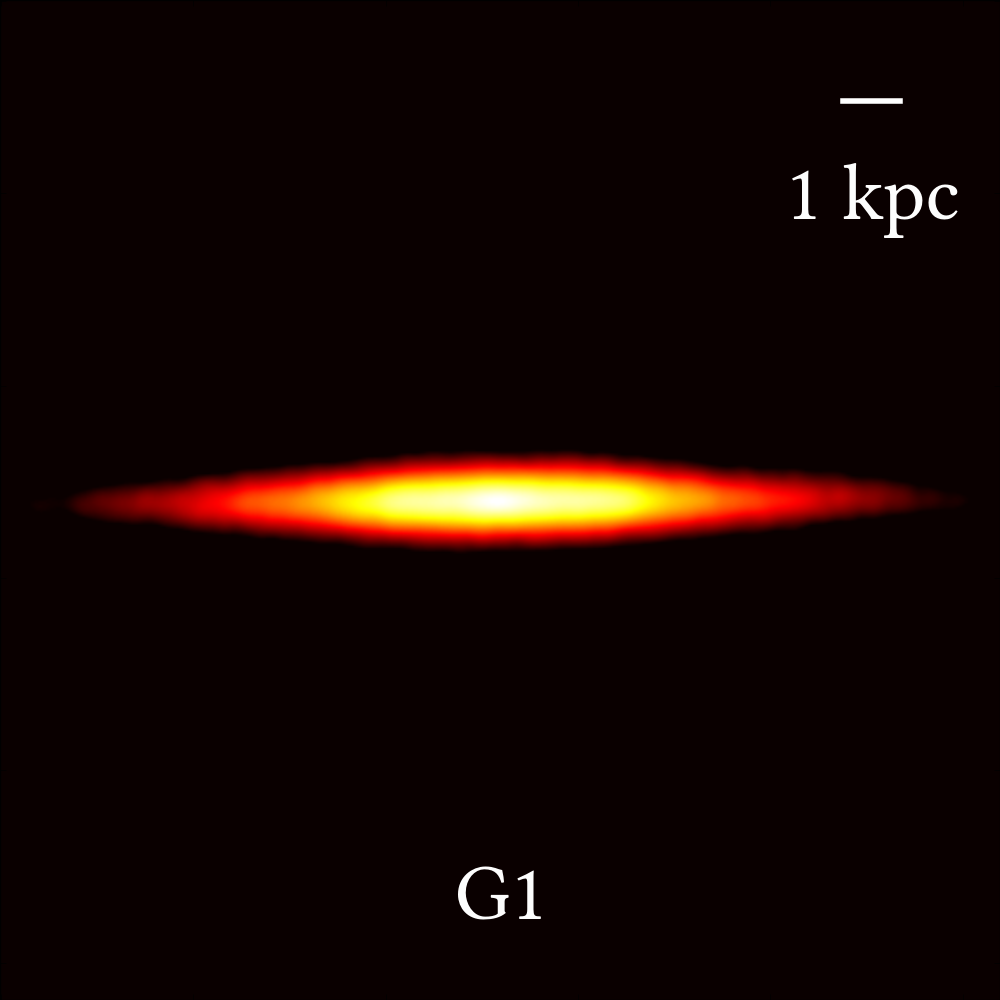}
\end{minipage}
\caption{Edge-on stellar density maps after 3 Gyr of evolution.
  Our galaxies do not form bulge-like central concentrations of matter.
  This fact is confirmed by the lack of central peaks in the circular velocity
  profiles of the stellar component $V_\star$ in Fig.~\ref{fig:Teo_RCs}.}
\label{galaxies_edgeon}
\end{figure}

We also checked that our galaxies are consistent with the baryonic TF
relation of \citet{Stark2009}, which was calibrated specifically using
small, gas-rich galaxies (mainly dwarf galaxies and LSBs).
Additionally, we require $V_{\rm flat}$ to stay below 130 km s$^{-1}$,
following the selection criterion of \citet{Spekkens2005} to
characterize a galaxy as a dwarf.  We note that galaxy models D1 to D4
retain well-behaved discs throughout the simulation.  There are
neither detectable signatures of central bulge-like mass concentrations
nor formation of bars, spiral arms, or other baryonic substructures
(see Figs.~\ref{galaxies} and ~\ref{galaxies_edgeon}).

\begin{table*}
\begin{minipage}{\textwidth}
\caption{Properties of the six simulated galaxies.}
\label{tab:simulations}
\begin{center}
\begin{tabular*}{\textwidth}{c @{\extracolsep{\fill}} cccccccccccc}
\hline
\hline
Model   & $M_{\rm halo}$$^\mathrm{a}$ & $N_{\rm halo}$$^\mathrm{b}$ &
                                                                      $c^\mathrm{c}$
  & $M_{\star}$$^\mathrm{d}$ & $N_{\star}$$^\mathrm{e}$      &
                                                               $h_0$$^\mathrm{f}$
  & $z_0$$^\mathrm{g}$ & $M_{\rm gas}$$^\mathrm{h}$ & $N_{\rm
                                                      gas}$$^\mathrm{i}$
  & $h_{\rm gas}$$^\mathrm{j}$ & $f_g$$^\mathrm{k}$ & $V_{\rm flat}$$^\mathrm{l}$  \\
        & (${\rm M}_\odot$) & ($\times10^6$)  & & (${\rm M}_\odot$) & ($\times10^6$)        & (kpc)         & (kpc) &  (${\rm M}_\odot$) & ($\times10^6$) & (kpc) & & (km s$^{-1}$) \\
\hline
Dwarf1  & 3.0$\times10^{10}$    & 1.5   & 11& 1.2$\times10^{8}$   & 0.1 & 0.8   & 0.16  & 3.7$\times10^{8}$  & 0.3      & 2.4   & 0.76  & 57    \\
Dwarf2  & 7.0$\times10^{10}$    & 3.5   & 11& 6.0$\times10^{8}$   & 0.5 & 1.7   & 0.34  & 1.1$\times10^{9}$  & 0.9      & 5.1   & 0.65  & 78    \\
Dwarf3  & 1.2$\times10^{11}$    & 6.0   & 10& 1.2$\times10^{9}$   & 1.0 & 2.5   & 0.5   & 1.6$\times10^{9}$  & 1.4      & 7.8   & 0.57  & 89    \\
Dwarf4  & 2.6$\times10^{11}$    & 6.5   & 10& 6.0$\times10^{9}$   & 1.2 & 3.0   & 0.6   & 3.4$\times10^{9}$  & 0.7      & 9.0   & 0.36  & 119   \\
G0      & 5.1$\times10^{10}$    & 1.3   & 14& 1.0$\times10^{9}$   & 0.2 & 1.1   & 0.22  & 6.1$\times10^{8}$  & 0.1      & 3.3   & 0.38  & 67    \\
G1      & 2.0$\times10^{11}$    & 5.0   & 12& 5.0$\times10^{9}$   & 0.3 & 1.5   & 0.3   & 2.0$\times10^{9}$  & 0.1      & 4.5   & 0.29  & 103   \\
\hline
\label{tab:params}
\end{tabular*}
\end{center}
Note: $^\mathrm{a}$Halo mass; $^\mathrm{b}$Number of particles in DM
halo; $^\mathrm{c}$Halo concentration; $^\mathrm{d}$Stellar mass;
$^\mathrm{e}$Number of stellar particles; $^\mathrm{f}$Stellar disc
scale length; $^\mathrm{g}$Stellar disc scale height; $^\mathrm{h}$Gas
mass; $^\mathrm{i}$Number of gas particles; $^\mathrm{j}$Gaseous disc
scale length; $^\mathrm{k}$Gas fraction relative to baryonic mass; $^\mathrm{l}$Maximum rotation velocity.
\end{minipage}
\end{table*}

We also re-simulated two galaxies from the sample of
\citet{Cox2004}.  These models are, by construction, representative of
late-type galaxies in the local Universe, as their main properties, such as
disc size, dynamical mass, and gas fraction, are consistent with a
large set of observations \citep{Roberts1994}.  From the sample
of \citet{Cox2004}, we only included the
systems with $V_{\rm flat}<130$ km s$^{-1}$, G0
and G1. The DM haloes exhibit NFW density profiles with numerical
parameters tuned to match the baryonic TF relation from \citet{Bell2001}.
We note that these parameters are quite consistent with the scheme we
propose for Dwarfs 1 to 4 regarding the stellar mass-halo mass and
halo mass-concentration relations.  G0's halo mass is lower than the
abundance matching prediction, but it is consistent with observed
galaxies of similar mass, as already discussed for D1 and D2.  A
major difference between our models and \citet{Cox2004} is that we
omitted the bulge component in the initial set up.  To minimize
perturbations in the original target relations we redistribute the
bulge mass into the disc, conserving the total stellar mass.
Nevertheless, a central matter concentration builds up in the case of
G1, resulting in the formation of a strong bar (see Fig.~\ref{galaxies}).

\subsection{Simulation technique and numerical parameters} %
\label{subsec:Parameters}                                               %

The specific version of {\sc gadget-2} we use includes radiative cooling of
the gas and a sub-resolution multiphase model for the interstellar
medium that models the effects of star formation and stellar
feedback \citep{Springel2003}.  Neither black hole accretion nor AGN feedback is
included because we want to study highly symmetric galaxies with
non-perturbed kinematics, for which black hole growth is expected to be
small.

Each galaxy is simulated for a period corresponding to six billion
years, with snapshots (i.e.~time slices) stored every one hundred
million years, which is comparable to the orbital time for particles
inside the first few kiloparsecs.  This means that there is enough
time between snapshots for the galaxy to undergo some small-scale
morphological transformations, and therefore we expect our average
results not to be strongly biased by odd individual cases with
peculiar configurations.  We exclude the first seven snapshots of each
simulation in order to discard possible transient states during the
initial relaxation towards a stable rotational configuration (this can
arise because the initial conditions are not in perfect equilibrium);
we then still have a large number of snapshots per galaxy
($\sim$54) in order to identify and explore global trends.

A key parameter in N-body simulations is the gravitational softening
length, which is meant to keep two-body relaxation effects and orbital
integration cost under control.  Since gravitational forces are
smoothed for particles approaching shorter distances than the
softening length, the dynamics of structures smaller than this scale
is artificially modified by this approximation.
The spatial resolution of an N-body simulation is generally considered
to be two to three times its softening length. Given that we want to
investigate scales as small as 100 pc, we set the softening length to
25 pc in all our simulations. The baryonic particle (i.e.~gaseous and
stellar particles) masses are
chosen to ensure at least 8 particles per softening volume within the central
2 kpc inside the disc. The DM particle mass is set to guarantee
a minimum of 150 particles inside the inner 100 pc in order to have a
relatively smooth DM distribution.  Further, we impose an additional
constraint, namely $\sqrt{GM_{\rm part}/\epsilon_{\rm soft}} \ll 16$
km s$^{-1}$, which ensures that perturbations induced by two-body
encounters are below the typical velocity dispersion of a warm
interstellar medium. The resulting baryonic particle masses in our
simulations range from $1.2\times10^3{\rm M}_\odot$ to
$1.6\times10^4{\rm M}_\odot$, and the DM particle masses range from
$2\times10^4{\rm M}_\odot$ to $4\times10^4{\rm M}_\odot$ (see
Table~\ref{tab:params}).

{\sc gadget-2} uses smoothed-particle hydrodynamics (SPH) to solve the
hydrodynamic equations of the gas component. This computational method
simulates fluids as collections of point-like elements.
The SPH technique considers that each gas particle carries a certain amount of
every gas property, which is smoothed over a finite volume according
to a given kernel function.  The value of a quantity at an arbitrary
location is given by the sum of the smoothed contributions from all
those particles enclosing that point inside their kernel volumes.  The
kernel employed by {\sc gadget-2} is a spline function with one parameter, the
smoothing length $h$, such that the value of every quantity
outside a radius $h$ is zero. The numerical approximations made in SPH
have been shown to be sometimes inaccurate, especially for
representing fluid instabilities such as the Kelvin-Helmholtz
instability \citep[e.g.][]{Agertz2007}. However, we do not expect that
these hydrodynamical accuracy issues have a bearing on the questions
studied in this paper, especially because \citet{Hayward2014} found that
the results of {\sc gadget-2} simulations of idealised isolated disc galaxies
are very similar to simulations of the same galaxies performed with
the state-of-the-art moving-mesh hydrodynamics code {\sc arepo}
\citep{Springel2010}.

\section{Analysis methods}						%
\label{sec:Analysis}							%

To fully understand the systematic differences that can arise between
the real circular velocity profile of the DM halo and the rotation
curve that is actually inferred observationally, one has to go through
a long chain of intermediate steps. In order to disentangle the impact
of each approximation on the deduced cuspiness of the halo, our
general strategy is to apply the same analysis methods to different
circular velocity profiles, starting from the most ideal case and
adding one layer of approximation at a time. Sorted by the degree of
idealization involved, these circular velocity profiles are the following:
\begin{enumerate}
\item Dark-matter-only circular velocity ($V_{\rm dm}$)
\item Total-mass circular velocity ($V_{\rm tot}$) 
\item True circular motions of the gas ($V_{\rm cir}$) 
\item Observed rotation curve ($V_{\rm obs}$) 
\end{enumerate}

The quantity $V_{\rm dm}$ represents the circular velocity of a test
particle under the gravitational potential of the dark matter halo
alone. From a theoretical point of view, this velocity profile is the
only one that traces the exact DM distribution. $V_{\rm tot}$
represents the circular velocity profile generated by the total
gravitational potential (i.e. DM plus baryons) in the plane of the
disc. With this definition at hand, the minimum disc approximation can
be thought of as using $V_{\rm tot}$ instead of $V_{\rm dm}$ to
directly estimate the cuspiness of the halo, thus neglecting the baryonic
contribution to the potential. In both cases, $V_{\rm dm}$ and
$V_{\rm tot}$ do not measure velocities but rather gravitational
radial accelerations, as they translate into rotational velocities via
$V_{\rm rot}(r) = \sqrt{a_r r}$. 

In contrast, $V_{\rm cir}$ is a direct measure for the actual
(circular) motions of the gas. Notice that $V_{\rm cir}$ can lie below
$V_{\rm tot}$ if the hydrodynamical pressure pushes the gas outwards,
thereby lowering the effective radial acceleration towards the center that
needs to be balanced by centrifugal forces. Finally, $V_{\rm obs}$
refers to our mock observed rotation curves, which mimic several
processes of real observations, such as projection effects and finite
spatial resolution. This generic name comprises a large
set of curves for each snapshot because we mimic two types of
observations, long-slit rotation curves and 2D velocity fields, at
five different inclinations and four different spatial resolutions
(to mimick the effect of distance variations). It is worth
emphasizing that $V_{\rm dm}$, $V_{\rm tot}$, and $V_{\rm cir}$
correspond to perfect theoretical measurements from the simulations,
whereas $V_{\rm obs}$ accounts for the limitations of real data.
However, we stress that because our simulations are constructed
to be highly symmetric discs in perfect rotational equilibrium,
even $V_{\rm obs}$ does not fully capture the difficulties inherent
in inferring the DM profile shape from observations of real dwarf
irregulars and LSB galaxies. Instead, the analysis that we
present here should be considered a best-case scenario, at least given
current observational limitations.

\subsection{Snapshot preprocessing}                                 %
\label{subsec:preprocessing}                                            %
Before we extract the relevant information from the snapshots, we
process them to make sure that the center of the gravitational
potential and the galactic disc orientation are robustly determined.
This is very important because poorly constrained values can introduce
harmful effects in the forthcoming analysis. For example, note that an
error of $5^\circ$ at low inclinations ($\sim$$15^\circ$) can propagate
to an error as high as 50 per cent in the normalization of the observed
rotation curves and that an incorrect determination of the center's position
may lead to a spurious flattening of the inner part of 
a spherically averaged density profile.

We recenter the snapshots using an iterative version of the shrinking
spheres method described in \citet{Power2003}. We first calculate
the center of mass inside a large sphere containing everything in the
simulation and recenter all coordinates around this point. Next we
shrink the sphere by 1 per cent in radius, find the new center of mass, and
recenter the particles again. This is repeated until there are
less than 10 particles in the sphere.
We also rotate the frame of reference to make the net angular momentum of the
gas component coincident with the $z$-axis.

\subsubsection{Real density profiles}	                                %
\label{subsec:realdens}		                                        %

To determine the true DM density profiles and verify their steepness,
we first measure the DM cumulative mass as a function of radius and
then we compute
\begin{equation}\label{eq:dens_profile}
\rho = \frac{\frac{\rm d}{{\rm d} r} M_{\rm dm}(<r)}{4\pi r^2}.
\end{equation}
We use spherical shells equally spaced in logarithmic radius and a central
finite-differences scheme for the derivative. We test different steps,
namely 0.05, 0.075, 0.1, 0.15, 0.2, and 0.3, and we adopt $\Delta\log(r)=0.15$,
where $r$ is in kpc, which is the smallest interval that produces smooth
profiles and is still largely affected by Poisson noise in the central
region ($r < 150$ pc). We note that the measured cuspiness of
the halo does not depend on this specific choice as variations in the
density profiles are very subtle. 

\subsection{Theoretical circular velocity profiles}                  	%
\label{subsec:theoreticalRCs}                                           %

We refer to measurements that are calculated from detailed information
in the snapshots as theoretical quantities, disregarding the issue of
whether or not they can be actually assessed observationally. These
quantities include the DM density profiles described earlier as well
as a subset of the circular velocity profiles introduced at the
beginning of Sec.~\ref{sec:Analysis}, namely $V_{\rm dm}$, $V_{\rm tot}$,
and $V_{\rm cir}$.

\paragraph*{(i) Dark-matter-only circular velocity,
  $V_{\rm dm}$:} Particles rotating at a constant speed
satisfy $V_{\rm rot}(r) = \sqrt{a_r r}$, so it suffices to determine
the mean radial acceleration attributed to the DM halo as a function
of radius to determine its equivalent circular velocity profile. To do
this, we export all DM particles in the snapshot to a separate initial
conditions file and run {\sc gadget-2} for a single time-step to calculate the
gravitational forces. Then, we measure the mean radial acceleration in
thin spherical shells and compute
\begin{equation}\label{eq:Vdm}
  V_{\rm dm}(r) = \left.\sqrt{\left<{a_r(r)}\right> r}\right|_{\rm gravity, ~dm ~only} .
\end{equation}

\paragraph*{(ii) Total-mass circular velocity, $V_{\rm tot}$:}
This velocity profile is related to the gravitational potential of the
whole system, i.e. dark matter plus baryons. It is interesting
because a joint analysis with the DM-only circular velocity profile
allows one to assess
the validity of the minimum disc approximation without mixing in any
other effect. Once again, we compute the mean radial accelerations due
to gravity using {\sc gadget-2}. However, this time we cannot assume spherical but
rather axial and vertical symmetries instead. Therefore, we compute the radial
accelerations using thin cylindrical shells in the $xy$-plane
($|z|\leq100$ pc) and then determine the circular velocity
profile using
\begin{equation}\label{eq:Vtot}
V_{\rm tot}(r) = \left.\sqrt{\left<{a_r(r)} \right> r}\right|_{\rm gravity, ~all ~particles} .
\end{equation}

\paragraph*{(iii) True circular motions of the gas, $V_{\rm cir}$:}
In this case, we select the gas particles in the equatorial plane
($|z| \leq100$ pc) and measure their circular (tangential)
velocities, taking the mean value in small radial bins:
\begin{equation}\label{eq:Vcir}
V_{\rm cir}(r) = \left. \left<{V_\phi(r)}\right>\right|_{\rm gas ~particles}.
\end{equation}
We emphasize that for axisymmetric systems in rotational equilibrium,
any difference between $V_{\rm cir}$ and $V_{\rm tot}$ must be due to
the fact that in addition to gravity, gaseous media also experience
hydrodynamical forces. The above velocity profile is one step closer
to reality because instruments do not detect gravitational potentials but
rather velocities; thus, $V_{\rm cir}$ can be thought of as the rotation curve
that a perfect instrument under perfect observational conditions would detect
(neglecting projection effects).

Additionally, we calculate the circular velocity profiles associated
with the gravitational potentials of the stellar and the gaseous
components separately, using the same procedure as for $V_{\rm tot}$.
These curves are denoted as $V_\star$ and $V_{\rm gas}$ in the text. This
complementary information is interesting for understanding the difference
between $V_{\rm dm}$ and $V_{\rm tot}$, as we have that
\begin{equation}\label{eq:quadsum} 
V_{\rm tot}^2 =
  V_{\rm dm}^2+V_\star^2+V_{\rm gas}^2 . 
\end{equation}

The theoretical rotation curves were calculated using shells of 100 pc
width, linearly spaced every 100 pc, the first of which is centered at
125 pc in order to exclude particles inside 3 times the softening
length. Except for when calculating $V_{\rm dm}$, we only consider particles in the
midplane because the assumption of vertical symmetry breaks down as
one moves above or below the equatorial plane, and this is likely to make
particles rotate slower \citep{Rhee2004}. We check that there is a
large enough number of particles in all bins, resulting in smooth
velocity profiles without visible signs of shot noise. When we explore the
effects of spatial resolution associated with the distance to the
galaxies, we use re-sampled versions of the theoretical rotation curves
matching the radial positions of the corresponding mock observations.
The error bars on the theoretical rotation curves are fixed to 1 km
s$^{-1}$ where necessary.

\subsection{Mock kinematic observations}				%
\label{subsec:MockRCs}							%
We mimic different kinds of observations of the gas component,
including optical long-slit rotation curves and 2D velocity maps. We
consider two different cases, labeled as H$\,${\sc i} and
H~$\alpha$, which represent 
some physical properties of the 21-cm and the H~$\alpha$
emissions, respectively. Example data products are shown in Fig. 3. 

We \emph{observe}
each galaxy at five different inclinations from $15^\circ$ to $75^\circ$
in steps of $15^\circ$, the smallest one being the closest to a face-on
view. We also try four different distances, namely 10, 20, 40, and 80 Mpc.
Note that the truly important quantity is the spatial resolution
of the observations, but because we use typical values for the instrumental
angular resolutions, these distances are useful indicators. For our fiducial
H~$\alpha$ PSF of 2 arcsec, the corresponding spatial resolutions are
$\sim$100, $\sim$200, $\sim$400, and $\sim$800 pc. Our H$\,${\sc i}
resolution is six times poorer, but as we use the radio observations only
in the outer part of the rotation curves, the optical data is most important for differentiating
cusps from cores.
In the following, we first describe the basic concepts and
common approximations employed to create our mock observations
and then move on to more specific details as necessary.

We assume that some of the gas particles emit radiation through a
spectrum composed of a single emission line with a gaussian profile. 
In the H$\,${\sc i} case, we
consider all gas particles in the simulation to be radio emitters,
and we make the amplitude of the 21-cm emission line
proportional to the mass of the gas particle. The broadenings are
given by the intrinsic velocity dispersions, $\sigma$, self-consistently
calculated from the temperature of each particle as 
\begin{equation}\label{eq:sigmah1}
  \sigma = \sqrt{\frac{kT}{\mu m_H}},
\end{equation}
where $k$ is the Boltzmann constant, $T$ is the temperature of the gas
particle, $\mu$ is the mean molecular weight, and $m_H$ the mass of a hydrogen nucleus.
For moving particles, we Doppler-shift the emission line without altering its
width. This means that the intensity of the emission from a gas element with a mean
velocity $\overline{v}_{\rm los}$ along the line of sight, detected at a
different velocity (frequency) $v$, is given by
\begin{equation}\label{eq:emih1}
  I_{\raisebox{-1pt}{\textrm{{\tiny HI}}}}(v) \propto \frac{M_{\rm
      part}}{\sigma}\,
  {\rm e}^{-\frac{(v-\overline{v}_{\rm los})^2}{2\,\sigma^2}} .
\end{equation}

In the H~$\alpha$ case, we only include gas particles with ongoing
star formation activity, which is a good proxy for the spatial distribution
of the ionized gas in H$\,${\sc ii} regions, though it may miss the
warm ionized gas pervading the rest of the ISM. The mock H~$\alpha$
observations defined in this way have a smaller radial extent than our
fiducial mock H$\,${\sc i} data, consistent with observations. Additionally,
the disc of star-forming gas is thinner than our mock H$\,${\sc i} disc,
which makes our choice conservative,
as our mock H~$\alpha$ data will be less affected by
projection effects and will trace the disc kinematics more faithfully.
We make the H~$\alpha$
intensity proportional to the mass of the gas particles
times their current star formation rate ({\small SFR}), which should
roughly correlate with the amount of ionizing radiation available from
young, massive stars. The H~$\alpha$ intensity versus
frequency is thus given by
\begin{equation}\label{eq:emiha}
  I_{\raisebox{-1pt}{\textrm{{\tiny H$\alpha$}}}}  (v) \propto
  {\rm SFR}\times \frac{M_{\rm part}}{\sigma}\,
  {\rm e}^{-\frac{(v-\overline{v}_{\rm los})^2}{2\,\sigma^2}} .
\end{equation}

Regarding the observational process itself, we simplify it to a
combination of a spatial and a spectral sampling of the emitted
radiation field. The spatial sampling mimics the pixels of the
detector. To be consistent with the SPH approximation,
the radiation flux that a gas particle contributes to a certain pixel is
inversely proportional to the projected distance between the particle
and the center of the pixel, using the same kernel and smoothing
lengths as in the SPH simulation. Adding contributions from all neighbouring
particles, we obtain a resulting spectrum per pixel made up of individual
shifted Gaussians. Then, we sample each spectrum in narrow velocity
(frequency) channels to mimic the spectral resolution of a
given instrument. At the end of this process, we obtain a datacube
containing spatial and kinematic information about the gas component
in the simulation box. Each slice of this datacube is an
intensity map of the H~$\alpha$ or H$\,${\sc i} radiation in a specific velocity
channel. Finally, a 2D Gaussian convolution is implemented across each
slice to mimic the effect of the optical seeing or the radio beam,
and the cube is collapsed along the spectral axis by means of a simple
intensity weighted mean (IWM) scheme to determine the observed
line-of-sight velocity at each pixel,
\begin{equation}\label{eq:iwm}
  \langle v \rangle = \frac{\sum_{\rm ch}{I_{\rm ch}v_{\rm
        ch}}}{\sum_{\rm ch}{I_{\rm ch}}} 
\end{equation}

For the 2D observations, mock intensity and velocity
dispersion maps are also generated from the data cubes following \citet{Walter2008}

\begin{equation}\label{eq:intensitymap}
  I	= \sum_{\rm ch}{I_{\rm ch}}
\end{equation}

\begin{equation}\label{eq:dispermap}
  \sigma = \sqrt{\frac{\sum_{\rm ch}{I_{\rm ch}\times(v_{\rm ch} - \langle v \rangle)^2}}{\sum_{\rm ch}{I_{\rm ch}}}}.
\end{equation}

Interesting comparisons of the IWM algorithm with
alternative ways of defining a velocity map from its parent data cube
are discussed by \citet{deBlok2008} and \citet{Oh2011}.
They show that IWM velocities are potentially biased in the
presence of non-circular motions as the emission lines become
asymmetric or the spectrum may exhibit secondary peaks (i.e., additional
velocity components). Notwithstanding, our galaxies
are strongly dominated by rotational motions; thus, we find the IWM method
good enough to trace the disc rotation (see Fig.~\ref{fig:Obs_RCs}).
Moreover, we will show in
Sec.~\ref{sec:Results} that for nearby galaxies, the final results obtained
from our mock observations agree remarkably with those obtained from perfect
theoretical measurements of the gas kinematics. Therefore, there is
no need to employ more complex algorithms than the IWM.
The same applies to more sophisticated schemes, as modelling 
the H$\,${\sc i} distribution and kinematics directly from
from the datacubes \citep[e.g.][]{Bouche2015,DiTeodoro2015,Kamphuis2015}.
Even though a comparison of these methods through our mock data set
would be interesting, it lies beyond the scope of this work.

We also address several numerical artifacts detected
during the experiments. For example, we noticed that in low-intensity
pixels, outliers moving at arbitrary velocities can dominate the
velocity estimation and introduce spurious and sometimes enormous
fluctuations in the final rotation curve. To avoid this effect, we define an
arbitrary luminosity threshold of 10$^{-4}$ times the maximum detected
intensity, which works well in suppressing the fluctuations while
filtering out just a few pixels. Another potentially harmful effect that we
detected is a systematic underestimation of the velocity at the last
measured points. We find the main cause of this to be related to the
use of the SPH approach, as this implies that in principle, one can
define velocities at points that are beyond all gas particles. To
remove this artifact, we impose the maximum radius for velocity
estimations to be the radius enclosing 99 per cent of the emitting particles.
A further comparison between the mock observations and the theoretical rotation curves confirmed
that imposing this maximum radius solved the problem. We only report
rotation curves for radii beyond 75 pc plus the seeing/beam such that
possible contamination from inner points affected by the softening is
virtually absent even after the {\small PSF}/beam convolution.
This places the first kinematic measurement at approximately 0.17, 0.27,
 0.46, and 0.85 kpc for galaxies viewed at 10, 20, 40, and 80 Mpc, respectively.

\subsubsection{H~$\alpha$ long-slit rotation curves}                     %
\label{subsec:longslit}                                                 %

In this case, we only consider pixels inside a virtual slit placed
along the major axis of the disc. We use square pixels and enough
spectral channels to sample the whole range of line-of-sight
velocities. We choose instrumental parameters consistent with the
majority of observational studies. The slit width is 1.4 arcsec, the pixel size is
0.7 arcsec, the FWHM of the PSF is 2 arcsec, and the spectral resolution is 47 km s$^{-1}$,
equivalent to a 20 km s$^{-1}$ channel separation. The spatial
resolution corresponds to $\sim$ 100 pc at a distance of 10 Mpc, matching
the size of the radial bins that we used to define the theoretical rotation curves.
For mock observations at 20, 40, and 80 Mpc, the corresponding physical resolutions
are $\sim$200, $\sim$400, and $\sim$800 pc, respectively.
Given that the slit width
is resolved into two pixels, we add them together before collapsing the
cube to have a single velocity estimate at each position along the
slit. Next, we fold the rotation curve to put the approaching and receding
sides together, and we average both the radial and velocity information
in bins of 1 arcsec ($\sim$ 50 pc at 10 Mpc), i.e. using 2 points per
seeing, as is common practice for this kind of data. We adopt
the standard deviation of individual velocity pixels in the bin as the
error bar, and we check that this is perfectly consistent with taking the
difference between the approaching and receding sides. 

Similar to \citet{deBlok2002}, we impose a minimum error of 2 km
s$^{-1}$ (they used 4 km s$^{-1}$). Finally, we de-project the observed
line-of-sight velocities to get the actual rotation curve,
\begin{equation}\label{eq:vlos}
v_{\rm rot}(r) =   \frac{v_{\rm los}(r)}{\sin{(i)}},
\end{equation}
where $i$ represents the galaxy inclination. We use the true inclination
and major axis of the gaseous disc, and we assume the galactic center
to coincide with the center of the overall gravitational potential. In
a forthcoming paper, we will present mock photometric data of these galaxies
and demonstrate that classical photometric estimators can recover the
center of the gravitational potential with an accuracy of 1 arcsec (half the
spatial resolution) and the inclination and position angle with typical
errors of less than $10^\circ$. As we commented in Sec.~\ref{subsec:preprocessing},
small inclination errors can induce large differences in the normalization
of the rotation curve for discs that are near to face-on. However, it is
not clear whether such errors can cause the inferred cuspiness
of the halo to be incorrect or they can be safely ignored; regardless, errors of this
magnitude would certainly
be considered acceptable in actual observational studies. In any case, in this work,
we focus on the best case and do not include the effects of errors in
the geometrical parameters.

\begin{figure}
\begin{minipage}[b]{0.1\linewidth}
\centering
\includegraphics[width=0.8\textwidth,height=5.15\textwidth]{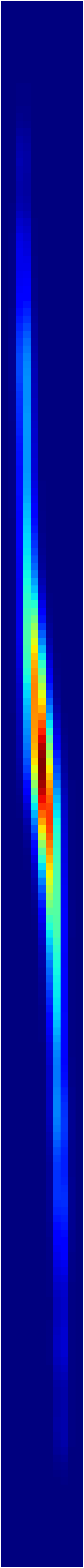}
\vspace{3.cm}
\end{minipage}
\begin{minipage}[b]{0.8\linewidth}
\centering
\includegraphics[width=\textwidth]{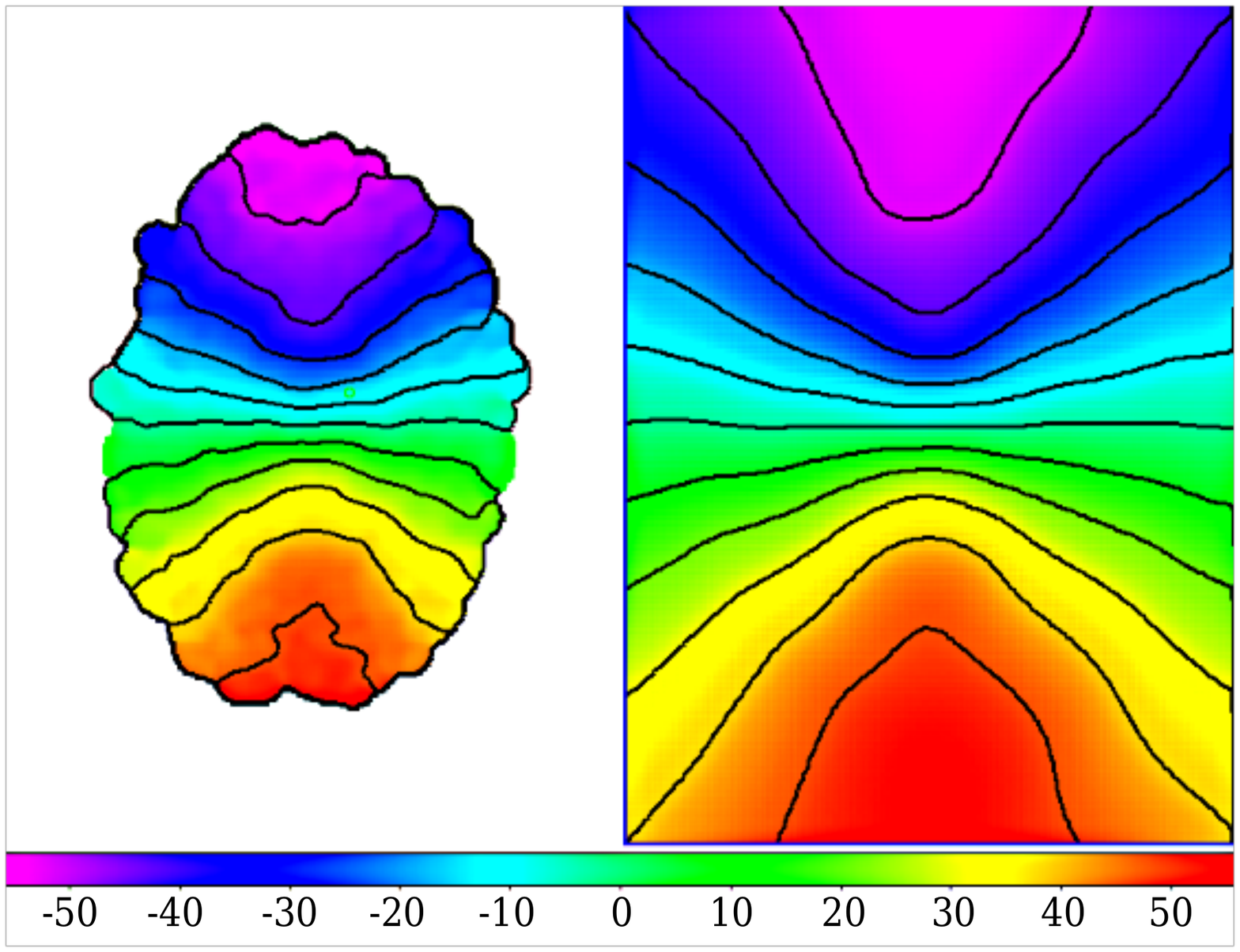}
\end{minipage}
\vspace{-2.15cm}
\caption{Example data products obtained from G0 at 3.0 Gyr,
  $45^\circ$ inclination, and 10 Mpc distance. From left to right, we
  present the optical long-slit spectrum, the H~$\alpha$ velocity
  field, and the H$\,${\sc i} velocity field. The H$\,${\sc i}
  velocity map extends over 7 kpc, reaching a maximum radius of
  3.5 kpc into the receeding and the approaching sides, which is roughly
  the optical radius. The H~$\alpha$ map barely reaches $\sim$3 kpc,
  corresponding to the region with non-negligible emission.
  Velocities are colour-coded equally in both maps according to the
  shown scale. Iso-velocity contours are drawn every 10 km s$^{-1}$.
  The long-slit spectrum also extends over 7 kpc. Pixels in the horizontal
  direction represent the spectral axis, coloured according to
  the intensity of the emission in each velocity channel.}
\label{fig:data_products}

\end{figure}

\subsubsection{H~$\alpha$ velocity fields}	                        %
\label{subsec:velocityfields}                                           %
To construct H~$\alpha$ velocity fields, we map the entire
H~$\alpha$ emission across the galaxy.
The instrumental parameters (pixel scale, seeing, and spectral resolution)
are the same as above. The velocity field
is computed using the IWM scheme defined in equation (\ref{eq:iwm}) and
the rotation curves are extracted using
the {\sc kinemetry} software package \citep{Krajnovic2006}. {\sc kinemetry} performs
a harmonic decomposition of the line-of-sight velocity field in
elliptical rings as a function of the position angle $\theta$,
performing a least-squares minimization of
\begin{equation}\label{kine2}
V_{\rm los}(\theta) = \sum_{j}{A_j \sin(j\theta)}+\sum_{j}{B_j \cos(j\theta)}
\end{equation}
to find the best set of coefficients for each ring. 
In the case of pure circular motions, we have
\begin{equation}\label{eq:pure_circ} V_{\rm los}(\theta) = V_{\rm
    rot}\sin(i) \cos(\theta). \end{equation}
Therefore, in the {\sc kinemetry} expansion, the net amount of rotation would
be proportional to the coefficient $B_1$, with all other components representing
non-circular motions. We calculate the rotation curves in radial bins
of 1 arcsec and adopt as final errors the formal errors of the fit reported
by {\sc kinemetry}. Even though {\sc kinemetry} is in principle
able to treat the ellipticity and position angle of the ellipses as
free parameters during the fitting, we kept them constant and fixed to
their real values in order to avoid one additional source of potential
error.

\subsubsection{H$\,${\sc i} velocity fields}                               %
\label{subsec:HIvelocityfields}                                           %

We also create H$\,${\sc i} velocity fields following the same prescription.
The instrumental parameters are chosen according to the H$\,${\sc i}
survey THINGS \citep{Walter2008}, which, along with LITTLE THINGS \citep{Hunter2012},
represents the best-quality H$\,${\sc i} data
ever used in the cusp-core context. We use a
$5.2\, {\rm km\,s^{-1}}$ channel separation, a pixel scale of
1.5 arcsec/pixel, and a Gaussian beam of 12 arcsec at FWHM, such that the spatial
resolution is $\sim$~580~pc for galaxies at 10 Mpc. Final rotation
curves are determined using {\sc kinemetry} in elliptical rings of 6 arcsec
width. We note that the spatial resolution of THINGS is
better by a factor of 2 in the velocity maps reduced with the
\emph{robust weighting} scheme instead of the \emph{natural weighting}
\citep[see][for details]{Walter2008}. This fact is exploited to
emphasize the advantages of using THINGS velocity maps for
cusp-core studies, although it is sometimes not properly stated. For
example, \citet{Oh2011} mentioned the higher spatial resolution in their
introduction, but then the poorer resolution ($\sim$12 arcsec) was actually used.
This mixup was propagated into \citet{Oh2011b}, in which a spatial
resolution of 6 arcsec was used to construct the mock kinematic observations of the
simulated galaxies. On the other hand, the data cubes used in those
studies had a spectral resolution of 2.6 km s$^{-1}$ (a
factor of 2 better than ours), but most of the velocity maps in a
related study by \citet{deBlok2008} had 5.2 km s$^{-1}$
resolution. Regardless, we have checked that
measured rotation curves are insensitive to the assumed spectral
resolution as long as the Gaussian emission lines are properly sampled.

\subsubsection{Hybrid rotation curves}                                  %
\label{subsec:hybrids}		                                        %

Optical rotation curves have a better spatial resolution than radio
observations, but the latter usually cover much larger radii. For
this reason, it is very convenient to mix both kinds of data when they
are available for the same object, using the H~$\alpha$ measurements in
the inner region and appending the H$\,${\sc i} velocities in the outer part. In
this manner, one minimizes concerns about beam smearing while hopefully
reaching the flat part of the rotation curve with a high enough number of
points in order to put meaningful constraints on the overall shape of the DM
halo using rotation curve fitting methods
(Sec.~\ref{subsec:RCfitting}). 

We create hybrid rotation curves by joining the H$\,${\sc i} data to both the
long-slit and the {\sc kinemetry} H~$\alpha$ rotation curves. Given
that H$\,${\sc i} rotation curves are clearly affected by beam smearing in their
inner parts, we implemented an automatic algorithm to discard the first
H$\,${\sc i} data points, as needed to get a continuous rotation curve with a
positive, monotonically decreasing radial derivative at the
H~$\alpha$/H$\,${\sc i} interface. The hybrid rotation curves are truncated at $\sim$2 
times the optical radius, which is approximated as $\sim$3.2 times the 
scale-length of the stellar disc \citep{Persic1995}, thus yielding radial extents between 5
and 19 kpc, which are representative of real observations.

\subsection{Differentiating cusps from cores via rotation curve fitting}                                     %
\label{subsec:RCfitting}                                                %
When there is kinematic information available beyond the rising part
of the rotation curve, the typical method to differentiate cusps from
cores is fitting different analytic models to the dataset and
choosing the best one on the basis of the minimum $\chi_\nu^2$ of the
fits. Among the variants presented in the literature, the most common
models are the cuspy NFW model (see equations~\ref{eq:nfw_dens}
and \ref{eq:vnfw_v200}), for which
$\rho_{\rm inner}\sim r^{-1}$, and the cored pseudo-isothermal sphere
(hereafter ISO),
for which $\rho_{\rm inner}\sim r^{0}$ and the full profile is given by
\begin{equation}\label{eq:piso_dens}
  \rho_{\raisebox{-2pt}{\textrm{{\tiny ISO}}}}(r) =
  \frac{\rho_0}{1+(r/R_c)^2} ,
\end{equation}
\begin{equation}\label{eq:piso_vel}
  V_{\raisebox{-2pt}{\textrm{{\tiny ISO}}}}(r) = \sqrt{\frac{4\pi
      G\rho_0
      R_c^3}{r}\left[\frac{r}{R_c}-\tan^{-1}\left(\frac{r}{R_c}\right)\right]} ,
\end{equation}
where $\rho_0$ represents the central DM density and $R_c$ the core
radius.

We note that regarding the structure of DM haloes, another criticism
of $\Lambda$CDM simulations is that the few acceptable
NFW fits to observations tend to violate the tight cosmological
relation between the free parameters ($c$, $V_{200}$) predicted by
the simulations themselves. Some authors explicitly use the
cosmological mass-concentration relation as a constraint for the NFW
fits in order to highlight this additional facet of the cusp-core
problem \citep{Kuzio2008}. We do not impose such a restriction.
Instead, we let both parameters vary freely and later investigate the range
of values covered by the fits.

When we fit the models to the theoretical velocity profiles
$V_{\rm dm}$, $V_{\rm tot}$, and $V_{\rm cir}$, we first re-sample
these curves to the same positions at which the mock hybrid rotation curves are
measured so that we can properly compare the results and interpret
the differences. These radial positions depend on the spatial resolution and the
extents of the H~$\alpha$/H$\,${\sc i} components of each galaxy. Note that
at 10 Mpc, the inner theoretical velocity profiles are oversampled
because they were created in steps of $\sim$100 pc, whereas the mock
H~$\alpha$ rotation curves are defined every $\sim$50 pc (2 points per seeing).
However, this is not a concern
because we find the results at 10 and 20 Mpc to be essentially identical,
as we will show in Sec.~\ref{subsec:RCfit_results} and Sec.~\ref{subsec:distincl}.
We assume constant error bars of 1 km s$^{-1}$ when fitting the
theoretical curves.

We recall that this work is focused on the minimum disc approximation;
thus, we do not attempt to explicitly account for the baryonic
contribution to the rotation curves. In other words, we fit the
different analytic models proposed for the DM halo directly to the
\emph{observed} data or to the theoretical circular velocity profiles.
To avoid spurious results, we do not try fits to rotation curves with less than 8 points,
which in practice only filters out the velocity profiles from the Dwarf1 and G0
simulations at 80 Mpc.

\section{Results}							%
\label{sec:Results}							%

Here, we describe the theoretical velocity profiles of our simulated
galaxies, their mock kinematic observations, and some statistics regarding
the fits with the NFW and the ISO analytic models. We give special
attention to the differences amongst the various velocity profiles and
to the effects of spatial resolution and inclination. When reporting
on general trends based on observations at all inclinations, we use the
term $V_{\rm kin}$ to refer to hybrid rotation curves where the H~$\alpha$
part was extracted with {\sc kinemetry} from the velocity maps,
and $V_{\rm ls}$ to refer to those hybrids where the H~$\alpha$ portion uses
the mock \emph{long-slit} data.

\subsection{Density profiles}                                           %
\label{subsec:densities}                                               	%

\begin{figure}
  \includegraphics[width=\linewidth]{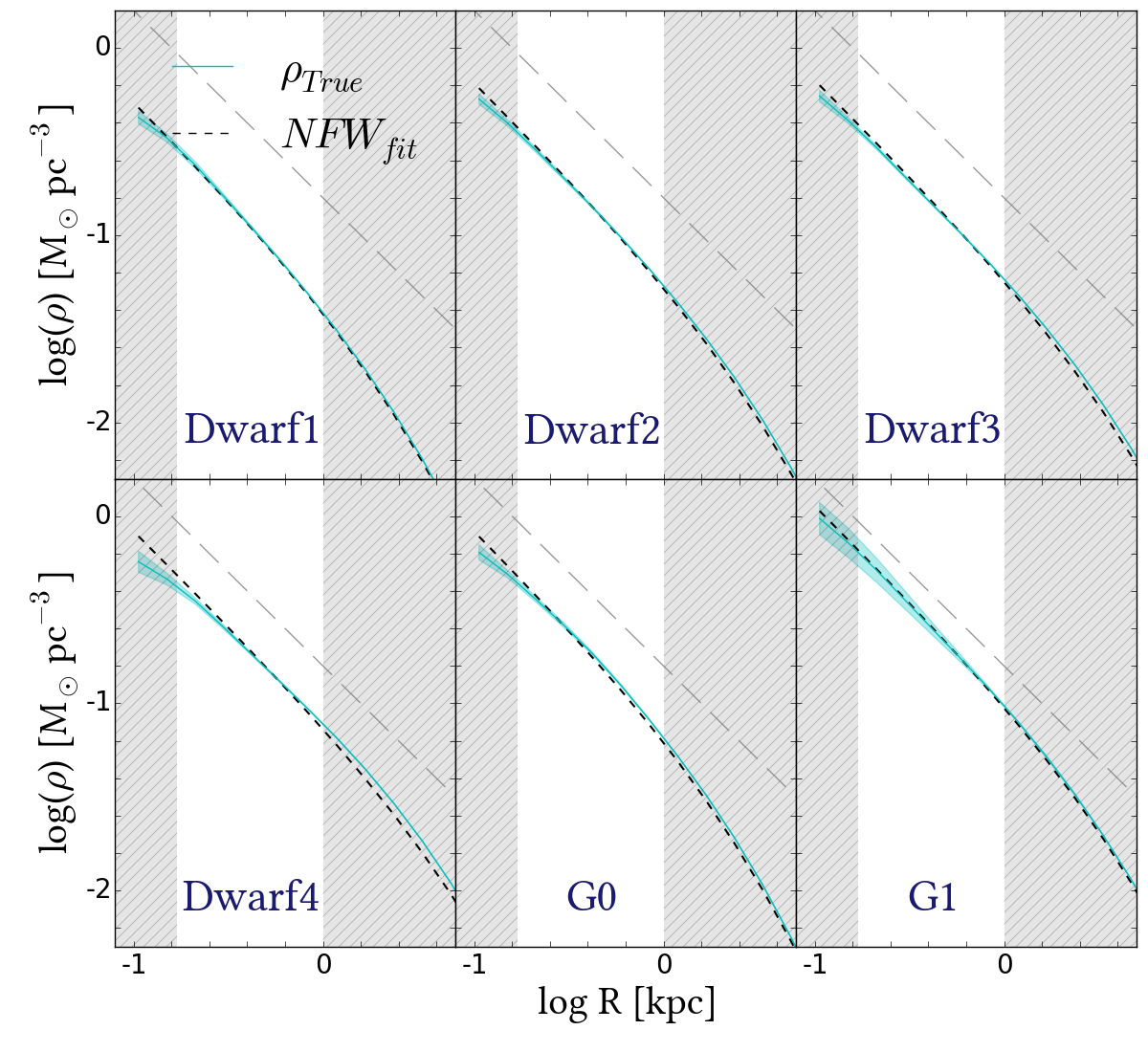}
  \caption{Real DM density profiles in the simulations. The solid cyan
    lines represent the mean of all the snapshots, and the shaded cyan regions
    denote the 1-$\sigma$ scatter. In each panel, the short dashed black line is the mean
    of the best NFW fits to the real density profiles. The hatched, grey shaded regions
    indicate the central region between 0.17 kpc (the
    position of our first velocity measurement) and 1 kpc, where the
    cusp-core discrepancy has been more debated. Straight diagonal lines (long dashed) with a
    slope of $-1$ (i.e.~the cusp profile expected for an NFW profile)
    are shown for comparison. All of the simulated galaxies clearly retain
    central DM cusps throughout their evolution.}
\label{fig:densities}
\end{figure}

In Fig.~\ref{fig:densities}, we present a compilation of the real DM
density profiles in our simulations. Remarkably, we see that our DM
haloes remain basically unchanged\footnote{This is likely the case because our
simulations use the \citet{Springel2003} effective equation of state
to model the effects of supernova feedback and do not include explicit
feedback-driven outflows; as a result, the simulated
galaxies form stars steadily, in contrast with the very bursty star formation
exhibited by simulations that explicitly include multi-channel stellar
feedback \citep[e.g.][]{Hopkins2014,Sparre2015}. As detailed in Sec.~
\ref{sec:Introduction}, strong bursts of star formation and the associated
supernova feedback-driven outflows may be able to transform cusps into
cores. Thus, it is desirable \emph{not} to include this still-uncertain effect
in our simulations because we wish to test whether cusps can be mistaken
for cores, not whether cusps can be transformed into cores via baryonic
processes.}, even in the case of G1, which
develops a relatively strong bar. 
The simulated galaxies' DM profiles can be accurately represented
with the NFW formula, and there is no ambiguity about their cuspy
nature. We emphasize this point by including in each panel an inclined
straight line with a slope of $-1$ to facilitate a visual comparison,
recalling that a core would appear here as a horizontal line.

\begin{figure*}
  \includegraphics[width=\textwidth,height=7.5cm]{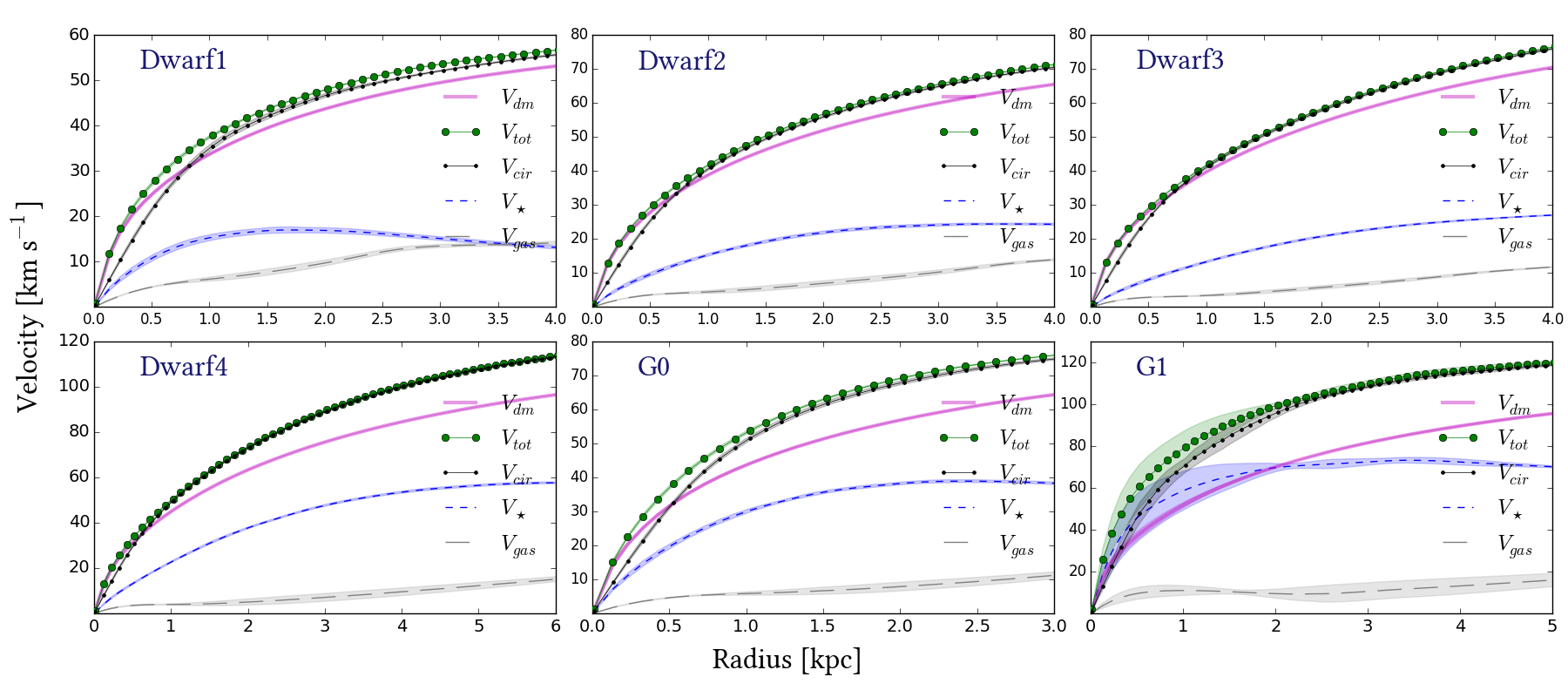}
  \caption{Inner parts of the theoretical rotation curves as defined in
    Sec.~\ref{sec:Analysis}. The thick magenta line represents $V_{\rm dm}$,
    solid green line with circles is $V_{\rm tot}$, solid black line
    with dots is $V_{\rm cir}$, the short-dashed blue line is $V_{\star}$,
    and long-dashed gray line is for $V_{\rm gas}$. In each case the main 
    line is the mean curve from all snapshots, and the shaded
    regions indicate the 1-$\sigma$ scatter.
    Except for G1, the simulated galaxies are DM-dominated at
    all radii. The differences between $V_{\rm tot}$ and $V_{\rm cir}$
    in the central $\sim1$ kpc, which are typically $\sim 4-5$ km s$^{-1}$,
    are due to pressure support; below, we shall see that this difference
    can have important consequences for the DM profile shape inferred
    from the rotation curves.}
\label{fig:Teo_RCs}
\end{figure*}

\subsection{Rotation curves}						%
\label{subsec:RC_results}						%

In Fig.~\ref{fig:Teo_RCs}, we present a compilation of the
theoretical velocity profiles for all galaxies and snapshots. Recall
that $V_{\rm gas}$, $V_{\star}$, and $V_{\rm dm}$ represent the
circular velocity profiles generated by the gravitational potential of
the individual components, $V_{\rm tot}$ is their sum in quadrature,
and $V_{\rm cir}$ is the actual circular speed of the gas in the disc.
The negligible scatter demonstrates that our target configurations
remain highly stable for all galaxies but G1. Apparently the formation
of the bar induces time evolution in the azimuthally averaged
velocity profiles, even if the DM structure does not change much. A
visual inspection of the face-on stellar maps confirms a smooth,
stable morphology without substructures for galaxies D1 to D4. Some
minor distortions in the central part of G0 are observed, but they are
not a concern because the initial potential-velocity structure remains the
same. In the following, we present results for the whole sample,
excluding G1 at first, and we then comment on this system afterwards.

Fig.~\ref{fig:Teo_RCs} shows that baryons are dynamically
sub-dominant in all cases. Inside the first kiloparsec,
$V_{\rm tot}$ exceeds $V_{\rm dm}$ by less than $6\,{\rm km\, s^{-1}}$
for galaxies D1 to D4 and less than $10\,{\rm km\, s^{-1}}$ for G0.
The mean difference $\left<V_{\rm tot}-V_{\rm dm}\right>$ is less
than $3\,{\rm km\, s^{-1}}$ in this inner region.
Beyond the first kiloparsec, the baryonic
contribution to $V_{\rm tot}$ is $\sim$10 per cent in galaxies D1, D2, and D3
and $\sim$20 per cent in D4 and G0.

It is also apparent in Fig.~\ref{fig:Teo_RCs} that $V_{\rm cir}$
closely follows $V_{\rm tot}$ in the outer region, but it is slightly
less than $V_{\rm tot}$ in the center. The general agreement demonstrates that our
systems are in a rotational equilibrium that is mainly sustained by
gravity. Nevertheless, as the rotational speed of the gas nearly
coincides with the expectation from its radial acceleration
(see Sec.~\ref{subsec:coreillusion}), the small
differences in the centres of the galaxies imply that the radial
acceleration experienced by the gas component is smaller than that
expected from the gravitational potential, which can be interpreted as
evidence for pressure support. The residual $(V_{\rm tot}-V_{\rm cir})$
peaks between 6 and 8 km s$^{-1}$ at 0.17 kpc and linearly
decreases to $\sim$1-2 km s$^{-1}$ at 1 kpc, which implies a mean pressure
support correction of $\sim\,$4-5 km s$^{-1}$ to be added to
$V_{\rm cir}$ over this radial range. The
importance of this small difference in the cusp-core scenario will
become clear in the next sections.

\begin{figure*}
  \includegraphics[width=\textwidth,height=7.5cm]{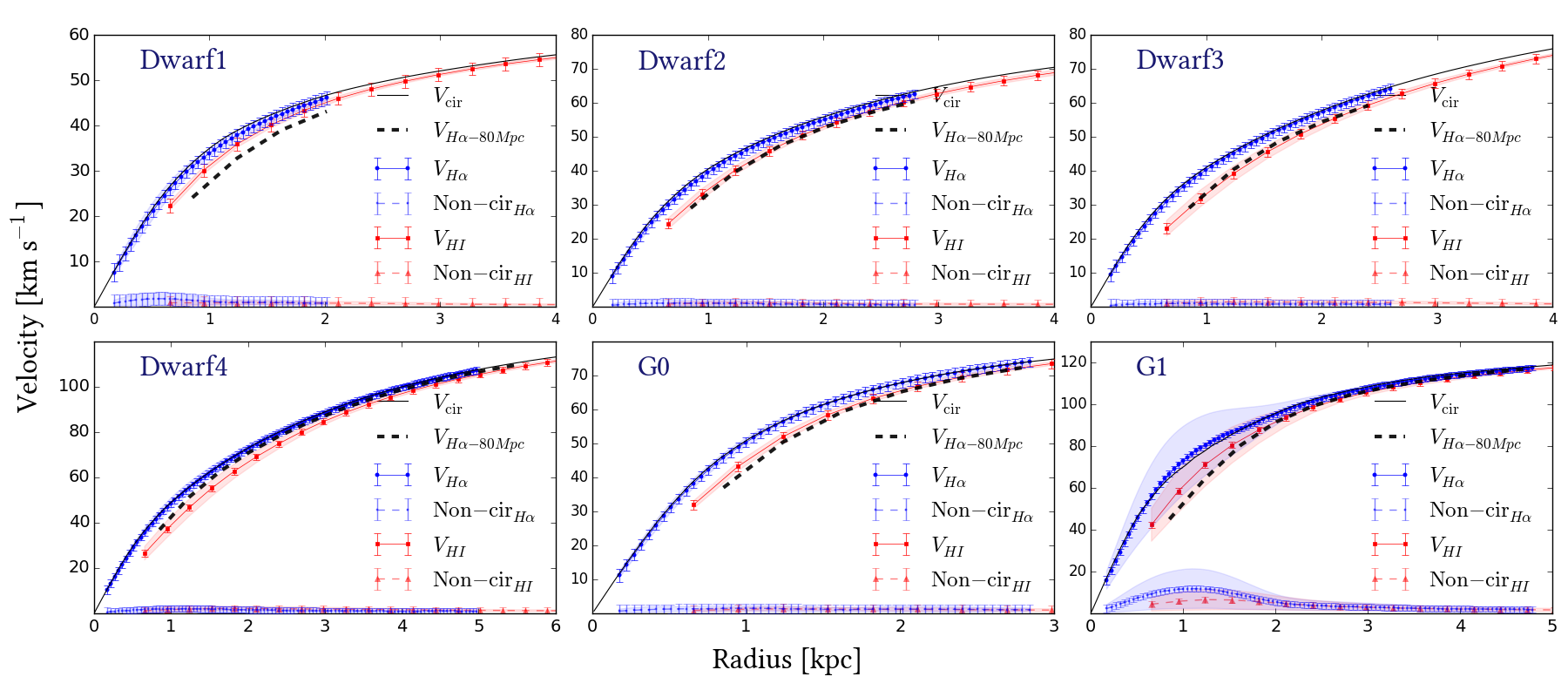}
  \caption{Inner parts of the mock observed rotation curves and net amount of
    non-circular motion as inferred by {\sc kinemetry} for the simulated
    galaxies placed at 10 Mpc and viewed at an inclination of $45^\circ$.
    H~$\alpha$ observations appear in blue and H$\,${\sc i} in red. In
    each case, the rotation curve is represented with a solid line and 
    the non-circular motion profile with a dashed line.
    For each of the above, the plotted curve represents the mean curve from all
    snapshots, the shaded region indicates the 1-$\sigma$ scatter, and the
    error bars represent the typical error bar at each radius on individual
    rotation curves. The theoretical $V_{\rm cir}$ is also shown. The thick
    black dashed line in each panel is the H~$\alpha$ rotation curve at 80 Mpc.
    The rotation curves are dominated by circular motion at all radii. The
    H~$\alpha$ rotation curves almost perfectly trace the true circular motions of the gas 
    for galaxies placed at 10 Mpc but underestimate them by as much as $\sim 5$
    km s$^{-1}$ when the galaxies are placed at 80 Mpc (and even more so for
    G1 because of the non-negligible amount of non-circular motion in the central
    2 kpc). The H$\,${\sc i} rotation curves underestimate the true circular velocity
    in the central few kiloparsecs because of beam smearing.}
\label{fig:Obs_RCs}
\end{figure*}

\begin{figure}
  \includegraphics[width=\linewidth]{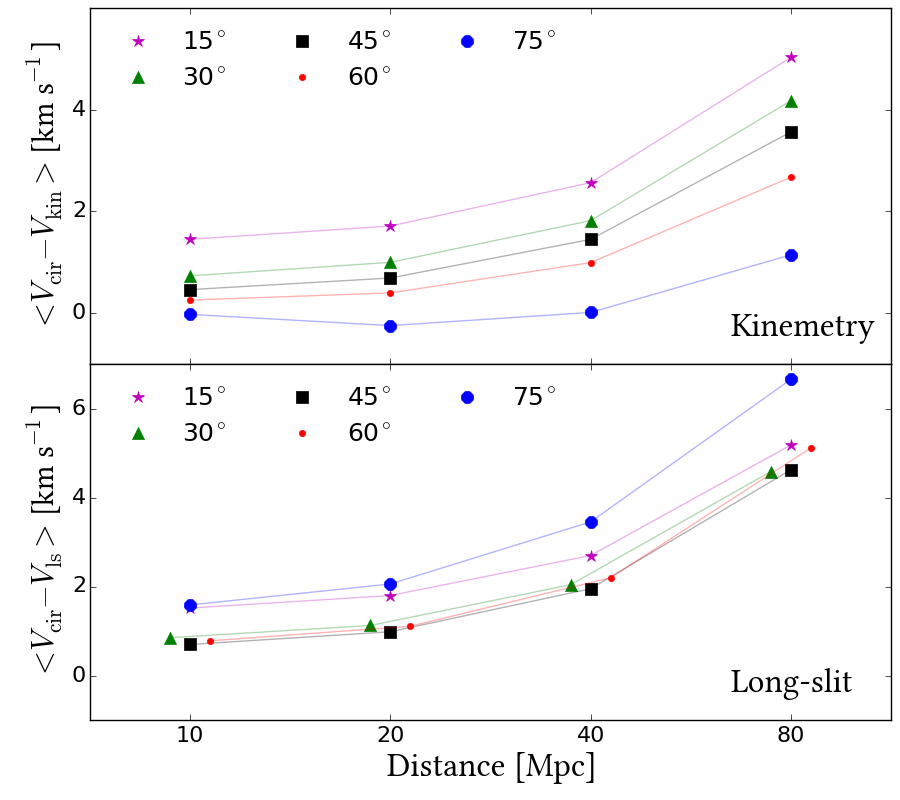}
  \caption{Mean H~$\alpha$ velocity residuals
    $\left<V_{\rm cir}-V_{\rm kin}\right>$ (top), and $\left<V_{\rm cir}-V_{\rm ls}\right>$ (bottom)
    as a function of distance (spatial resolution) and inclination for all galaxies and snapshots.
    Some markers are slightly shifted horizontally to make the plot more readable.
    The residual increases as the spatial sampling gets coarser, 
    reaching as much as $\sim$5 km s$^{-1}$.
    The rotation curves from the {\sc kinemetry} analysis exhibit larger residuals
     in galaxies that are viewed more face-on. In the long-slit case, the largest residual
    occurs in discs inclined at $75^\circ$.}
\label{fig:mean_residuals}
\end{figure}

In Fig.~\ref{fig:Obs_RCs}, we present some of the rotation curves and
the non-circular motion profiles extracted from the mock H~$\alpha$ and
H$\,${\sc i} velocity maps using {\sc kinemetry} for the simulated 
galaxies placed at 10 and 80 Mpc and viewed at an inclination
of $45^\circ$. We also show the theoretical $V_{\rm cir}$ to facilitate the
comparison. It is clear from the figure that our galaxies are completely
dominated by rotational motions
at all radii (except for G1). Fig.~\ref{fig:mean_residuals}
shows the mean (taken over all simulations and snapshots)
$\langle V_{\rm cir} - V_{\rm kin} \rangle$ residual as a
function of spatial resolution for the simulated galaxies viewed from multiple inclinations,
considering only the inner region with H~$\alpha$ emission.
These figures demonstrate that H~$\alpha$ observations of the simulated galaxies
at a 100-pc spatial resolution (placed at a distance of $\sim$10 Mpc) almost perfectly trace the
actual circular motions of the gas ($\langle V_{\rm cir} - V_{\rm kin}
\rangle < 2$ km\nobreakspace s$^{-1}$). The residual increases as the 
spatial resolution diminishes and reaches as much as $\sim 5$ km
s$^{-1}$ for a 800-pc resolution ($\sim$80 Mpc distance) and inclination of $15^\circ$.
The H$\,${\sc i} rotation curves (the red lines in Fig.~\ref{fig:Obs_RCs})
underestimate the true rotational velocities in the central
few kiloparsecs of the galaxies (even when the galaxies are placed at 10 Mpc)
but are in agreement with $V_{\rm cir}$ at larger radii.
The effect of beam smearing is
stronger with distance, but because we use the H$\,${\sc i} data only to extend the
optical rotation curves into the flat part, this is not a primary concern for our
discussion.

G1 qualitatively differs from the other simulations. In this simulation,
the stellar component dominates the gravitational
potential within the first 2 kpc. As a result, $V_{\rm tot}$
exceeds $V_{\rm dm}$ by $27 \pm 5$ km s$^{-1}$ at
1 kpc (see Fig.~\ref{fig:Obs_RCs}), which represents a discrepancy
of $\sim$50 per cent. In the
outer region, this excess is about 30-40 per cent. The residual
$(V_{\rm tot}-V_{\rm cir})$ is $15 \pm 5$ km s$^{-1}$ at
0.17 kpc, $8\pm5$ km s$^{-1}$ at 1 kpc, and
2 km s$^{-1}$ at $\sim 2.5$ kpc.
However, in this case, the difference is not 
simply attributable to the
effect of pressure support because of the presence of the bar and its
associated non-circular motions,
which invalidate the axisymmetric
approximation. 
We emphasize that G1 resembles
classical dwarf galaxies and LSBs in terms of several properties, such as the
maximum circular velocity and stellar mass; for this reason, we keep it in our
sample even though it is not close to the ideal case. Modelling the complex
kinematics of barred potentials is far from simple and beyond the
scope of this work.

The mock H~$\alpha$ rotation curves obtained from the 2D velocity maps underestimate the
circular motions of the gas more severely in less-inclined (i.e.~more face-on) galaxies.
Despite the differences being small, they are systematic, as
shown in Fig.~\ref{fig:mean_residuals}. On a
case-by-case basis, some scatter and random small-scale fluctuations
are visible. In particular, the rotation curves for galaxies viewed at an
inclination of $15^\circ$ often exhibit
prominent shape distortions that are not seen at other inclinations.

We find the same trend with spatial resolution but not with inclination in the
mock long-slit data. In this case, the rotation curves of galaxies
viewed at an inclination of $75^\circ$ exhibit the most
underestimated circular velocities, followed by galaxies at $15^\circ$;
for other inclinations, the long-slit rotation curves are effectively
independent of inclination (see Fig.~\ref{fig:mean_residuals}).
The long-slit rotation curves are a bit
noisy and exhibit more scatter than those from the 2D velocity maps,
mainly at low inclinations and high spatial samplings (small distances). Nevertheless, the
average difference $\left<V_{\rm cir}-V_{\rm ls}\right>$ is comparable
to the previous case. The only exception is G1 viewed at  an inclination of
$75^\circ$, for which the average velocity underestimation at 80 Mpc increases to
$\sim 11$ km s$^{-1}$ because of the galaxy's lack of symmetry during
the second half of the simulation.

\subsection{Rotation curve fitting}                                     %
\label{subsec:RCfit_results}                                            %

\begin{figure} \centering
  \includegraphics[width=\linewidth]{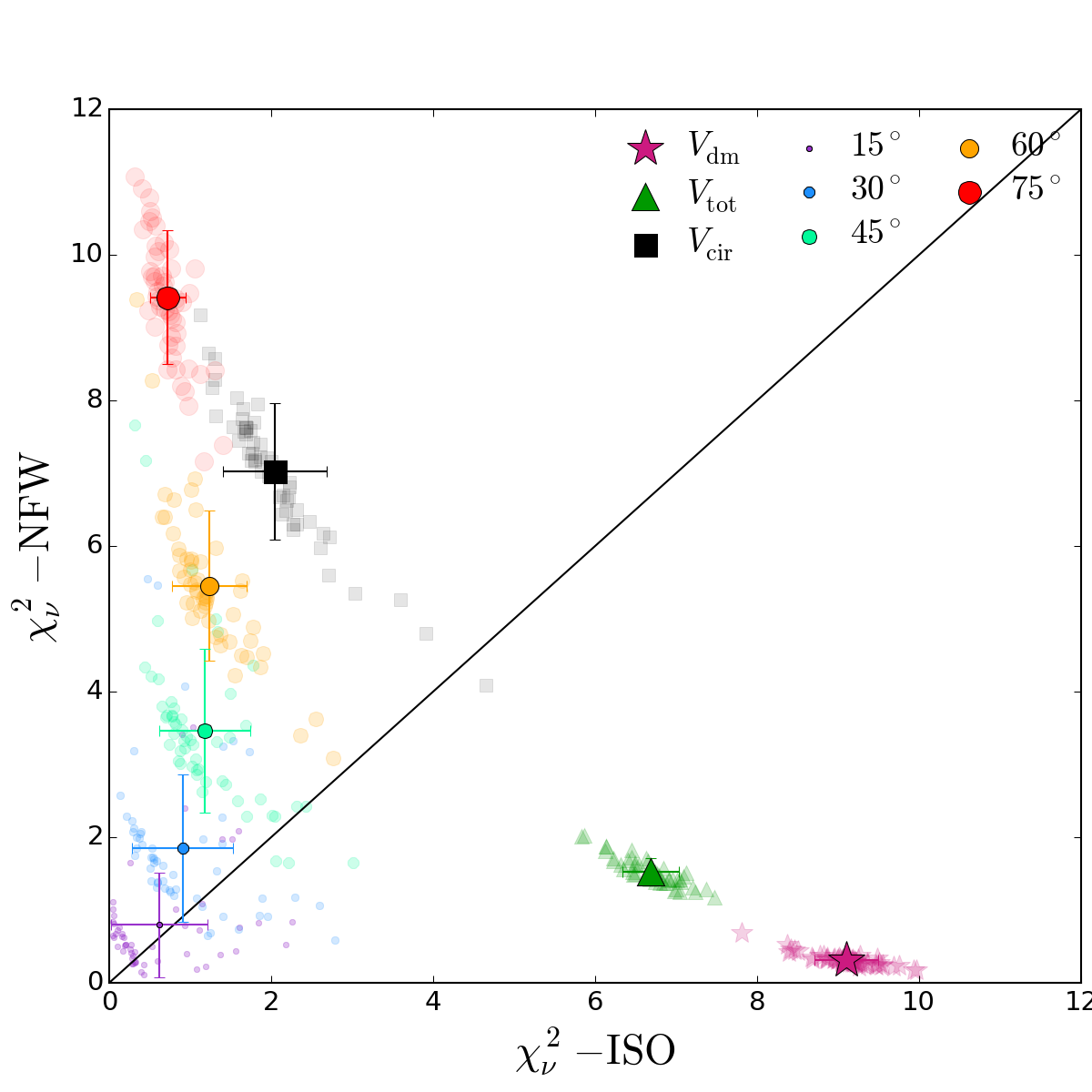}
  \caption{Distribution of $\chi_\nu^2$ obtained by fitting the NFW and
    ISO models to the various types of rotation curves for the Dwarf2
    simulation placed at 10 Mpc. The shapes, colours, and sizes of the markers are coded according
    to the specific rotation curve used, as detailed in the legend. 
    Background semi-transparent symbols correspond to the results for individual snapshots, the solid
    symbols in the front denote the centroids of the corresponding clouds of points, and
    the error bars represent the 1-$\sigma$ scatter in the horizontal and vertical directions.
    The black diagonal line represents equality between the goodness of
    the fits. For this simulation, the theoretical rotation curves $V_{\rm dm}$ and
    $V_{\rm tot}$ are better fit by the cuspy NFW model. However, the theoretical
    rotation curve $V_{\rm cir}$ is better fit by the ISO model
    because pressure support causes $V_{\rm cir}$ to be less than $V_{\rm tot}$
    in the central $\sim 1$ kpc. The mock rotation curves obtained using {\sc kinemetry}
    are generally better fit with the ISO model, especially when the galaxies are viewed
    at inclinations of $45^\circ$ or greater. These results demonstrate that rotation
    curve fitting can indicate the presence of a core when the true DM profile is cuspy.}
\label{fig:X2_behav}
\end{figure}

In Fig.~\ref{fig:X2_behav}, we compare the reduced $\chi_\nu^2$ of the
best-fitting NFW and ISO models for the various types of rotation curves
for the D2 simulation placed at $10\,{\rm Mpc}$. This is a good
example of the general trends in the whole sample, so we use it to
introduce our main findings before going into a more detailed
analysis.

Points that lie below the one-to-one line in Fig.~\ref{fig:X2_behav},
represent better agreement of the data with
the NFW model, and points above the diagonal indicate that the ISO model
provides a better description of the data. We see that the NFW profile
fits the theoretical $V_{\rm dm}$ and $V_{\rm tot}$ curves much
better, as expected because by construction the central potential of the galaxy is dominated
by the DM, which obeys an NFW profile. The
results from fitting $V_{\rm tot}$ are somewhat closer to the line of equality 
between models than those from $V_{\rm dm}$ because of the effect of
the baryonic contribution to the rotation curve. Surprisingly,
$V_{\rm cir}$ is better fit by the ISO model than by the NFW model. The fits to
the mock rotation curves tend to favour the ISO
model, especially when the galaxies are viewed at high inclinations.

\begin{figure} \centering
  \includegraphics[width=\linewidth]{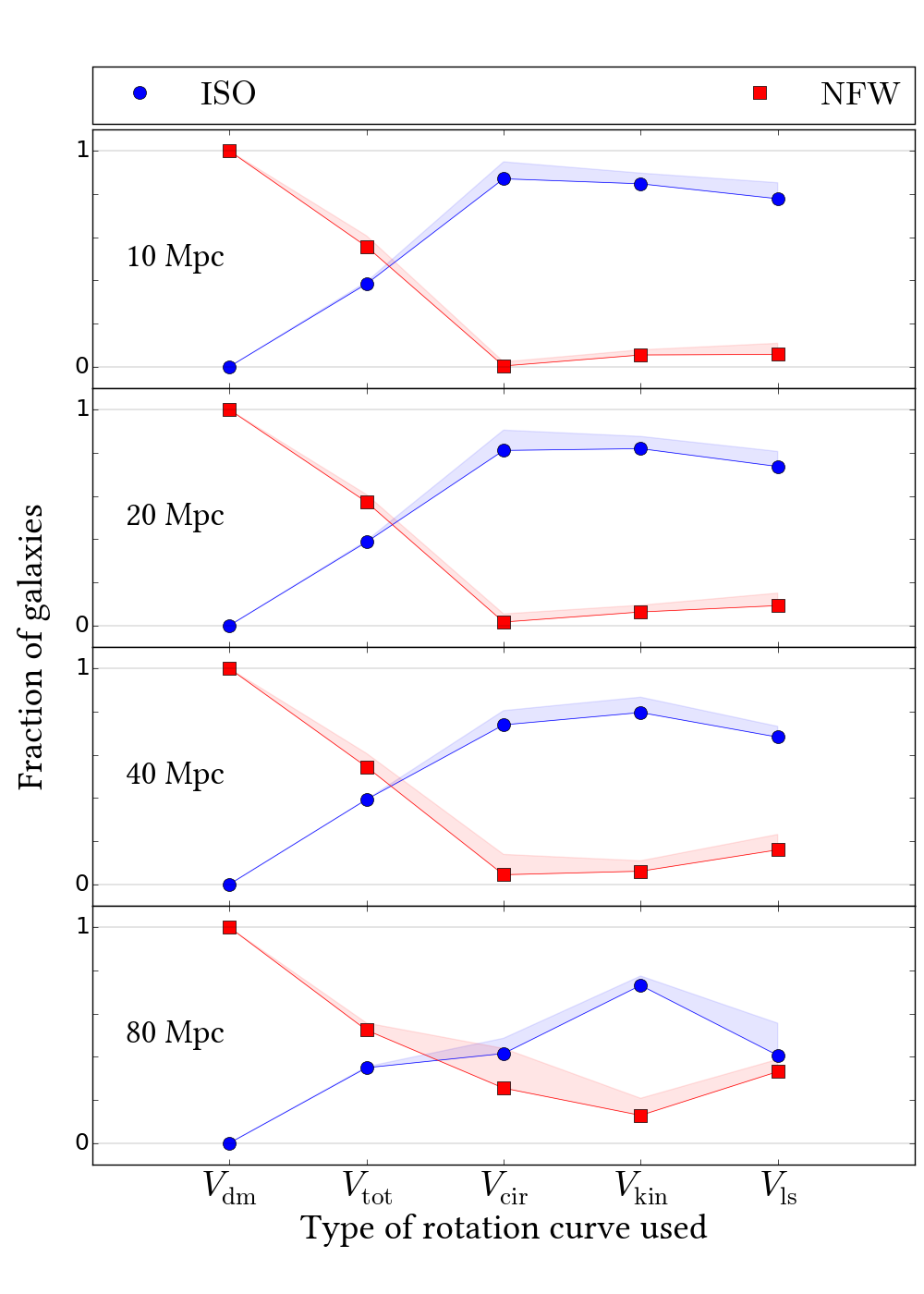}
  \caption{Fraction of galaxies that are better represented by the NFW
    or ISO models based on the best-fit $\chi_\nu^2$ values. Solid
    lines are established by demanding that one of $\chi_\nu^2$ values
    is at least 1.5 times bigger than the other one. The upper limit
    of the shaded regions represents the upper limits on the NFW/ISO
    fractions, which were obtained using a threshold of 1.1 instead of 
    1.5. $V_{\rm dm}$ is always better fit by the NFW model, as expected
    because the DM profiles of the simulations obey an NFW profile. In contrast,
    $V_{\rm tot}$ is sometimes better fit by the ISO model because of the baryonic
    contribution to the potential. In the vast majority of cases, the cored
    ISO model provides a better fit to the theoretical circular velocity profile $V_{\rm cir}$
    and the mock rotation curves, despite the simulations containing cuspy dark matter profiles
    by construction.}
\label{fig:X2_fractions}
\end{figure}

We now discuss the results for all of the galaxies and
snapshots. In Fig.~\ref{fig:X2_fractions} and Table~\ref{tab:X2}, we
show the fraction of cases in which one of the models is preferred
over the other for each type of circular velocity profile.\footnote{We remind the reader
that the theoretical rotation curves ($V_{\rm dm}$, $V_{\rm tot}$, and
$V_{\rm cir}$) are resampled to the same positions of the mock observations;
consequently, the results depend on the assumed distance.
Also recall that we \emph{observe} the curves at a rate of two points per seeing.
So, at a distance of 80 Mpc, the spatial resolution of the mock observations
is $\sim$800 pc, but they are sampled at a rate of $\sim$1/400 pc$^{-1}$, and this
is what matters for the theoretical velocity profiles.} We
estimate a lower limit on these fractions by demanding one $\chi_\nu^2$
to be at least 1.5 times smaller than its counterpart and stating
that both models are equally good otherwise. For this reason, the sum of
the NFW and ISO fractions is not always unity. Upper limits are obtained
by relaxing our threshold on the $\chi^2_{\nu}$'s ratio from 1.5 to 1.1.
We find 100 per cent of the $V_{\rm dm}$ rotation curves to be better represented
by the cuspy NFW model for all the tested spatial samplings
($\leq$~400 pc for $D\leq80$ Mpc)
in perfect agreement with the underlying DM distributions in the simulations.

Using $V_{\rm tot}$, we find that between 52 and 61 per cent of the rotation curves
are better fit with the NFW profile, whereas between 35 and 39 per cent of the cases
are better fit by the ISO model.
This means that the signature of the NFW haloes is still
detectable, although the inclusion of the baryonic contribution to the
potential without an explicit correction during
the fit introduces considerable errors. In particular, we note that the NFW
model is always better for D1, D2, and D3, but it is disfavoured for D4 and G0.
In the case of G1, both models provide comparable fits to $V_{\rm tot}$.
Strikingly, the $V_{\rm cir}$ rotation curves for simulated galaxies viewed
at $D\leq40$ Mpc, i.e. whose inner parts
are sampled every $\leq$~200 pc,
are better fit with the ISO model for 
74-95 per cent of the sample. Such a result would be
typically interpreted as proof of the ubiquitous presence of cores
in the central region of galactic haloes, but because there are no cores in
our simulations, this conclusion would be incorrect.
This fact is particularly shocking because $V_{\rm cir}$ is a
perfect theoretical measurement of the gas rotational velocity, and
it occurs even when the galaxies are viewed at 10 Mpc (i.e. the
spatial resolution is optimal). When viewed at 80 Mpc, the difference between models 
decreases, although the illusion of DM cores does not vanish entirely;
the NFW profile is preferred in 25-44 per cent of the cases, and the ISO 
model provides a better fit in 42-49 per cent of the cases.

Regarding the mock observations and considering all distances and
inclinations, the ISO model provides a better fit to $V_{\rm kin}$ for
73-90 per cent of the sample, and the NFW model is preferred for only
5-21 per cent of the sample. Thus, this type of
analysis applied to our \emph{observed} rotation curves would provide
\emph{strong evidence} of the widespread existence of cores in the
DM haloes of dwarf galaxies and LSBs, which is in tension with the
cuspy nature of the DM haloes in our simulations. This
effect is more misleading for more nearby galaxies. It is worth
emphasizing that whatever the reason for the ISO model being a better
fit to the mock observations, the effect is already evident when
we analyse the theoretical circular motions of the gas, $V_{\rm cir}$.
We note the same trend in the results from long-slit data as for the
2D velocity maps. The only difference occurs at 80 Mpc, where 
the ISO model is preferred by a much smaller margin and the number
of unresolved cases is larger when the long-slit rotation curves are used.

\begin{figure} \centering
  \includegraphics[width=\linewidth]{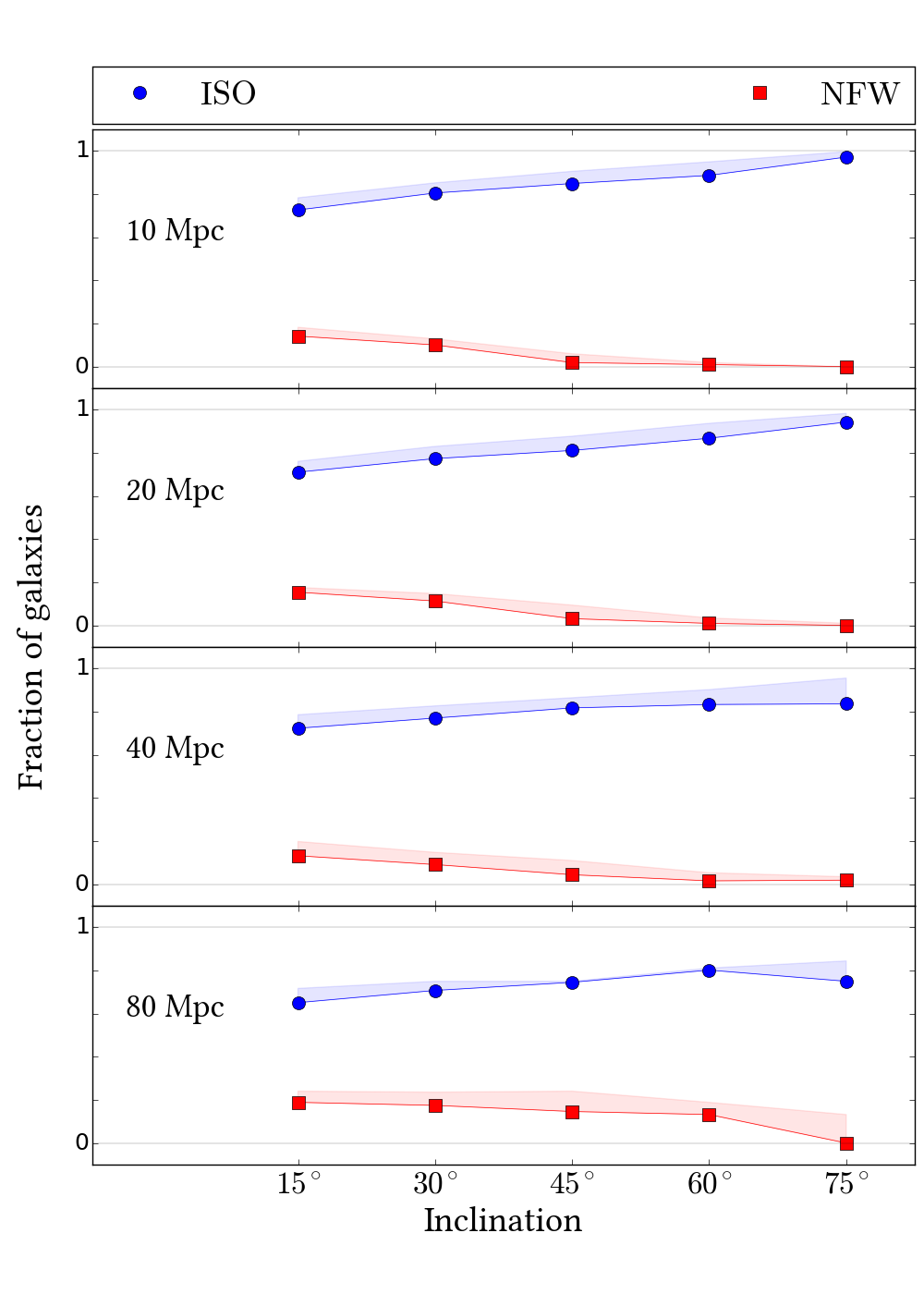}
  \caption{Fraction of mock rotation curves that are better described by the
    NFW/ISO models as a function inclination for 
    the mock 2D observations analysed with {\sc kinemetry}. The different rows
    correspond to different assumed distances (from top to bottom: 10, 20, 40, and 80 Mpc).
    The symbols and colours are the same as in Fig.~\ref{fig:X2_fractions}. Independent of
    inclination and distance, the rotation curves derived from the 2D velocity maps are always
    better fit by the cored ISO model than by the cuspy NFW model.}
\label{fig:kin_X2_fractions}
\end{figure}

In Fig.~\ref{fig:kin_X2_fractions} and Table~\ref{tab:X2_obs}, we
categorize the results from the 2D mock observations according to the
inner spatial resolution
and inclination. We see that the NFW profile is virtually never
preferred when the galaxies are viewed at an inclination of $75^\circ$.
Considering specific
combinations of distance and inclination, we see that the NFW model
sometimes provides better fits, but these are limited to a maximum of
24 per cent of the sample, whereas the ISO profile fits the data
better in 65-99 per cent of the cases. In Fig.~\ref{fig:kin_X2_fractions},
note that the blue (ISO) and red (NFW) regions never overlap, reflecting
the fact that the fraction of rotation curves that are better fit with the ISO model is
always greater than the fraction better fit with the NFW model. The difference
between the two fractions increases with inclination.
Finally, we note that the gap between the coloured stripes becomes narrower
at 80 Mpc, as the samplig gets poorer.

\begin{table}
\caption{Percentages of rotation curves that are better represented by the NFW or ISO models
according to the type of rotation curve and the spatial resolution (assumed distance).}
\label{tab:X2}
\begin{center}
\begin{tabular}{cccccc}
\hline
\multicolumn{2}{c}{H~$\alpha$ PSF (pc)} & $\sim$100 & $\sim$200 & $\sim$400 & $\sim$800 \\
\hline
\multicolumn{2}{c}{D (Mpc)} & 10 & 20 & 40 & 80 \\
\hline
 \multirow{3}{*}{$V_{dm}$} & NFW & 100 (100)  & 100 (100)  & 100 (100)  & 100 (100)  \\ 
 & ISO & 0 (0)  & 0 (0)  & 0 (0)  & 0 (0)  \\ 
 & Both & 0 (0)  & 0 (0)  & 0 (0)  & 0 (0)  \\ 
\hline
 \multirow{3}{*}{$V_{tot}$} & NFW & 56 (61)  & 57 (61)  & 54 (61)  & 52 (56)  \\ 
 & ISO & 38 (39)  & 39 (39)  & 39 (39)  & 35 (35)  \\ 
 & Both & 6 (0)  & 4 (0)  & 6 (0)  & 13 (9)  \\ 
\hline
 \multirow{3}{*}{$V_{cir}$} & NFW & 0 (2)  & 2 (5)  & 4 (14)  & 25 (44)  \\ 
 & ISO & 87 (95)  & 81 (91)  & 74 (81)  & 42 (49)  \\ 
 & Both & 13 (3)  & 17 (4)  & 22 (6)  & 33 (8)  \\ 
\hline
 \multirow{3}{*}{$V_{kin}$$^\mathrm{a}$} & NFW & 5 (8)  & 6 (9)  & 6 (11)  & 13 (21)  \\ 
 & ISO & 85 (90)  & 82 (88)  & 80 (87)  & 73 (77)  \\ 
 & Both & 10 (2)  & 12 (3)  & 14 (2)  & 14 (2)  \\ 
\hline
 \multirow{3}{*}{$V_{ls}$} & NFW & 6 (11)  & 9 (15)  & 16 (23)  & 33 (39)  \\ 
 & ISO & 78 (85)  & 74 (81)  & 68 (73)  & 41 (56)  \\ 
 & Both & 16 (4)  & 17 (4)  & 16 (4)  & 26 (6)  \\ 
  \hline
\end{tabular}
\end{center}
Note: The fiducial NFW/ISO fractions are estimated by demanding that one of the
$\chi_\nu^2$ values to be at least 1.5 times smaller than the other
one. Values in parentheses require a minimum ratio of 1.1 between
the $\chi_\nu^2$ values. The rows labeled `Both' correspond to cases
for which the $\chi_\nu^2$ values of the NFW and ISO fits differ by less than a
factor of 1.5 (1.1).
$^\mathrm{a}$ The values for $V_{\rm kin}$ and $V_{\rm ls}$ are for the mock rotation curves at all
inclinations.
\end{table}

\begin{table}
\caption{Similar to Table~\ref{tab:X2}, but for the rotation curves obtained by
analysing the mock 2D velocity maps with {\sc kinemetry} ($V_{\rm kin}$)
and classified according to the assumed distance and inclination.}
\label{tab:X2_obs}
\begin{center}
\begin{tabular}{cccccc}
\hline
\multicolumn{2}{c}{H~$\alpha$ PSF (pc)} & $\sim$100 & $\sim$200 & $\sim$400 & $\sim$800 \\
\hline
\multicolumn{2}{c}{D (Mpc)} & 10 & 20 & 40 & 80 \\
\hline
 \multirow{3}{*}{$15^{\circ}$} & NFW & 14 (18)  & 15 (18)  & 13 (20)  & 19 (24)  \\ 
 & ISO & 73 (78)  & 71 (76)  & 72 (79)  & 65 (72)  \\ 
 & Both & 13 (3)  & 14 (6)  & 14 (2)  & 16 (4)  \\ 
\hline
 \multirow{3}{*}{$30^{\circ}$} & NFW & 10 (13)  & 11 (15)  & 9 (15)  & 17 (24)  \\ 
 & ISO & 81 (85)  & 77 (83)  & 77 (83)  & 71 (75)  \\ 
 & Both & 9 (2)  & 11 (2)  & 14 (3)  & 12 (1)  \\ 
\hline
 \multirow{3}{*}{$45^{\circ}$} & NFW & 2 (6)  & 3 (9)  & 4 (11)  & 15 (24)  \\ 
 & ISO & 85 (91)  & 81 (88)  & 82 (86)  & 75 (75)  \\ 
 & Both & 13 (3)  & 16 (3)  & 14 (3)  & 11 (1)  \\ 
\hline
 \multirow{3}{*}{$60^{\circ}$} & NFW & 1 (2)  & 1 (3)  & 2 (5)  & 13 (19)  \\ 
 & ISO & 89 (95)  & 87 (94)  & 83 (90)  & 80 (81)  \\ 
 & Both & 10 (3)  & 12 (3)  & 15 (4)  & 7 (0)  \\ 
\hline
 \multirow{3}{*}{$75^{\circ}$} & NFW & 0 (0)  & 0 (1)  & 2 (3)  & 0 (13)  \\ 
 & ISO & 97 (100)  & 94 (98)  & 84 (96)  & 75 (84)  \\ 
 & Both & 3 (0)  & 6 (1)  & 14 (1)  & 25 (2)  \\ 
\hline
\end{tabular}
\end{center}
\end{table}

\begin{figure} \centering
  \includegraphics[width=\linewidth]{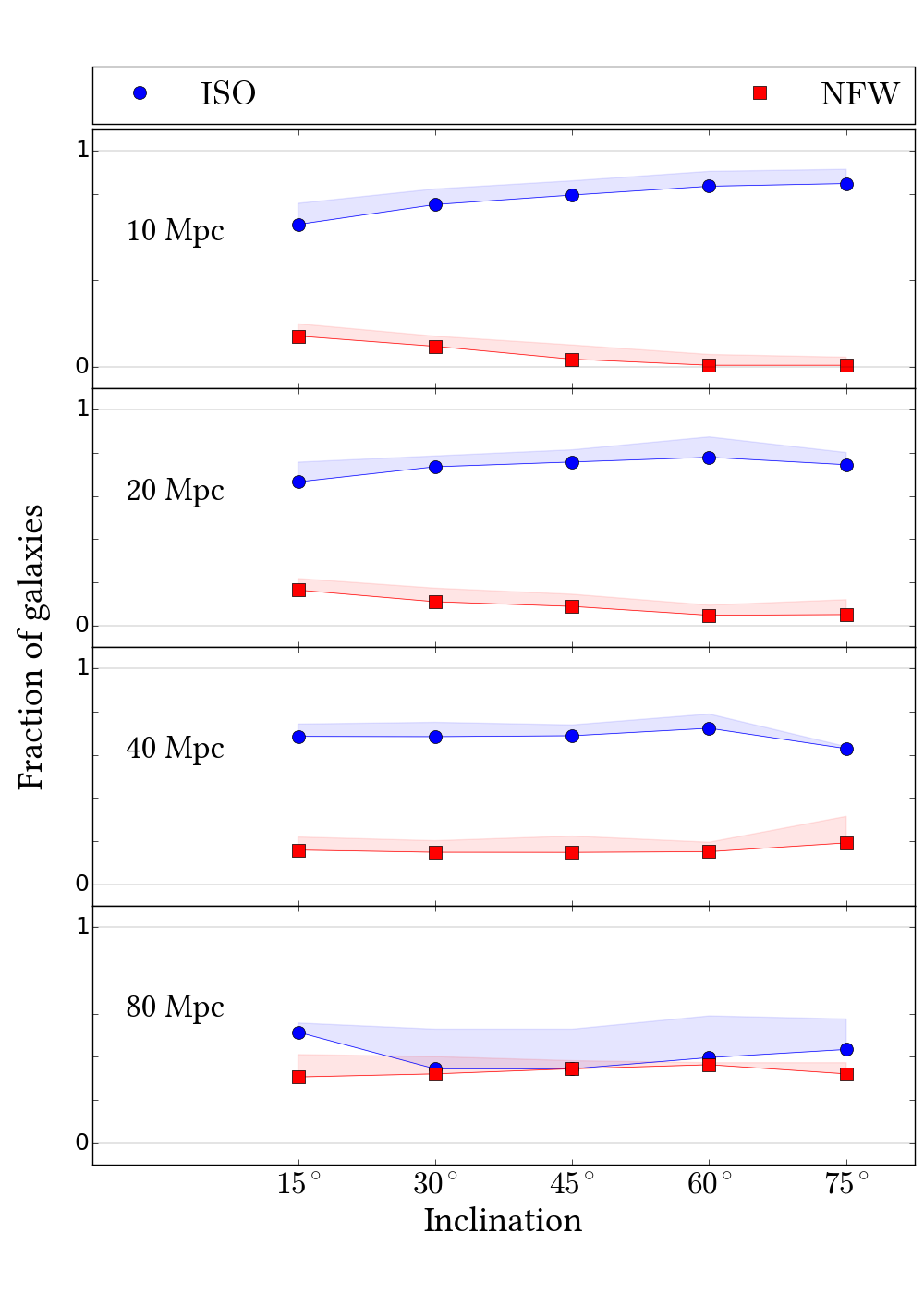}
  \caption{Fraction of mock rotation curves that are better described by the
    NFW/ISO models as a function of distance and inclination for the hybrid rotation curves
    using mock H~$\alpha$ long-slit data. The symbols and colours are the same
    as in Fig.~\ref{fig:X2_fractions}. Except for an assumed distance of 80 Mpc, the
    ISO model is incorrectly preferred in the majority of cases. For
    80 Mpc, of order half of the rotation curves are better fit with the cored
    ISO model despite the galaxies having cuspy profiles.}
\label{fig:LS_X2_fractions}
\end{figure}

In Fig.~\ref{fig:LS_X2_fractions} and Table~\ref{tab:X2_LS}, we
repeat the exercise for the long-slit data. At 10 Mpc ($\sim$100 pc inner spatial resolution), we see the same
trend with inclination as before, i.e. more-inclined galaxies appear
more cored. This tendency becomes weaker at 20 Mpc, and it does
not hold beyond that. Regarding the trend with distance, the
ambiguity between models increases more rapidly compared with when the 2D
velocity maps were used. In particular, when the galaxies are placed at 80 Mpc,
of order half the sample appears to contain cores and a major
fraction of the other half appears to contain cusps.
Finally, it is worth noting that these general results are
not biased by the inclusion of G1, i.e. they do not change
substantially if we exclude this simulation.

\begin{table}
\caption{Similar to Table~\ref{tab:X2} but for $V_{\rm ls}$, the hybrid
rotation curves using the mock H~$\alpha$ long-slit data,
classified according to the assumed distance and inclination.}
\label{tab:X2_LS}
\begin{center}
\begin{tabular}{cccccc}
\hline
\multicolumn{2}{c}{H~$\alpha$ PSF (pc)} & $\sim$100 & $\sim$200 & $\sim$400 & $\sim$800 \\
\hline
\multicolumn{2}{c}{D (Mpc)} & 10 & 20 & 40 & 80 \\
\hline
 \multirow{3}{*}{$15^{\circ}$} & NFW & 14 (20)  & 16 (22)  & 16 (22)  & 31 (41)  \\ 
 & ISO & 66 (76)  & 67 (76)  & 69 (74)  & 51 (56)  \\ 
 & Both & 20 (4)  & 17 (3)  & 16 (4)  & 18 (3)  \\ 
\hline
 \multirow{3}{*}{$30^{\circ}$} & NFW & 9 (14)  & 11 (17)  & 15 (20)  & 32 (40)  \\ 
 & ISO & 75 (82)  & 74 (79)  & 68 (75)  & 34 (53)  \\ 
 & Both & 15 (3)  & 15 (4)  & 17 (5)  & 33 (7)  \\ 
\hline
 \multirow{3}{*}{$45^{\circ}$} & NFW & 3 (10)  & 9 (14)  & 15 (22)  & 34 (38)  \\ 
 & ISO & 80 (86)  & 76 (81)  & 69 (74)  & 34 (53)  \\ 
 & Both & 17 (4)  & 15 (4)  & 16 (4)  & 31 (9)  \\ 
\hline
 \multirow{3}{*}{$60^{\circ}$} & NFW & 1 (6)  & 5 (9)  & 15 (19)  & 36 (37)  \\ 
 & ISO & 84 (91)  & 78 (87)  & 72 (79)  & 40 (59)  \\ 
 & Both & 16 (4)  & 17 (3)  & 13 (2)  & 24 (4)  \\ 
\hline
 \multirow{3}{*}{$75^{\circ}$} & NFW & 1 (4)  & 5 (12)  & 19 (31)  & 32 (37)  \\ 
 & ISO & 85 (92)  & 75 (80)  & 63 (64)  & 43 (58)  \\ 
 & Both & 14 (4)  & 20 (8)  & 18 (5)  & 25 (5)  \\ 
\hline
\end{tabular}
\end{center}
\end{table}

\section{Discussion}							%
\label{sec:Discussion}							%

So far we have presented strong evidence that rotation curve
fitting can yield qualitatively incorrect conclusions regarding the
inner curvature of the density profiles of
galactic DM haloes. Here, we discuss the origin of the errors based on
the differences amongst the theoretical velocity profiles, check the
dependence of the results on the radial range covered by the rotation
curves, and compare the coefficients of our best fits to a
collection of results from the literature to demonstrate that our
mock observed rotation curves are in fact representative of real galaxies.
We also compare our main findings against other works which analysed
synthetic observations from numerical simulations in the cusp-core context.

\subsection{Unraveling the illusion of DM cores}    		        %
\label{subsec:coreillusion}                                             %

\begin{figure*}
  \includegraphics[width=\linewidth]{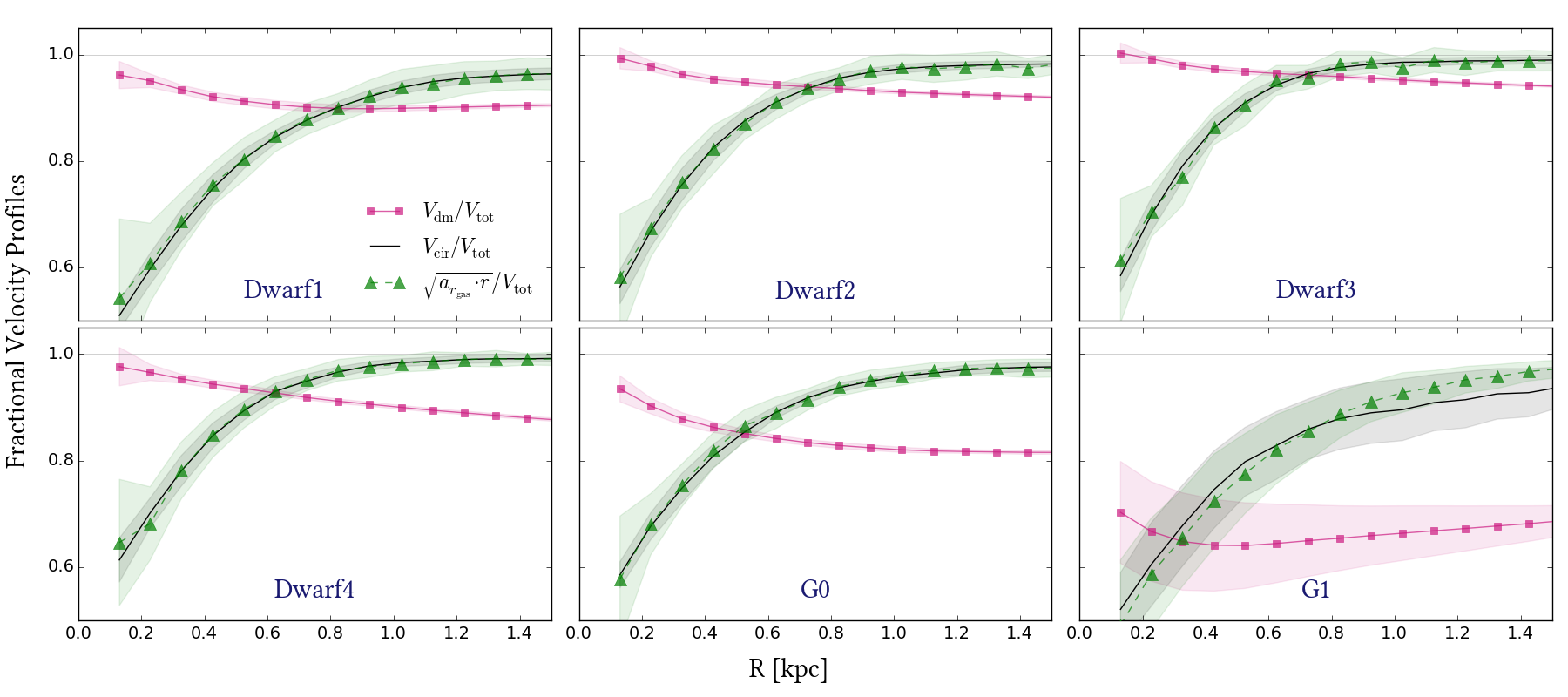}
  \caption{Theoretical velocity profiles within the central 1.5 kpc expressed as a fraction
    of $V_{\rm tot}$. Black solid line is for $V_{\rm cir}$ and magenta solid
    line with square markers for $V_{\rm dm}$.
    In addition, we show with green triangles joined by a dashed line the
    velocity profiles inferred from the radial acceleration of the
    gaseous particles in our simulations, i.e.
    $\sqrt{a_{r_{\rm gas}}\, r}$.
    Each line represents the mean taken over all snapshots, and the shaded 
    regions represent the 1-$\sigma$ scatter amongst snapshots.
    The fact that $V_{\rm dm}/V_{\rm tot}$ decreases with radius
    implies that the DM profile is cuspier than the total mass profile and thus the minimum
    disc assumption does not place an upper limit on the cuspiness of the DM profile.
    Moreover, both $V_{\rm cir}$ and the circular velocity corresponding to the radial acceleration
    experienced by the gas component, which are almost identical, are generally less than $V_{\rm tot}$ because
    of pressure support, and the fractional difference is greater at smaller radii.
    This systematic beaviour implies that the matter distribution inferred from $v_{\rm cir}$
    will look flatter than that inferred from $V_{\rm tot}$.}
\label{fig:residuals}
\end{figure*}

First, we focus on the real shapes of the different velocity profiles,
i.e. we refer to rotation curves of simulated galaxies placed at 10 Mpc,
for which the spatial resolution is high
($\sim 100$ pc). The explicit effect of poorer spatial
resolution is commented on in Sec.~\ref{subsec:distincl}. We recall
that 100 per cent of the $V_{\rm dm}$ profiles are better described by the
NFW model. Consequently, in principle, rotation curve fitting can
recognise a cuspy DM distribution from its true circular velocity
profile given sufficient spatial sampling ($\leq$~400 pc).

As for $V_{\rm tot}$, we have seen that NFW profiles are more favoured 
in the sample as a whole, although the ISO model fits the velocity
profiles from D4 and G0 better, and for G1, the two models
provide superior fits for similar fractions of the rotation curves.
Additionally, for D1, D2, and D3, the
confidence with which the rotation curve fitting detects the cuspy haloes, as judged
based on the $\chi_\nu^2$ values
of the fits to $V_{\rm tot}$, is somewhat reduced compared with $V_{\rm dm}$.
These results challenge the widespread assumption that the contribution
of baryons to the potential will tend to make the DM profile inferred under
the minimum disc assumption cuspier than
the true DM profile.\footnote{This is
usually expressed in other words, i.e. stating that the minimum disc
approximation puts an upper limit on the steepness of the DM halo
density profile \citep[e.g.][]{deBlok2001}.}
These results also imply that the minimum
disc approximation is unacceptable for some dwarf galaxies.
To better understand this result, in Fig.~\ref{fig:residuals} we plot the inner
velocity profiles expressed as fractions of $V_{\rm tot}$. 
Noting that the circular velocity profile of a spherical halo
is given by the enclosed mass as a function of radius, which can be
estimated from the density profile by integration, it can be shown that
$\rho_{\rm inner}\propto r^0$ cores result in linear inner circular velocity profiles,
$V_{\rm inner}\propto r^1$, whereas $\rho_{\rm inner}\propto r^{-1}$ NFW cusps lead to
$V_{\rm inner}\propto r^{0.5}$. In other words, cuspier density profiles produce
more curved velocity profiles characterized by smaller exponents, but always
inside the interval $\lbrack0.5,1\rbrack$ for the limits we are considering.
Because the fraction $V_{\rm dm}/V_{\rm tot}$ systematically decreases with radius, this
quotient has a negative exponent when expressed as a power law, which implies that the exponent of 
$V_{\rm tot}$ is larger (closer to 1) than that of $V_{\rm dm}$.
Consequently, the dark matter distributions inferred from
$V_{\rm tot}$ are less cuspy than those inferred from $V_{\rm dm}$.
In other words, the contribution of baryons to the potential tends to make
the circular velocity profile flatter rather than cuspier.
This result may not hold for all galaxies, but the salient point is that there is no
justification to claim an universal ability of the minimum disc
approximation to make the inferred dark matter density profiles appear cuspier than they truly are.
More likely, the resulting cuspiness
of $V_{\rm tot}$ will depend on the relative curvatures and
normalizations of the dark and luminous components. Also note that in
Fig.~\ref{fig:residuals}, the inner, negative slope of the 
$V_{\rm dm}/V_{\rm tot}$ profile is shallower in D1, D2, and D3 than in the other
simulations, and these simulated galaxies have the smallest baryonic-to-DM ratios
amongst our sample. Consequently, the inner curvatures of $V_{\rm dm}$ and $V_{\rm tot}$
must differ the less in these galaxies.

In the case of $V_{\rm cir}$, the rotation curve fitting analysis is completely
misleading because the cored ISO model provides better fits to most of
the rotation curves. Thus, the core-like shape that we infer from
the mock rotation curves is already imprinted in $V_{\rm cir}$ and is
therefore not (only) a consequence of observational errors (e.g. projection
effects). Fig.~\ref{fig:residuals} provides
insight into this issue. We see that in all galaxies,
$V_{\rm cir}$ is approximately $\sim 50$ per cent of $V_{\rm tot}$ in the very
center and progressively increases until it reaches a fractional
contribution of order unity just outside the first kiloparsec. This
systematic trend is opposite to the one that we found in the case of
$V_{\rm dm}$ and necessarily implies that the inner curvature of
$V_{\rm cir}$ is much less pronounced than that of $V_{\rm tot}$
(i.e. the mass distribution inferred from $V_{\rm cir}$ is less cuspy
than the true mass distribution).

Why is $V_{\rm cir}$ typically less than $V_{\rm tot}$, especially
at small radii? Our tests
indicate that it is the result of pressure support, i.e.~an effective
decrease in the radial acceleration because of an outward force of the gas
on itself, which is the result of a negative radial pressure gradient.
To illustrate this point, we include in
Fig.~\ref{fig:residuals} the circular velocity corresponding to the
radial acceleration experienced by the gas component,
i.e.~$\sqrt{a_{r_{\rm gas}}\, r}$, which closely corresponds to the
actual circular velocity of the gas, $V_{\rm cir}$. This correspondence demonstrates
that the gas component is in rotational equilibrium but also shows
that within the first kiloparsec, the radial acceleration experienced by
the gas particles is less than that experienced by the stellar ones
(which is used to define $V_{\rm tot}$). Because the overall gravitational
potential is the same for gas and stars, the only physical phenomenon
we identify that is capable of causing the difference is pressure support
associated with the high gas density in the galactic centres and the injection
of thermal energy from stars into the interstellar medium (ISM).
In Sec.~\ref{subsec:correction} we present further evidence in
this regard and address the (im)possibility of accurately correcting the
mock rotation curves for pressure support using observationally accessible
information.

Regarding the mock observations, we know that for galaxies
placed at 10 Mpc, the mock rotation curves
perfectly follow the theoretical $V_{\rm cir}$ curve
(Fig.~\ref{fig:Obs_RCs}); thus, their cored shapes simply reflect
the curvature of the true circular motions of the gas, as determined by
the full hydrodynamics and not only by gravity. Moreover, a
careful inspection of Fig.~\ref{fig:X2_fractions} reveals that the
gap between the ISO and NFW fractions is less for $V_{\rm ls}$
than for $V_{\rm kin}$. After a thorough review of the fitting
procedure, we concluded that this difference is a result of the size of the error
bars, which are larger for the long-slit observations than for the rotation curves 
obtained from the 2D velocity maps with {\sc kinemetry}. Consequently,
the individual points of the mock H~$\alpha$ long-slit rotation curves,
which determine the inner curvature of the profiles, have a smaller weight during
the fits than the outer points from the mock H$\,${\sc i} rotation curves, thus reducing the significance of
the core detections.

\subsection{Dependence on spatial resolution and inclination:
  beam-smearing and projection effects}%
\label{subsec:distincl}                                                 %
Recalling that our theoretical rotation curves for galaxies at $D > 10$ Mpc
are simply re-sampled versions of the 10-Mpc curves and noting that the NFW
model fits better all the $V_{\rm dm}$ profiles independent of
distance, we conclude that rotation curve fitting methods can distinguish the
signature of an ideal cuspy NFW profile even at a 
sampling rate of $\sim$400 pc in the rising part (1 arcsec
at 80 Mpc) with the first measurement at $\sim 0.9$ kpc, at least
for the $(c,V_{200})$ parameter space covered by our models.

The situation is a bit different for cores. 
We know from the former sections that the intrinsic shape of
$V_{\rm cir}$ is close to the ISO profile, which
is favoured in more than 80 per cent of the fits for galaxies placed
at 10 Mpc. However, the fraction of cores
steadily declines as the galaxies are placed further away,
with a simultaneous increase in the number of preferred NFW fits and
ambiguous cases. At 80 Mpc, the number of galaxies catalogued as cusps
is comparable to the number of galaxies catalogued as cores. 
This is partially explained by the small sizes of the fake cores
($\sim$1 kpc), whose signature is progressively washed out as the
sampling gets poorer and the first measured point moves to larger radii.
Yet we note that this is only true for D2, D3,
and mildly for G1. Fits to $V_{\rm cir}$ in the other galaxies
favour the ISO model at all distances, showing that the outcome
of the fits depends on more subtleties than just the sampling rate.

It is important to note that undersampling the curves is not sufficient
to fully model the impact of spatial resolution. Beam smearing, i.e.~the
smoothing of the velocity gradients in the rising part of the rotation curves
caused by averaging over large PSF/beam areas, is also a resolution effect.
The measured velocities are further lowered by projection effects
related to the fact that, in inclined discs of finite thickness,
any line of sight targeting an inner position mixes information
from tracers at larger radii, whose line-of-sight velocity component
is very small because of the inherent geometry of the rotating
disc \citep{Rhee2004}. Note that the volume
of gas crossed by a single telescope pointing increases as the spatial
resolution becomes coarser and as the galaxy inclination and disc thickness
are increased; thus, in order to
assess these entangled effects properly, we must confront our mock
kinematic observations with the theoretical velocity profiles.

Comparing the results from $V_{\rm kin}$ with those from $V_{\rm cir}$
(Fig.~\ref{fig:X2_fractions}), we see that they are very similar at
$D\leq 40$ Mpc, 
i.e.~the effect of beam smearing and projection effects is mild for
inner spatial resolutions $\leq400$ pc, though they effectively introduce
the dependence with inclination discussed in Sec.~\ref{subsec:RCfit_results}.
At 80 Mpc, in contrast, beam-smearing and projection
effects cause a significant difference between the results from $V_{\rm kin}$ and 
$V_{\rm cir}$, as a large fraction of
the mock observations are better
explained by a cored profile. In this case, the signature of the fake
core does not vanish with the poor sampling in any case because the additional
effects considerably lower the inner measured velocities, thereby reinforcing
the illusion of a DM core (Fig.~\ref{fig:Obs_RCs}).

As can be seen in Fig.~\ref{fig:kin_X2_fractions} and Table~\ref{tab:X2_obs},
the effect of distance on $V_{\rm ls}$ is to progressively reduce the
fraction of cases that are better fit by the ISO model, while augmenting
those which are better fit by the NFW profile. The increasing ambiguity,
which is more noticeable than in the case of $V_{\rm kin}$,
results from the larger size of the
H~$\alpha$ long-slit error bars in comparison to the H$\,${\sc i}
error bars reported by kinemetry and illustrates the sensibility of the
rotation curve fitting analyses to the specific details in the treatment
of the data. Using $V_{\rm ls}$, we only observe systematic effects with
inclination at 10 and 20 Mpc; we suspect they are strongly dependent on the
existence of inner measurements and their relative weights during the
fits.

\subsection{Dependence on the extent of the rotation curve}             %
\label{subsec:Npoints}                                                  %

\begin{figure}
  \includegraphics[width=\linewidth]{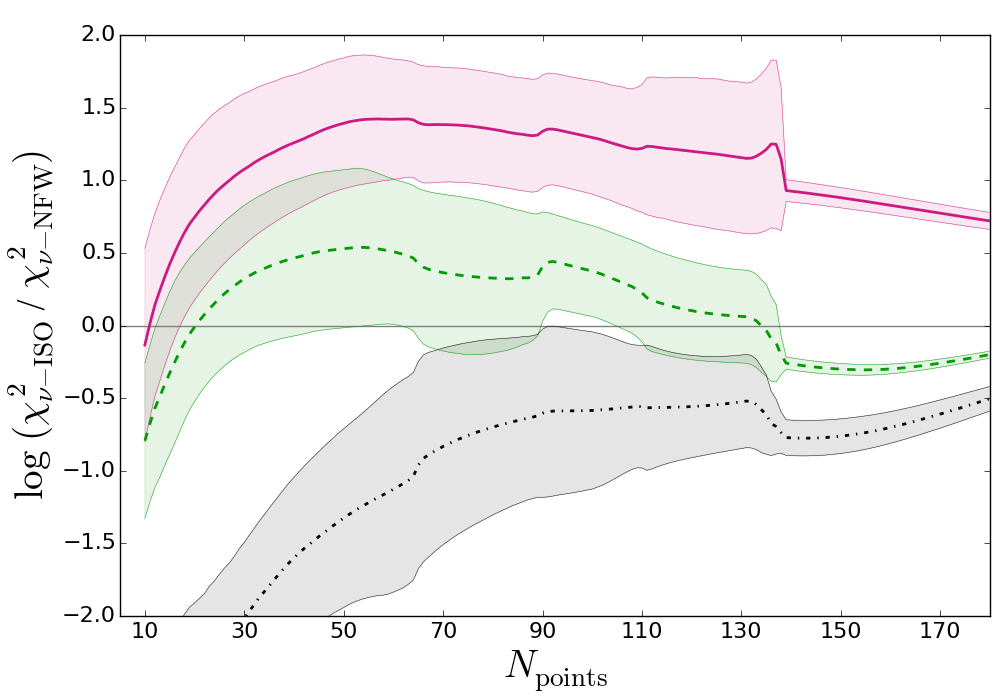}
  \caption{Dependence of the $\chi^2_\nu$ ratio on the extent of
    the rotation curves. $N_{\rm points}$ represents the number of points
    considered for the fit. The horizontal line represents equality
    between the models. We show the results for $V_{\rm dm}$ (magenta, solid line),
    $V_{\rm tot}$ (green, dashed line), and $V_{\rm cir}$ (black,
    dot-dashed line) for galaxies placed at 10 Mpc ($\sim$100 pc inner resolution).
    The reported lines denote the mean trends from all galaxies (or those reaching
    a given radial extent), whereas the shaded regions represent
    the 1-$\sigma$ scatter. Beyond 134 points the only contributor is D4,
    which is why the scatter is much smaller than at smaller radii.
    This plot demonstrates that the conclusion that
    $V_{\rm dm}$ and $V_{\rm tot}$ are better fit with the NFW model (in the sample as a whole), whereas
    $V_{\rm cir}$ (and the mock observed rotation curves, which are not shown) is
    better fit with the ISO model, is effectively independent of the truncation radius
    employed.}
\label{fig:X2_profile}
\end{figure}

In Fig.~\ref{fig:X2_profile}, we show the average value of the ratio
$\chi^2_{\nu_{\rm ISO}}/\chi^2_{\nu_{\rm NFW}}$ after truncating the 10-Mpc
theoretical velocity profiles at different radii and repeating the fits. We
quantify the extent of a curve by the number of points inside the
truncation radius, $N_{\rm points}$, which we vary from 10 to approximately
3.3 times the optical radius, roughly corresponding to 65 points in Dwarf1
and to 182 points in Dwarf4, the most extended galaxy in our sample.
According to Fig.~\ref{fig:X2_profile},
$V_{\rm dm}$ is always better represented by the NFW formula
independently of the truncation point. $V_{\rm tot}$ is also better
represented by the NFW profile but still marginally consistent with
the ISO model. In contrast, $V_{\rm cir}$ is better fit by the ISO profile for any
truncation radius, although for small values of $N_{\rm points}$, the
difference between models is more extreme because the fake inner core
dominates the fits. We do not plot equivalent profiles for the mock
observations to avoid overcrowding the figure, but we note that
they behave in the same manner as $V_{\rm cir}$. Thus, the
underestimation of the first points of the rotation curves determines
the outcome of the rotation curve fitting analysis (i.e. the detection
of spurious DM cores), regardless of the extent of the kinematic
observations. Consequently, our specific choice to truncate the velocity
profiles at approximately two times the optical radius does not affect the main
conclusions of this work. At larger distances, we observe exactly the
same qualitative results and the same shape for the
$\chi^2_{\nu_{\rm ISO}}/\chi^2_{\nu_{\rm NFW}}$ radial profiles.

We remade the main plots presented in Sec.~\ref{sec:Results}
but with the truncation radii of all the curves varying between
$\sim$1 and $\sim$3 times the optical radius, and we observed
only minor differences in the results. In particular,
we noted that when using more extended rotation curves, the fits to
$V_{\rm ls}$ less often preferred the ISO model than when the fiducial
truncation radius was used. Once again, this is due to the fact that in the
long-slit case, the H~$\alpha$ error bars are larger than the H$\,${\sc i} error
bars, such that the effect of the inner core on the fit is progressively diminished as more
outer points are used in the $\chi^2_\nu$ calculation.

\subsection{Comparison with the literature}				%
\label{subsec:literature}                                               %

\subsubsection{Comparison with observational results}

\begin{figure*}
\begin{minipage}[b]{0.48\linewidth}
\flushleft
  \includegraphics[width=\textwidth]{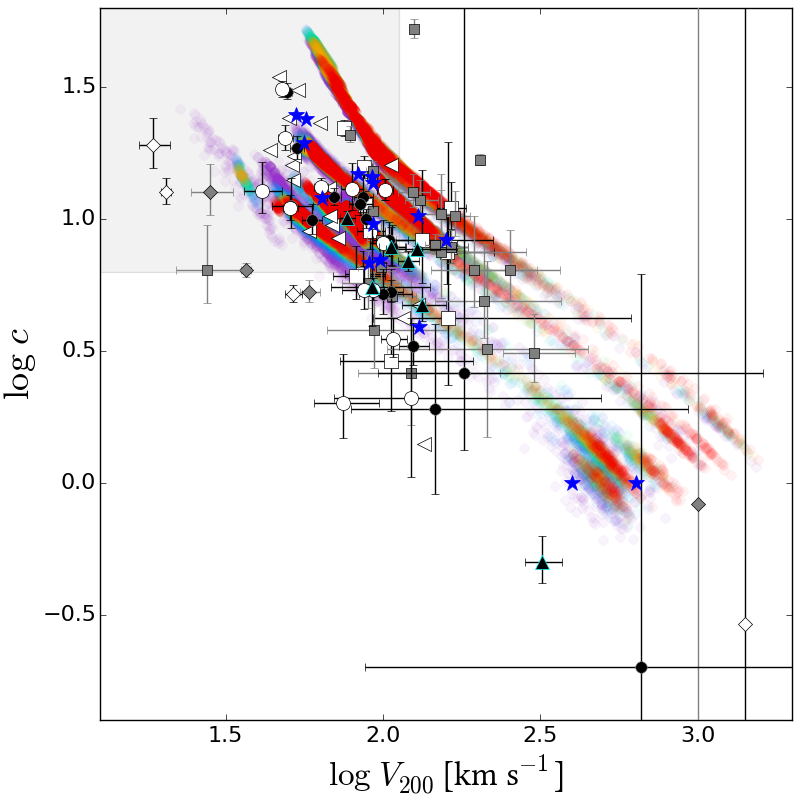}
\end{minipage}
\begin{minipage}[b]{0.48\linewidth}
  \centering
  \includegraphics[width=\textwidth]{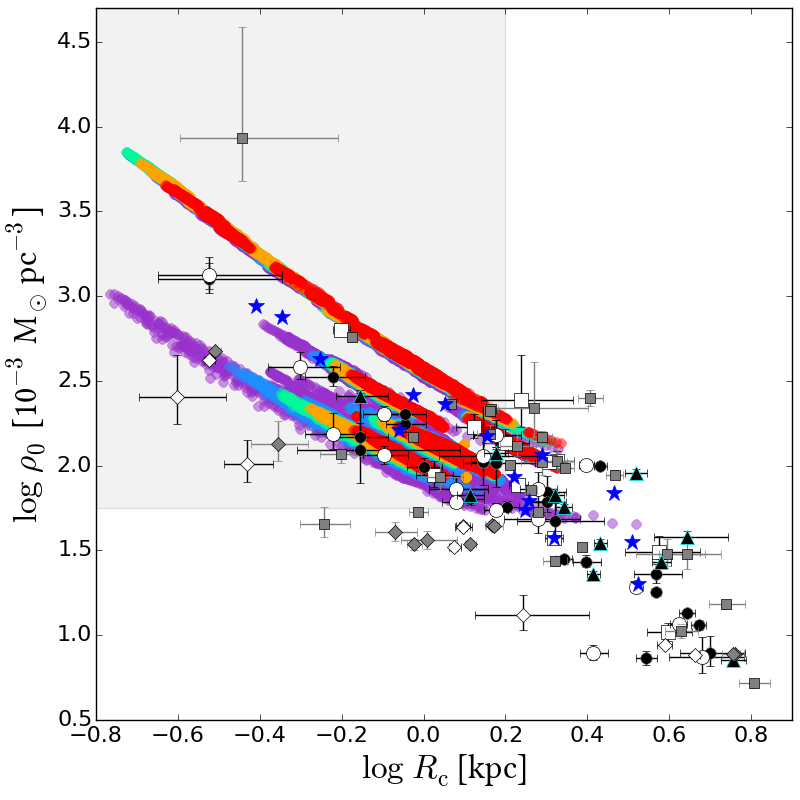}
\end{minipage}
\caption{Coefficients of the best fits to our mock rotation curves (NFW models
    on the left, ISO models on the right) compared with observational results
    from the literature.
    We used different extents for the mock rotation curves,
    ranging from 1 to 3 times the optical radius. Results are colour-coded
    by inclination as in Fig.~\ref{fig:X2_behav}.
    The following background markers represent observational results from
    the literature: circles \citep{deBlok2002}, triangles \citep{Kuzio2006},
    stars \citep{Swaters2003}, squares \citep{deBlok2001}, left-pointed triangles
    \citep{VandenBosch2001}, and diamonds \citep{Oh2011}. White symbols represent fits for which
    the gas contribution was subtracted from the observed rotation curve before
    the fitting (but a mass-to-light ratio of zero was still assumed for the stellar
    component). We include error bars whenever they are reported.
    Note that \citet{deBlok2001,deBlok2002,Oh2011} estimated ($c$,$V_{200}$)
    in alternative ways for some galaxies when the best-fit $c$ 
    value was negative or too close to zero; we do not include those galaxies in our comparisons.
    The positions of the simulated galaxies in the $(c,V_{200})$ plane depend on the
    distance to the galaxy and the radial extent of the mock data.
    At 10 Mpc, NFW fits to curves with fiducial extent ($2\times R_{\rm opt}$) lie in
    a subset of the parameter space spanned by observations,
    roughly ${\rm log}(V_{200}) \leq 2$ and ${\rm log}(c) \geq 0.8$, 
    as indicated with the gray square in the plot. Only when a larger diversity of
    extents and distances is allowed we see the NFW fits almost covering the full
    parameter space that observations do.
    A similar trend occurs for the ISO model, though it appears less dramatic. At 10 Mpc,
    fits to curves with the fiducial extent lie on the shaded gray region, and slightly
    larger, less-dense cores are only recovered when we try different extents and distances.
    Nevertheless, none of the fits to our mock data is able to reproduce the largest, least dense
    cores observed in some galaxies.
    Interestingly, as the ISO fits to our simulated data represent illusory cores,
    and they reproduce the correlation between $\rho_0$ and $R_c$ that has been
    obtained for real galaxies, this may be an artefact of the fitting process rather
    than a physical relation obeyed by real DM haloes.}
\label{fig:fit_coeff}
\end{figure*}

In Fig.~\ref{fig:fit_coeff} (left), we present the coefficients ($c$, $V_{200}$)
of the best NFW fit obtained for each velocity profile in
our study, along with a collection of observational results from the literature
based on high-resolution rotation curves.
We fit our rotation curves trying all possible extents from 1 to 3 times
the optical radius (typical of observations) in order to approximate
the stochasticity of real data.
This is equivalent to maximum extents of between 2.6 and 29 kpc and $N_{\rm points}$
between 42 and 180, depending on the size of the galaxy.

We note that the coefficients $c$ and $V_{200}$ from the fits appear
correlated both in our simulations and in the observational data and, more importantly,
that our fits are very consistent with the parameter space spanned by observations.
Strikingly, this is only true when our results consider different
truncation radii and different distances together, as the fits
to our fiducial mock data (truncated at 2 times the optical radius) at 10 Mpc cover
just a narrow subset of the whole parameter space, roughly
${\rm log}(V_{200}) \leq 2$ and ${\rm log}(c) \geq 0.8$,
as indicated with the shaded gray region in the plot.
The agreement between observations and the fits to our mock data is
enhanced by the inclusion of fits to less-extended versions
of our rotation curves, probably because many observational studies
use only H~$\alpha$ information, and they often do not reach
the same radial extent as our fiducial mock dataset.
This is related to another facet of the cusp-core problem pointed out by
observational studies, according to which the few reasonable
NFW fits to real galaxies prefer large values of $V_{200}$ and very
low concentrations, which are not compatible with the
mass-concentration relation predicted by cosmological simulations.
As our results illustrate, this might originate in the large dependence
of the observational analyses on subtle details and thus may not represent
a genuine discrepancy between $\Lambda$CDM cosmological simulations and observations.

We repeat the same experiment with the ISO model and present the results
and a comparison with the literature in Fig.~\ref{fig:fit_coeff} (right).
The shaded region encloses the subset of the parameter space
covered by the fits to galaxies at 10 Mpc using the fiducial extent
of the rotation curves ($2\times R_{\rm opt}$). The full set of results
includes fits to the mock rotation curves truncated at different radii,
ranging from 1 to 3 times $R_{\rm opt}$. Our fits populate a region
of the $\rho_0$-$R_c$ parameter space that is consistent with the
observational works, although they do not fully cover the low
$\rho_0$-high $R_c$ end of the observed relation. This time the
disagreement is only slightly alleviated by the use  of different
distances and truncation radii, i.e., despite the fact that the effects
we discuss in this work offer a plausible explanation for the cusp-core
discrepancy in terms of logarithmic density slopes, they are not
sufficient to explain the largest cores inferred from ISO fits to real
galaxies, associated with the large diversity of dwarf galaxy rotation
curves, as discussed by \citet{Oman2015}.
It is also interesting to note that our fits naturally
produce a correlation between $\rho_0$ and $R_c$ that agrees
extremely well with that evident from observations.
More specifically, both
``free'' parameters lie on a straight line with a slope close
to $-1$, which corresponds to a relation of
inverse proportionality (the product $\rho_0\times R_c$ is
approximately constant). This observation has been used to argue
that DM haloes are cored and exhibit a constant inner surface density
as some sort of universal property \citep[see][and references therein]
{Spano2008,Kormendy2016}; however, the physical meaning of this conclusion is
clearly put into question by our results\footnote{Note that 
\citet{Spano2008} and \citet{Kormendy2016} did not use the
pseudo-isothermal sphere model but a variant known as the non-singular
isothermal sphere, but \citet{Kormendy2016} found the core radius
and the central surface density estimated from one model 
to be proportional to the corresponding quantities in the other model,
so, our discussion regarding the alleged $\sim$constancy of
$\rho_0\times R_c$ is straightforward.}.

\subsubsection{Comparison with simulation works employing similar methodologies}

We now put our work into context by means of a brief comparison with
other studies that have investigated how well the inner shape of DM profiles
can be extracted from synthetic observations of simulated galaxies\footnote{Here we
refer to N-body numerical simulations, not to synthetic observations generated
via analytic models. The advantage of the former is that they are more
suitable for studying the three-dimensional dynamical evolution of galactic
systems in a self-consistent way.}. We start with the pioneering work of
\citet{Rhee2004}, which found that bars, small bulges, and projection
effects induce underestimates of the inner circular velocities
that can prevent the detection of cuspy DM haloes, making them look cored.
The authors mention that the individual effects lead to velocity errors of only
a few km s$^{-1}$, but their cumulative effect can result in qualitatively
incorrect conclusions. 

This result is consistent with our study, but we note
that the reason for this agreement is not straightforward because \citet{Rhee2004}
did not simulate the gaseous phase. The velocity underestimation they
observe is partially due to the dynamics of the stellar component
(the so-called asymmetric drift effect, see Sec.~\ref{subsubsec:pressVsdrift})
and to projection effects.
Fig. 19 of \citet{Rhee2004} suggests that in
axisymmetric systems, the inferred DM density profile slightly
flattens, although it does not present a strong core. In contrast, in the
presence of small bulges and bars, the combination of the former effects
and non-circular motions makes the inferred
density profiles artificially cored.

Interestingly, \citet{Rhee2004} show with convincing observational
evidence that stars and ionized gas rotate similarly in the inner
region, contrary to the general belief that the gas rotates faster
because it is a dynamically colder component.  The authors list possible
explanations for this behaviour that have been suggested in the literature, including
pressure support. In our simulations, the rotational
velocities of the stellar and the gaseous particles inside the first
kiloparsec are indeed similar, but not identical. In general we find
the gaseous rotation curves to be more lowered in the very centre, 
whereas beyond $\sim 0.5-1$ kpc, the stellar component
rotates more slowly. This means that the matter
distribution recovered from the gaseous $V_{\rm cir}$ will tend to be flatter,
explaining why we infer the presence of spurious cores even in systems
without bulges or bars. 

\citet{Valenzuela2007} further investigated the impact of systematic
effects and presented detailed mass modelling of the dwarf galaxies
NGC 3109 and NGC 6822. They compared collisionless
N-body simulations and full hydrodynamical runs with and without star
formation and stellar feedback and found that without feedback,
the gas rotational speed is similar to the true circular velocity of
the gravitational potential. In contrast, once injection of
thermal energy from stellar processes into the ISM
is included, the gaseous phase rotates considerably more slowly as a
result of pressure support. This effect creates a notable flattening
of the inferred density profile, which is further accentuated by the
presence of a small bar and projection effects.
\citet{Valenzuela2007} clearly illustrated how non-circular
motions related to small asymmetries in the baryonic matter
distribution, which might be easily overlooked, can bias the measured
rotational velocities towards core-like profiles.
An exhaustive analysis of asymmetric drift and
pressure support corrections is also presented; \citet{Valenzuela2007}
conclude that it is possible to recover the true circular velocity from the
observed gaseous rotation curve in their models, but this requires very careful,
detailed corrections for numerous systematic effects, even for
axisymmetric discs.  They also emphasize that observing low-rms
velocity dispersions ($\sim$10 km s$^{-1}$) in a galaxy does not mean
that such corrections can be safely neglected.
Overall, we find our results to be in very good agreement
with those of \citet{Valenzuela2007} and \citet{Rhee2004}.

\citet{Kuzio2011} analysed mock observations of simulated galaxies
embedded in cuspy, cored, and triaxial DM haloes.
They generated realistic mock observations by 
choosing the spatial resolution, spatial coverage, and inclination
according to the observational sample studied in
\citet{Kuzio2006} and \citet{Kuzio2008}.
\citet{Kuzio2011} found that typical
rotation curve analyses are able to efficiently recognise the
cuspiness of a DM halo and that it should be possible
to observe the characteristic signature of a DM cusp in velocity maps.
Particularly, in cored spherical haloes, the iso-velocity contours appear parallel in the
center of the galaxy, whereas in spherically symmetric cuspy DM haloes, the central velocity
contours are ``pinched''. The authors present examples of mock
velocity fields in which such features are evident.
They also found that stellar feedback had little effect on the
mock velocity fields and observed rotation curves.

Why these results are so different from those referenced earlier is
difficult to understand, especially noting that \citet{Kuzio2011} used the
same code as \citet{Valenzuela2007}. In the latter, stellar feedback
was proven to be an efficient source of pressure support; thus,
we speculate that the specific feedback implementation might have been
different in \citet{Kuzio2011}, leading to weaker pressurisation of the ISM.
Alternatively, the simulations of \citet{Kuzio2011}, which were initially composed only
of gas and DM, might not have formed a sufficient number of stars for stellar 
feedback to pressurize the ISM significantly. We highlight the fact that using a
different code with an independent stellar feedback implementation, our own
simulations, which represent real galaxy populations in terms of the amount
of gas and stars in the disc, reproduce the previous findings of
\citet{Valenzuela2007} regarding the role of  pressure support.
We also note that our velocity fields are not pinched in the centre; instead
the iso-velocity contours appear parallel there (see Fig.~\ref{fig:data_products}).

\citet{Oh2011b} analysed mock velocity fields and images of two dwarf
galaxies formed in cosmological simulations in the same manner
as a sample of dwarfs previously studied in \citet{Oh2011}.
The authors did not use the minimum disc approximation but rather
attempted to subtract the contributions of gas and stars from the
mock rotation curve. They were able to approximately
recover the true DM density distribution from
the mock observations using typical tilted-ring modelling, and they state
that pressure support effects and non-circular motions did not hamper
this recovery.
Nevertheless, one of their two recovered DM density profiles underestimated
the true density by a factor of 3 in the central region, but they associated
this to errors in the estimation of the gravitational potential of the gas.
\citet{Oh2011b} found good agreement between
their simulation results and observations. They
concluded that DM cores likely exist in the sample of real galaxies
and that the mechanisms which transform
DM cusps into cores in their cosmological simulations (i.e.~violent
outflows of gas caused by repetitive, intense starburst episodes)
are plausible. However, an important cautionary note is that
\citet{Oh2011b} only analysed mock observables when the
DM haloes were already cored; therefore, this work does not
lend insight into the possibility of identifying DM cusps via the same type of
analysis. Regardless of whether it is possible to infer the presence of cores
via rotation curve fitting, our results (and others mentioned above)
suggest that cusps may be mistaken for cores and thus the cusp-core
problem may be an illusion.

\subsection{Can pressure support effects be corrected?}
\label{subsec:correction}                                               %
Now that we have demonstrated the crucial role of pressure support in
cusp-core studies, we address the matter of whether its effect
may be corrected using observationally accessible information.

\subsubsection{Pressure support vs. asymmetric drift}
\label{subsubsec:pressVsdrift}                                          %

First, we would like to clarify the theoretical concepts
related to the physics of pressure support, as there is some
ambiguity in the literature. Our main concern is the use of the
terms \emph{pressure support} and \emph{asymmetric drift} as
interchangeable expressions despite the fact that they represent different physical properties.
This common mix up was recently pointed out by \citet{Dalcanton2010},
but it still deserves further attention, as we will demonstrate below.

One can start by stating what pressure support truly means and its
proper formulation. Consider the Euler momentum conservation equation
for a gas element in an external gravitational field,
\begin{equation}\label{eq:euler}
\frac{{\rm d}\boldsymbol{V}}{{\rm d}t} = \boldsymbol{a}_{\rm grav} - 
\frac{1}{\rho}\boldsymbol{\nabla} P .
\end{equation}

If this gas element moves on a circular orbit in the midplane of a
system with axial and vertical symmetries, the three terms of
equation (\ref{eq:euler}) are aligned in the radial direction, leading to
the scalar relation
\begin{equation}\label{eq:support}
\frac{v_\phi^2}{r} = \frac{v_c^2}{r}+\frac{1}{\rho}\frac{{\rm d} P}{{\rm d}r} ,
\end{equation}
where $v_\phi$ represents the rotational speed of the gas; $v_c$ is the
expected circular velocity from the gravitational potential; and the second
term on the right is the radial acceleration due to pressure gradients,
$\rho$ being the gas density and $P$ its pressure. The sign inversion on the
right hand side is because centripetal accelerations point inwards, i.e.~in
the negative $\boldsymbol{r}$ direction. Note that if pressure
falls off as a function of radius (as it is often the case in galactic
discs), the gradient becomes negative, making $v_\phi < v_c$; this
is pressure support. Also note that equation~(\ref{eq:support}) is valid
in the equatorial plane for any combination of spherical and disc-like
components, so the approximation of a spherical potential ($a_{r_{\rm grav}}
\approx \frac{GM(<r)}{r^2}$) used by \citet{Dalcanton2010} is not necessary.

The sources of pressure typically mentioned (if any) in the cusp-core
literature are thermal pressure and turbulence. Other possible sources,
such as magnetic fields or cosmic rays, are rarely mentioned, although they may be
important for the dynamics of the ISM  (\citealt{Boulares1990,Ferriere2001}; but
cf. \citealt{Su2016}).
\citet{Dalcanton2010} highlight that turbulence dominates the random
motions of the gas and for that reason the pressure relates to the
1-dimensional velocity dispersion as
\begin{equation}\label{eq:pressure}
P = \rho\sigma^2 .
\end{equation}

Substituting equation~(\ref{eq:pressure}) into equation~(\ref{eq:support}),
rearranging terms, and expressing the velocities in our notation, the
pressure support correction can be written as
\begin{equation}\label{eq:correc}
V_{\rm tot}^2 = V_{\rm cir}^2 - \sigma^2
\frac{{\rm d}\log{\left(\rho\sigma^2\right)}}{{\rm d}\log{r}} .
\end{equation}

Furthermore, assuming that the vertical structure of the disc does not
depend on radius, one can express the logarithmic derivative in terms of
the mass surface density of the gas $\Sigma$, which is the observable quantity
\begin{equation}\label{eq:correction}
V_{\rm tot}^2 = V_{\rm cir}^2 - \sigma^2 \frac{{\rm d}\log{\left(\Sigma\sigma^2\right)}}{{\rm d}\log{r}}.
\end{equation}

This is equivalent to equation (11) of \citet{Dalcanton2010}. We have
omitted the subindex $r$ in the velocity dispersion to emphasize that
turbulence is classically treated as an isotropic source of pressure\footnote{
This seems to be a good approximation when the dominant contribution comes
from microturbulence at much smaller scales than the size of the region
being considered (\citealt{MacLow2004}; but cf.
\citealt{Elmegreen2004})}.
\citet{Dalcanton2010} argue that the distinctive boundary conditions
in the vertical direction and in the equatorial plane likely invalidate
the assumption of isotropy in the velocity dispersion, which is
indeed a relevant concern for the macroscopic scales probed by
observations. Nevertheless, in practice, $\sigma$ is
in general treated as isotropic, mostly due to the difficulty of
disentagling its radial component from the observed projection along the
line of sight.

Asymmetric drift, on the other hand, is a phenomenon
  experienced by collisionless particles. Its origin is succinctly
  presented in \citet{Binney2008}, to which we refer the reader for
  details. In brief, the Jeans equations for a stellar population
  rotating in the equatorial plane of a smooth potential with axial
  and vertical symmetries yield
\begin{equation}\label{eq:asymmetric}
v_{\rm c}^2 - \overline{v_\phi}^2 = \sigma_\phi^2 - \overline{v_r^2} - 
\frac{r}{\nu}\frac{\partial{(\nu\overline{v_r^2})}}{\partial{r}} - 
r \frac{\partial{\left(\overline{v_r v_z}\right)}}{\partial{z}} ,
\end{equation}
where $v_{\rm c}$ represents the circular velocity associated
with the gravitational potential, $v_r$,
$v_\phi$, and $v_z$ are the three components of the actual stellar velocity, $\sigma_\phi$
is the azimuthal velocity dispersion, and $\nu$ is the probability of finding
a star at a certain position. The last quantity is proportional to the mass density
of the tracers and can be replaced by the mass surface density $\Sigma$
if the vertical structure of the stellar distribution does not vary with radius.
Assuming no net radial motions ($\overline{v_r^2}=\sigma_r^2$) and using our
notation, equation (\ref{eq:asymmetric}) becomes
\begin{equation}\label{eq:asymmetric2}
V_{\rm tot}^2=  V_{\rm cir}^2 + \sigma_\phi^2 - \sigma_r^2 - 
\sigma_r^2\frac{{\rm d}\log{(\Sigma\sigma_r^2)}}{{\rm d}\log{r}} - 
r \frac{\partial{\left(\overline{v_r v_z}\right)}}{\partial{z}}   .
\end{equation}

The similarity of equations (\ref{eq:asymmetric2}) and (\ref{eq:correction})
is remarkable, but further assumptions are needed if one wants to make
them look identical. These are (1) that there are no tilts in the velocity ellipsoid,
so the last term of equation (\ref{eq:asymmetric2}) vanishes, and (2) that
the velocity dispersion of the stars is isotropic. Neither of
these assumptions is straightforward, so, in spite of the fact that the
Euler and the Jeans equations can be obtained from the Boltzmann equation
in similar fashions, they are not physically equivalent. 

Some authors explicitly suggest that ionized gas might actually
experience the effects of asymmetric drift if it is clustered into
individual clouds which dynamically behave like collisionless particles
\citep[e.g.][]{Cinzano1999,Verdoes2000,Weijmans2008}. Invariably, all
the studies addressing this hypothesis are focused on early-type galaxies which
are dynamically hot, exhibiting internal velocity dispersions of hundreds
of km s$^{-1}$ that are hard to reconcile with thermal agitation and small-scale
turbulence.
On the other hand, late-type galaxies such as those regularly studied in the cusp-core literature
are comparatively cold systems, with reported H$\,${\sc i} velocity dispersions
of the order of $\sim$12 km s$^{-1}$. These small velocity dispersions can be
naturally explained by thermal and turbulent pressures. Therefore,
there is no need to invoke alternative scenarios.
Note also that a large fraction of the 21-cm emission comes from the warm
(6000$-$10000$^\circ$K) neutral gas in the disc, which, in contrast with the clumpy, cold
(100$^\circ$K) H$\,${\sc i} phase, is a diffuse medium with a large spatial extent
\citep{Ferriere2001}. The same applies for the warm ($\sim$8000$^\circ$K)
ionized medium responsible for the diffuse H~$\alpha$ emission
outside of H$\,${\sc ii} regions \citep{Cox2005}.
In light of these facts, it seems more likely that the dominant source of
non-centrifugal support in gaseous discs of late-type galaxies
is pressure support, not asymmetric drift. Incidentally, we note that 
equation~(\ref{eq:correction}), which describes pressure support, is the
universal recipe employed to correct gaseous rotation curves affected by
random motions in cusp-core studies, but it is often improperly
named ``\emph{the asymmetric drift correction}''. Sometimes equation
(\ref{eq:asymmetric2}) is invoked, but then the precise supplementary
approximations to reduce it to (\ref{eq:correction}) are invariably
assumed.

The trouble with conflating pressure support
and asymmetric drift is that this may lead to an erroneous assessment of
the magnitude of the correction. Note that in equation (\ref{eq:asymmetric2}),
all the quantities refer exclusively to the stellar population that is
being traced, while in equation (\ref{eq:correction}), they refer to the
gaseous medium as a whole. One may argue that according to the classic
pictures of the ISM \citep{Field1969,McKee1977}, the different phases
are expected to coexist in thermal pressure equilibrium; if so, tracing
a single phase (for instance, the warm H$\,${\sc i}) provides information
about the global thermal state of the ISM. Nevertheless, this scenario is
an oversimplification, and thermal pressure imbalances have been observed
even in the local ISM \citep{Bowyer1995,Berghofer1998}. This implies
either that pressure equilibrium must include non-thermal sources
that are dynamically important, such as turbulence, magnetic fields, and
cosmic rays \citep{Ferriere2001,Cox2005}, or that pressure imbalances may
exist locally as a result of complex events such as recent supernova explosions,
for instance. Therefore, a comprehensive assessment of the dynamical state
of the gas through multiple observational tracers is highly desirable.

Some authors have expressed concerns
  about the estimation of the gas velocity dispersion and the gas
  surface density profiles from a single tracer, as this may bias the
  measurements in regions with significant fractions of other ISM
  components \citep[e.g.][]{Simon2003,Dalcanton2010}. Regrettably,
  these biases are most often simply overlooked, and the majority of
  the cusp-core literature considering pressure support only vaguely
  states that random motions may provide support to the gas when they
  are comparable to the rotation velocity. A fraction of this subset
  of studies argues that pressure support corrections are known to be
  unimportant, increasing the magnitudes of rotation curves by only a
  few km s$^{-1}$, and do not
  attempt any sort of correction. The remaining fraction applies
  equation (\ref{eq:correction}) to the data at hand, normally from
  only a single tracer, often without considering the uncertainties in
  this correction. As we will demonstrate below, effective corrections
  for pressure support effects may be much more challenging than has
  been traditionally assumed.

\subsubsection{Ideal and realistic pressure support corrections}
\label{subsubsec:idealcorrections}                                      %

We here study the feasibility of pressure support corrections in our
models, first based on ideal theoretical measurements and then using
observationally accessible information, i.e. the surface density profiles
($\Sigma$) and the velocity dispersion profiles ($\sigma$), obtained
from the corresponding mock maps by taking azimuthal averages
along the same elliptical rings used to analyse the velocity maps.

\begin{figure}
  \includegraphics[width=\linewidth]{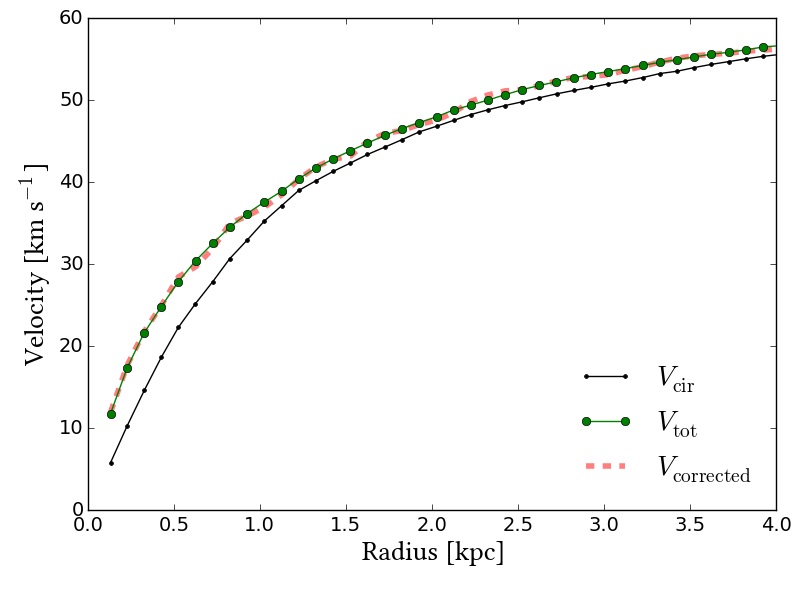}
  \caption{Theoretical verification of the pressure support
  effect in galaxy Dwarf1 at half the simulation time. The solid green line
  with circles represents $V_{\rm tot}$, the black solid line with points represents
  $V_{\rm cir}$, and the long-dashed red thick line in the background is the
  corrected version of $V_{\rm cir}$ after adding the term on the right-hand
  side of equation (\ref{eq:correc}). The correction exactly recovers
  $V_{\rm tot}$ from $V_{\rm cir}$, confirming pressure support as the cause
  of their difference.}
\label{fig:RC_corrected}
\end{figure}

We start by checking the validity of equation (\ref{eq:correc}) directly
in the simulations. We measure $\sigma$ and $\rho$ from all gas particles
in the equatorial plane using the same radial bins as for the theoretical
velocity profiles. The velocity dispersion is calculated using equation
(\ref{eq:sigmah1}), recalling that the code indirectly models the effect of the
turbulence induced by stellar and supernova feedback by employing an
effective equation of state that is stiffer than isothermal, thereby enhancing
the pressure and thus temperature in high-density, star-forming gas.
By means of equation (\ref{eq:correc}), we find
exceptional agreement between $V_{\rm tot}$ and the corrected version of
$V_{\rm cir}$ for all galaxies, as we illustrate with an example in
Fig.~\ref{fig:RC_corrected}. Moreover, the results of the NFW/ISO fits
to the corrected version of $V_{\rm cir}$ are in stunning agreement with 
those of $V_{\rm tot}$, as we report in Table~\ref{tab:X2_corrected}.
The correction is performed using the theoretical profiles 
in radial bins of 100-pc width. The fits reported at distances $D=20$
Mpc in Table \ref{tab:X2_corrected} are done after
resampling the corrected curves to the corresponding poorer spatial
resolution.

\begin{table}
\caption{Percentages of corrected rotation curves that are better represented
by the NFW or ISO models. For comparison, we include the results from $V_{\rm tot}$,
which were already presented in Table~\ref{tab:X2}. We do not correct the mock
observations at 40 or 80 Mpc because the H$\,${\sc i} spatial resolution is too low.}
\label{tab:X2_corrected}
\begin{center}
\begin{tabular}{cccc}
\hline
\multicolumn{2}{c}{H~$\alpha$ PSF (pc)} & $\sim$100 & $\sim$200 \\
\hline
\multicolumn{2}{c}{D (Mpc)} & 10 & 20 \\ 
\hline
 \multirow{3}{*}{$V_{tot}$} & NFW & 56 (61)  & 57 (61)  \\ 
 & ISO & 38 (39)  & 39 (39)  \\ 
 & Both & 6 (0)  & 4 (0)  \\  
\hline
 \multirow{3}{*}{$V_{\rm cir}$ (corrected)} & NFW & 57 (61)  & 58 (60)  \\ 
 & ISO & 37 (39)  & 36 (38)  \\ 
 & Both & 6 (0)  & 6 (2)  \\ 
\hline
 \multirow{3}{*}{$V_{kin}$ (corrected)} & NFW & 15 (21)  & 13 (20) \\ 
 & ISO & 67 (74)  & 69 (77)  \\  
 & Both & 18 (5)  & 17 (4)  \\ 
\hline
\end{tabular}
\end{center}
\end{table}

Notwithstanding, even though the pressure corrections work properly in the
ideal case, a number of difficulties prevent effective correction of
the mock observations. After applying equation (\ref{eq:correction}) to
our mock rotation curves using the \emph{observed} $\Sigma$, $\sigma$ profiles, the
cuspy dark matter haloes still appear disguised as cores to the rotation
curve fitting method, as we also show in Table~\ref{tab:X2_corrected}.
To understand this result, we need to discuss in detail several aspects
of the implementation of equation (\ref{eq:correction}).

\begin{figure}
\begin{minipage}{\linewidth}
\centering
\includegraphics[width=\linewidth]{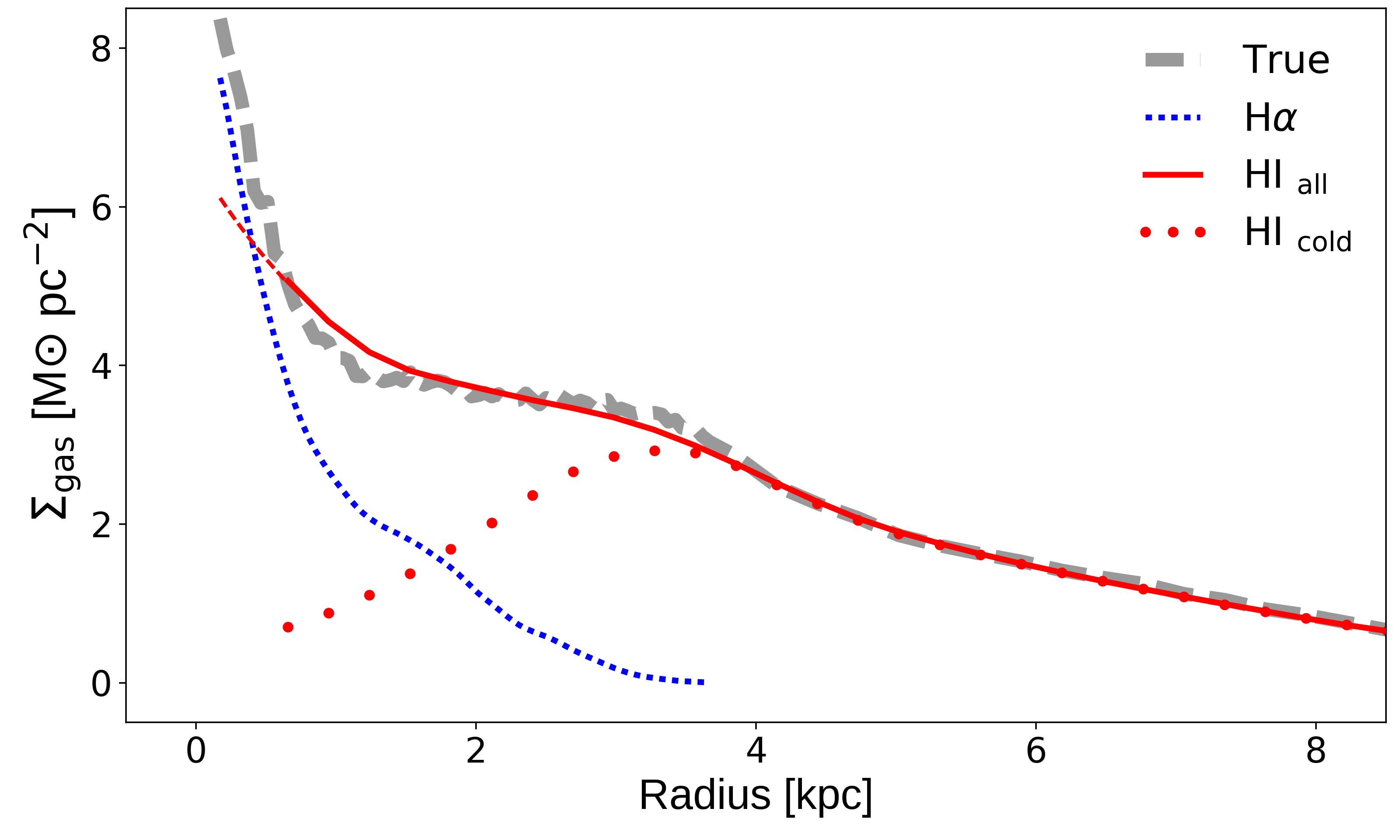}
\end{minipage}
\caption{Gas surface density profiles 
  for G0 at 3 Gyr, 45$^\circ$ inclination, and a distance of 10 Mpc. The thick
  dashed gray line represents the true mass surface density profile as measured
  from the simulation. The dotted blue line is the H~$\alpha$ surface
  brightness profile, re-scaled to have a central amplitude similar
  to that of the true mass profile. The solid red line shows our fiducial
  $\Sigma_{\rm HI}$
  mass surface density
  profile, extrapolated in the very center with an analytical function
  represented by a thin dashed line. The red dots represent the alternative
  observation from the cold gas alone, $\Sigma_{\rm HI-cold}$.
  The low spatial resolution smoothes the $\Sigma_{\rm HI}$ profile
  and leads to an underestimation of the central gas concentration.
  Observations from the cold H$\,${\sc i} phase would
  be more strongly biased in the center, where most
  of the gas in our simulations is hot and forming stars.}
\label{fig:dens_profiles}
\end{figure}

\begin{figure}
\begin{minipage}{\linewidth}
\centering
\includegraphics[width=\linewidth]{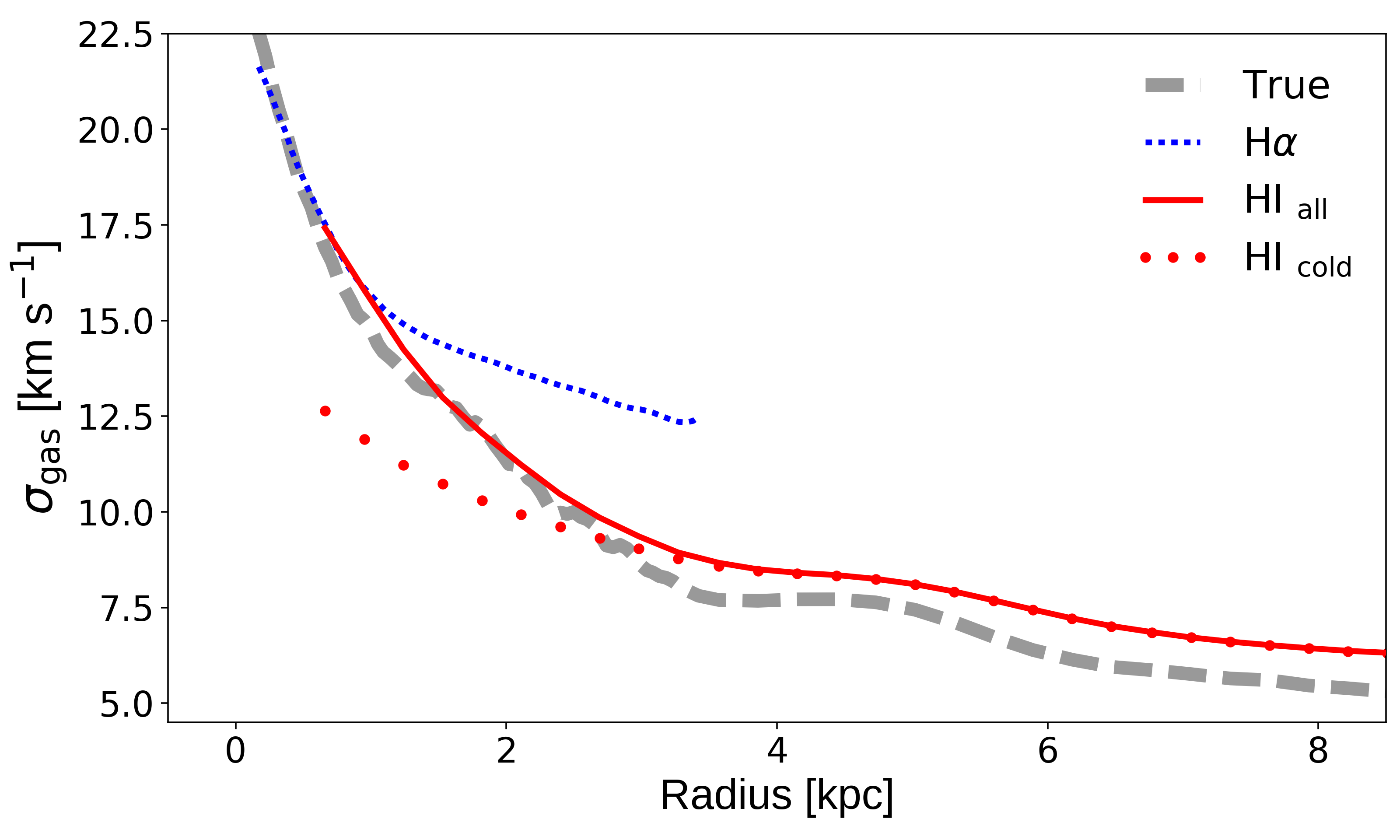}
\end{minipage}
\caption{Velocity dispersion profiles
  for G0 at 3 Gyr, 45$^\circ$ inclination, and a distance of 10 Mpc. Colours
  and symbols are the same as in Fig.~\ref{fig:dens_profiles}, identifying
  the true profile in the simulation and those inferred from our mock
  H~$\alpha$, H$\,${\sc i}, and ${\rm H\,I_{cold}}$ observations.
  This plot shows that in areas where hot and cold gas
  coexist, measurements based on a single tracer differ from the true
  values.}
\label{fig:disper_profiles}
\end{figure}

The first difficulty in correcting for pressure support
based on observed quantities is that virtually all the cusp-core studies
considering this effect use H$\,${\sc i} data alone, with very
few exceptions that use kinematic data from the ionized gas but
then lack the H$\,${\sc i} extension \citep{Simon2003,Chemin2016}.
Extrapolating a sort of ``standard'' observational correction for our
case, we correct the inner H~$\alpha$ velocities using the
$\sigma_{\raisebox{-1pt}{\tiny{\textrm{H}}$\alpha$}}$ velocity
dispersion and the outer H$\,${\sc i} velocities using
$\sigma_{\raisebox{-1pt}{\textrm{{\tiny HI}}}}$. For the gas surface
density profile, we take $\Sigma_{\rm HI}$ at all radii. Because the
H$\,${\sc i} spatial resolution is
coarser than that of the H~$\alpha$ observations, it is
necessary to extrapolate the $\Sigma_{\rm HI}$ profile in the center.
We achieve this by fitting a polynomial-plus-gaussian function to the
first 10 kpc of $\Sigma_{\raisebox{-1pt}{\textrm{{\tiny HI}}}}$,
checking that it follows the data well and that it exhibits reasonable
asymptotic behaviour in the inner region. Then, we calculate the
pressure support correction at each point of the rotation curves by
evaluating the derivatives in equation (\ref{eq:correction}) with a
finite-difference scheme. In Figs.~\ref{fig:dens_profiles} and
~\ref{fig:disper_profiles}, we present an example of the mass surface
density profile and of the velocity dispersion profile as traced by different
gas phases, along with the true quantities from all the gas in the
equatorial plane of the simulations. Because our mock H$\,${\sc i}
observations use all the gas particles regardless of their physical
state, we also show for comparison the profiles extracted from the cold
gas ($T < 10^{4}\phantom{.}^{\circ}$K) alone, which may be a better
proxy for real H$\,${\sc i} observations. We note that
unlike the H$\,${\sc i} case, our mock H~$\alpha$ emission
is not only proportional to the mass but also the star formation
rate of the emitting particles, making the conversion factor between
the intensity maps and the gaseous mass uncertain. For this
reason, we plot the mock H~$\alpha$
luminosity profile  in Fig.~\ref{fig:dens_profiles}, rescaled as necessary to facilitate the
comparison with the surface mass density profiles. These plots
are for a specific \emph{observed} galaxy, but similar trends are
observed for the rest of the sample.

Recalling that our mock H~$\alpha$ emission traces only the star-forming gas,
a joint analysis of Figs.~\ref{fig:dens_profiles} and
~\ref{fig:disper_profiles} reveals that there are three different regimes
in the disc. In the very inner region, within $\sim$0.6 kpc, virtually
all of the gas is forming stars, and both the 
$\Sigma_{\raisebox{-1pt}{\tiny{\textrm{H}}$\alpha$}}$ and
$\sigma_{\raisebox{-1pt}{\tiny{\textrm{H}}$\alpha$}}$ profiles follow
well the true theoretical quantities. At intermediate radii
(between $\sim$0.6 and $\sim$3.5 kpc), there is a mix of star-forming
and non star-forming gas, with the H~$\alpha$ emission steeply
going to zero while the cold gas density steadily grows. Taking into
account that the star-forming gas is systematically hotter than average,
this explains why in this intermediate regime,
$\sigma_{\raisebox{-1pt}{\tiny{\textrm{H}}$\alpha$}}$ increasingly
overestimates the true global velocity dispersion, while
$\sigma_{\raisebox{-1pt}{\tiny{\textrm{HI-cold}}}}$ stays below the
true $\sigma$ profile but systematically approaches it as the fraction
of cold, non-star forming gas starts to dominate.
We also note that in this intermediate region, the slope of
$\sigma_{\raisebox{-1pt}{\tiny{\textrm{H}}$\alpha$}}$, which will influence
the pressure correction term, is shallower than
that of the true total velocity dispersion profile.
Beyond $\sim$3.5 kpc, virtually all the gas in the simulation is
cold and passive, there is no mock H~$\alpha$ emission and the
measurements from the cold gas phase coincide with our fiducial H$\,${\sc i}
observations.
In general the fiducial mock H$\,${\sc i} observations follow the true
theoretical quantities well, though the low resolution makes the 
\emph{observed} profiles appear smoother and $\sigma$ slightly
overestimated. This smoothing propagates and introduces a bias 
during the extrapolation of $\Sigma_{\raisebox{-1pt}{\tiny{\textrm{HI}}}}$
to the very center, leading to an underestimation
of the true total surface density and also of its steepness.

\begin{figure}
\begin{minipage}{\linewidth}
\centering
\includegraphics[width=\linewidth]{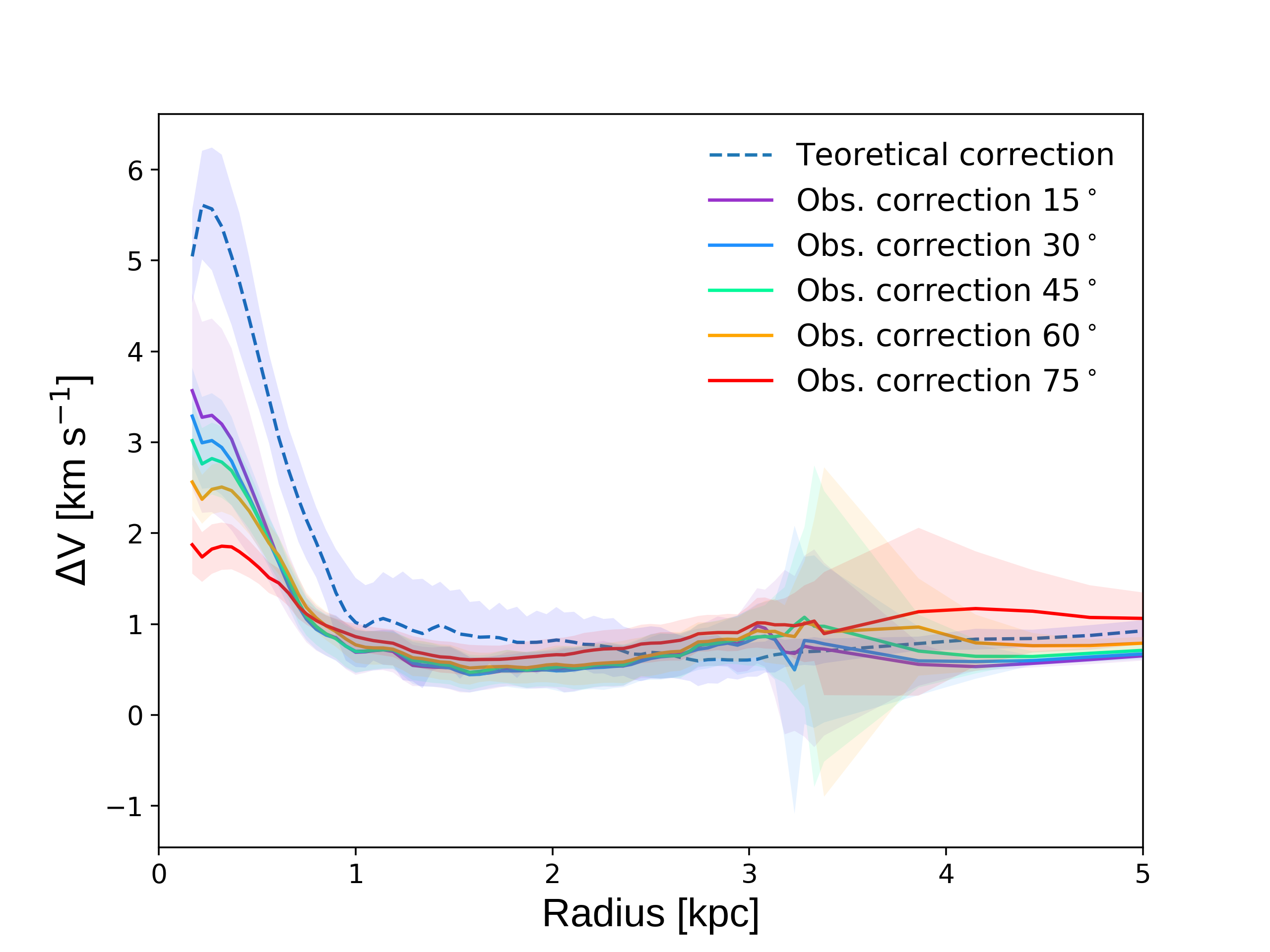}
\end{minipage}
\caption{Theoretical and observational pressure support corrections for
  galaxy Dwarf2 viewed at 10 Mpc. We plot the net velocity excess
  $\Delta V$ to be added to the observed rotation curves (or to
  $V_{\rm cir}$) to get their corrected versions. Solid lines represent
  the mean correction as a function of radius, and the shaded regions
  enclose the 1-$\sigma$ scatter of the profiles. The observational
  corrections are colour-coded by inclination as indicated in the
  legend. This plot
  shows that observational corrections underestimate the theoretical
  correction as a result of biased estimates for $\Sigma$ and $\sigma$;
  even though the difference seems small, it prevents faithful
  recovery of the inner curvature of the \emph{observed} rotation
  curves, which still appear more compatible with the ISO model
  after the attempted correction.}
\label{fig:corrections_profile}
\end{figure}

We have shown that there are biases in both the amplitude and the slope
of the observationally inferred $\Sigma$ and $\sigma$ profiles at different
radial ranges and from different tracers. Regarding posterior
cusp-core analyses, in which direction will these biases act? The answer
depends largely on the logarithmic slope of the product $\Sigma\sigma^2$,
i.e. on its curvature, which can not be simply inferred from our
qualitative analysis. After evaluating equations (\ref{eq:correc})
and (\ref{eq:correction}) we found that, for our particular experiment
and for the choices we made, the observational correction exhibits
the right radial profile but it falls behind the theoretical expected
value. We show this in Fig.~\ref{fig:corrections_profile}, where
we present the net difference between the corrected velocity profiles
and the original ones, $\Delta V$, as a function of radius and inclination.
The mean difference between the theoretical and the observational
corrections is never larger than 4 km s$^{-1}$, falling between 1 and 2 km
s$^{-1}$ over most of the first kiloparsec. However, this small difference
is enough for the intended observational correction to fail because
the curvature of the ``corrected'' rotation curves is still
more compatible with the ISO model in at least 67\% of the sample
(see Table~\ref{tab:X2_corrected}).
In spite of the fact that the tension is slightly less critical than
before the correction, the inferences from the rotation curve fitting
continue to be misleading compared with those from $V_{\rm tot}$,
largely overestimating the fraction of ``observed'' cores because of the
inaccurate pressure support corrections.

Fig.~\ref{fig:corrections_profile} and Table~\ref{tab:corr_inclD}
also demonstrate that the magnitude of the pressure support is more
underestimated, i.e. the correction is less effective, in more inclined discs.
This is as a result of projection effects and the low spatial resolution,
which make the $\Sigma$ and $\sigma$ profiles flatter, thus lowering
the inferred logarithmic slope of the product $\Sigma\sigma^2$.
We do not correct the mock observations at 40 or 80 Mpc because the
observables are extremely biased by the effect of the spatial resolution.
Particularly, note that at 40 Mpc, the first $\Sigma_{\rm HI}$ data point
lies at 2.4 kpc, ruling out any possibility of performing an accurate
correction inside the first kiloparsec.

\begin{table}
\caption{Percentages of rotation curves extracted from the 2D maps
that are better represented by the NFW or ISO models after the
observational pressure support corrections.}
\label{tab:corr_inclD}
\begin{center}
\begin{tabular}{cccc} 
\hline
\multicolumn{2}{c}{H~$\alpha$ PSF (pc)} & $\sim$100 & $\sim$200 \\ 
\hline
\multicolumn{2}{c}{D (Mpc)} & 10 & 20 \\ 
\hline
 \multirow{3}{*}{$15^{\circ}$} & NFW & 22 (31)  & 19 (28) \\ 
 & ISO & 55 (63)  & 57 (66)  \\ 
 & Both & 22 (6)  & 24 (6)  \\ 
\hline
 \multirow{3}{*}{$30^{\circ}$} & NFW & 24 (31)  & 20 (28)  \\ 
 & ISO & 57 (64)  & 61 (69)  \\ 
 & Both & 19 (5)  & 19 (3) \\ 
\hline
 \multirow{3}{*}{$45^{\circ}$} & NFW & 19 (26)  & 17 (24) \\ 
 & ISO & 61 (70)  & 63 (71)  \\ 
 & Both & 20 (4)  & 20 (5)  \\ 
\hline
 \multirow{3}{*}{$60^{\circ}$} & NFW & 8 (15)  & 8 (14) \\ 
 & ISO & 70 (78)  & 74 (82) \\ 
 & Both & 22 (7)  & 18 (4)  \\ 
\hline
 \multirow{3}{*}{$75^{\circ}$} & NFW & 1 (3)  & 0 (0)  \\ 
 & ISO & 93 (96)  & 98 (99)  \\ 
 & Both & 6 (1)  & 2 (1)  \\ 
\hline
\end{tabular}
\end{center}
\end{table}

It is very interesting to note that the observationally inferred
$\Delta V$ values peak at $\sim$3.5 km s$^{-1}$ and quickly drop to
$\sim$1 km s$^{-1}$, in agreement with typical corrections estimated
from observations, which are sometimes interpreted as a reflection of
the insignificant role of pressure support. This conviction is often
supported by the observed low velocity dispersions, typically
$\sim$10 km s$^{-1}$. However, note that this estimate
regularly comes from H$\,${\sc i} data alone, and that in general, the ionized gas
is expected to have a larger velocity dispersion, one of the reasons
why it tends to form thicker discs than the neutral gas \citep{Fathi2007}.
Fig~\ref{fig:disper_profiles} reveals that the cold dynamical component
in our simulations looks similar to many H$\,${\sc i} observations in
this regard, peaking at $\sim$13 km s$^{-1}$ and exhibiting a smooth
radial gradient; however, we know that this tracer is extremely biased
relative to the full velocity dispersion of the whole gas, which
is the necessary quantity for accurate pressure support 
corrections\footnote{Interestingly, at inclinations $\leq30^\circ$,
the $\sigma_{\raisebox{-1pt}{\tiny{\textrm{HI-cold}}}}$ profile is
even flatter and stays below 10 km s$^{-1}$.}.
The true value of $\sigma$ in our simulations peaks at
$\sim$25 km s$^{-1}$ and steeply decreases, dropping to $\sim$10 km s$^{-1}$
at $\sim$2 kpc. This is compatible with the few reported H~$\alpha$
observations we could find in cusp-core works. 
For example, \citet{Simon2003} observed a linewidth of 34 km s$^{-1}$
for the dwarf spiral NGC 2976, \citet{Epinat2010} mentions an
average velocity dispersion of 24 km s$^{-1}$ for a local sample of
153 nearby disc galaxies of mixed morphological types extracted from
the Gassendi H-Alpha survey of SPirals, GHASP \citep{Epinat2008,Epinat2008b}, and \citet{Chemin2016}
found a velocity dispersion profile peaking at 25 km s$^{-1}$ and
dropping to $\sim$20 km s$^{-1}$ at 2 kpc for the grand-design spiral M99.

Note also that according to Fig.~\ref{fig:dens_profiles}, the cold gas
strongly underestimates the true gas density in the center. Moreover,
the inferred $\Sigma_{\rm HI-cold}$ profile has a positive slope there,
which would partially reverse the sign of the pressure support correction
term if it were inserted in equation (\ref{eq:correction}).
Even though this modelling may be too
simplistic to explain the complexity of real H$\,${\sc i} observations,
it is interesting to note that central ``holes'' in the $\Sigma_{\rm HI}$
profiles are not rare, and they are treated in different ways when performing
pressure support corrections; some authors use $\Sigma_{\rm HI}$
as is, while others extrapolate the external exponential disc behaviour
to the center to try to compensate for the ionized
and molecular hydrogen mass contributions, which are very difficult to
assess.

It is not absolutely clear if the gas in our simulations may be
  over-pressurized compared with real systems, but we consider that
  the very good agreement between our mock data and real observations,
  as well as the agreement in the best-fit coefficients presented in
  Figure \ref{fig:fit_coeff}, should motivate further scientific
  discussion and careful review of some observational
  results. Independent of the possible differences between our
  models and real dwarf irregulars and LSBs, the exercise performed
  here suffices to demonstrate the intrinsic difficulties in
  properly assessing the effect of pressure support (amongst others),
  and it underlines the sensitivity of rotation curve
  fitting methods to very small errors or biases in the central parts of
  galaxies.

\section{Conclusions}
\label{sec:Conclusions}

Our results demonstrate that pressure support effects can easily make
DM cusps appear as cores in kinematic observations. Small
errors of $\sim5$ km s$^{-1}$ within the central kiloparsec are sufficient to
completely remove the signature of a DM cusp if they coherently decrease
the measured velocity. Thus, not correcting for pressure support can
be catastrophic in the cusp-core context, even for high-resolution
data generated from perfectly symmetric rotating discs. We highlight the fragility
of this type of rotation curve analysis: small errors can lead to qualitatively
incorrect conclusions regarding the shape of the inner DM profile.
Because multiple sources of errors (e.g. beam-smearing, non-circular motions,
small bulges, and projection effects) act in a similar manner as
pressure support (i.e. they tend to cause the observed circular velocity
to underestimate the true circular velocity in the center),
even if the amount of pressure support present in our simulations
differs in detail from reality, our main conclusion would still hold.

Strikingly, increasing the spatial resolution does not lead to more
reliable conclusions. Instead, the ISO
model is preferred more often when the simulated galaxies
are placed at smaller distances because the
signature of the fake core is better sampled. We also note that
the coefficients of the best-fitting NFW profile strongly depend
on the spatial extent of the rotation
curve. Best fits to data that do not extend into the flat part of
the rotation curve
(which is often the case for H~$\alpha$-only data) tend to prefer larger
$V_{200}$ values and lower concentrations. Our data are very consistent with
the literature once we account for the different rotation curve truncation radii
employed; this agreement pre-empts criticism regarding
possible inconsistency between the coefficients of the NFW fits of real galaxies
and the $\Lambda$CDM mass-concentration relation predicted by
N-body simulations.

Our best ISO fits also lie in the region of parameter space spanned
by the results of observational studies and naturally reproduce the observed
inverse proportionality between $\rho_0$ and $R_c$. This result
suggests that this correlation may be an artefact of the fitting process
rather than a real physical property of DM haloes.
It is interesting to note that beam-smearing can
dramatically impact H~$\alpha$ rotation curves of galaxies at $\sim$80
Mpc. This is not a surprise because \emph{linear} resolution depends
not only on the \emph{angular} resolution but also on the distance at
which a galaxy is located. Nonetheless, this result
must be emphasized because the high angular resolution of optical data
is often interpreted as providing sufficient protection against
beam-smearing. This is evidenced by the large amount of
galaxies more distant than 50 and even 100 Mpc that have been employed in
some cusp-core studies. Projection effects also play an important
role, making false detections of cores more likely in galaxies at high
inclinations, especially at larger distances.

Our model galaxies do not support the widely accepted claim
that the minimum disc approximation yields an upper
limit on the true steepness of a dark matter halo density profile.
In all our galaxies, whose properties are carefully modeled after observations
of real dwarf galaxies in the local Universe, the addition of baryons
to the gravitational potential of the dark matter haloes makes the true circular
velocity profiles flatter rather than cuspier. Thus, the minimum disc approximation
would cause one to infer that the dark matter profile is flatter than it actually is.

As noted above, not correcting for pressure support can
result in erroneous conclusions regarding the inner DM profile shape.
After carefully defining pressure support and distinguishing it from
asymmetric drift, we have examined whether it is possible to correct
for pressure support using observables generated from our simulations.
We find that even with the data available from the highest-quality
cusp-core studies, it is extremely challenging -- if not impossible --
to fully correct for pressure support, and, as noted above, even small
errors of a few km s$^{-1}$ can cause DM cusps to be disguised as
cores.

The analysis presented here has highlighted the difficulties involved
in reliably inferring the central dark matter structure in a rotation
curve analysis of observational data. In particular, we have shown
that it is comparatively easy to mistake a DM cusp for a DM core,
even with high-quality data. This certainly suggests that previous
observational claims of core detections need to be taken with a grain
of salt and should be followed-up further.

Our simulations have limitations, of course, the most important ones being that
they are idealised compound galaxy models that lack a self-consistent
cosmological context and that they do not include explicit supernova
`blast wave' feedback, which some simulations have suggested can
transform cusps into cores. It would thus be interesting to repeat a
similar analysis with galaxies extracted from full cosmological
hydrodynamic simulations of galaxy formation, which now have
sufficiently high resolution to study the cusp-core problem
\citep[e.g.][]{Vogelsberger2014, Marinacci2014,Hopkins2014,Onorbe2015,Chan2015,Schaye2015,Sawala2016}.
Such models could then also account for perturbations in the circular
motions of the gas due to central black holes, galaxy encounters, and
cosmological torques. However, we stress that the idealised nature
of our simulations (which contain strong, unperturbed NFW cusps by construction
and, with the exception of G1, are highly axisymmetric)
should make it \emph{easier} to detect cusps than would be the
case if fully cosmological simulations with explicit stellar feedback
were employed. For this reason, in reality, it may actually be
\emph{more difficult} to observationally identify cusps via rotation curve
fitting than suggested by our work.

\acknowledgments
{\small JCBP} and {\small CMdO} would like to thank Funda\c{c}\~{a}o de Amparo
\'{a} Pesquisa do Estado de S\~{a}o Paulo ({\small FAPESP}) for support fellowships
2012/21375-0, 2011/21678-0, and projects  2011/51680-6 and 2014/07684-5.
{\small CCH} is grateful to the Gordon and Betty Moore
Foundation for financial support. We thank Philippe Amram, Manoj
Kaplinghat, Simon White, and Giuliano Lorio for useful discussions
that helped us improve this manuscript. We also thank the referee,
Kyle Oman, for carefully reading the manuscript and providing
many useful comments that led us to improve the work.
{\small JCBP} deeply acknowledges the hospitality of  the Heidelberg
Institute for Theoretical Studies ({\small HITS}), and Laboratoire
d'Astrophysique de Marseille ({\small LAM}), where part of this work was done.
The Flatiron Institute is supported by the Simons Foundation.\\

\footnotesize{
\bibliography{mybib}
}
\label{lastpage}

\end{document}